\RequirePackage{lineno}   
\documentclass[12pt]{article}
\usepackage{epsfig}
\usepackage{epstopdf}
\usepackage{graphicx}
\usepackage{amsmath}
\usepackage{hhline}
\usepackage{amssymb}
\usepackage{times}
\usepackage{cite}
\usepackage{rotate}
\usepackage{multirow}

\usepackage{captcont}

\usepackage{subfigure}
\usepackage[pagebackref=false,colorlinks=true]{hyperref} 

\newlength{\dinwidth}
\newlength{\dinmargin}
\begin{document}  
\renewcommand{\arraystretch}{1.1}
\newcommand{\pom}{{I\!\!P}}
\newcommand{\reg}{{I\!\!R}}
\newcommand{\slowpi}{\pi_{\mathit{slow}}}
\newcommand{\fiidiii}{F_2^{D(3)}}
\newcommand{\fiidiiiarg}{\fiidiii\,(\beta,\,Q^2,\,x)}
\newcommand{\n}{1.19\pm 0.06 (stat.) \pm0.07 (syst.)}
\newcommand{\nz}{1.30\pm 0.08 (stat.)^{+0.08}_{-0.14} (syst.)}
\newcommand{\fiidiiiful}{F_2^{D(4)}\,(\beta,\,Q^2,\,x,\,t)}
\newcommand{\fiipom}{\tilde F_2^D}
\newcommand{\ALPHA}{1.10\pm0.03 (stat.) \pm0.04 (syst.)}
\newcommand{\ALPHAZ}{1.15\pm0.04 (stat.)^{+0.04}_{-0.07} (syst.)}
\newcommand{\fiipomarg}{\fiipom\,(\beta,\,Q^2)}
\newcommand{\pomflux}{f_{\pom / p}}
\newcommand{\nxpom}{1.19\pm 0.06 (stat.) \pm0.07 (syst.)}
\newcommand {\gapprox}
   {\raisebox{-0.7ex}{$\stackrel {\textstyle>}{\sim}$}}
\newcommand {\lapprox}
   {\raisebox{-0.7ex}{$\stackrel {\textstyle<}{\sim}$}}
\def\gsim{\,\lower.25ex\hbox{$\scriptstyle\sim$}\kern-1.30ex%
\raise 0.55ex\hbox{$\scriptstyle >$}\,}
\def\lsim{\,\lower.25ex\hbox{$\scriptstyle\sim$}\kern-1.30ex%
\raise 0.55ex\hbox{$\scriptstyle <$}\,}
\newcommand{\pomfluxarg}{f_{\pom / p}\,(x_\pom)}
\newcommand{\dsf}{\mbox{$F_2^{D(3)}$}}
\newcommand{\dsfva}{\mbox{$F_2^{D(3)}(\beta,Q^2,x_{I\!\!P})$}}
\newcommand{\dsfvb}{\mbox{$F_2^{D(3)}(\beta,Q^2,x)$}}
\newcommand{\dsfpom}{$F_2^{I\!\!P}$}
\newcommand{\gap}{\stackrel{>}{\sim}}
\newcommand{\lap}{\stackrel{<}{\sim}}
\newcommand{\fem}{$F_2^{em}$}
\newcommand{\tsnmp}{$\tilde{\sigma}_{NC}(e^{\mp})$}
\newcommand{\tsnm}{$\tilde{\sigma}_{NC}(e^-)$}
\newcommand{\tsnp}{$\tilde{\sigma}_{NC}(e^+)$}
\newcommand{\st}{$\star$}
\newcommand{\sst}{$\star \star$}
\newcommand{\ssst}{$\star \star \star$}
\newcommand{\sssst}{$\star \star \star \star$}
\newcommand{\tw}{\theta_W}
\newcommand{\sw}{\sin{\theta_W}}
\newcommand{\cw}{\cos{\theta_W}}
\newcommand{\sww}{\sin^2{\theta_W}}
\newcommand{\cww}{\cos^2{\theta_W}}
\newcommand{\trm}{m_{\perp}}
\newcommand{\trp}{p_{\perp}}
\newcommand{\trmm}{m_{\perp}^2}
\newcommand{\trpp}{p_{\perp}^2}
\newcommand{\alp}{\alpha_s}

\newcommand{\alps}{\alpha_s}
\newcommand{\sqrts}{$\sqrt{s}$}
\newcommand{\LO}{$O(\alpha_s^0)$}
\newcommand{\Oa}{$O(\alpha_s)$}
\newcommand{\Oaa}{$O(\alpha_s^2)$}
\newcommand{\PT}{p_{\perp}}
\newcommand{\JPSI}{J/\psi}
\newcommand{\sh}{\hat{s}}
\newcommand{\uh}{\hat{u}}
\newcommand{\MP}{m_{J/\psi}}
\newcommand{\PO}{I\!\!P}
\newcommand{\xbj}{x}
\newcommand{\xpom}{x_{\PO}}
\newcommand{\ttbs}{\char'134}
\newcommand{\xpomlo}{3\times10^{-4}}  
\newcommand{\xpomup}{0.05}  
\newcommand{\dgr}{^\circ}
\newcommand{\pbarnt}{\,\mbox{{\rm pb$^{-1}$}}}
\newcommand{\gev}{\,\mbox{GeV}}
\newcommand{\WBoson}{\mbox{$W$}}
\newcommand{\fbarn}{\,\mbox{{\rm fb}}}
\newcommand{\fbarnt}{\,\mbox{{\rm fb$^{-1}$}}}

\newcommand{\vtab}{\rule[-1mm]{0mm}{4mm}}
\newcommand{\htab}{\rule[-1mm]{0mm}{6mm}}
\newcommand{\photoproduction}{$\gamma p$}
\newcommand{\ptmiss}{$P_{T}^{\rm miss}$}
\newcommand{\epz} {$E{\rm-}p_z$}
\newcommand{\vap} {  $V_{ap}/V_p$}
\newcommand{\Zero}   {\mbox{$Z^{\circ}$}}
\newcommand{\Ftwo}   {\mbox{$\tilde{F}_2$}}
\newcommand{\Ftwoz}   {\mbox{$\tilde{F}_{2,3}$}}
\newcommand{\Fz}   {\mbox{$\tilde{F}_3$}}
\newcommand{\FL}   {\mbox{$\tilde{F}_{_{L}}$}}
\newcommand{\wtwogen} {W_2}
\newcommand{\wlgen} {W_L}
\newcommand{\xwthreegen} {xW_3}
\newcommand{\Wtwo}   {\mbox{$W_2$}}
\newcommand{\Wz}   {\mbox{$W_3$}}
\newcommand{\WL}   {\mbox{$W_{_{L}}$}}
\newcommand{\Fem}  {\mbox{$F_2$}}
\newcommand{\Fgam}  {\mbox{$F_2^{\gamma}$}}
\newcommand{\Fint} {\mbox{$F_2^{\gamma Z}$}}
\newcommand{\Fwk}  {\mbox{$F_2^{Z}$}}
\newcommand{\Ftwos} {\mbox{$F_2^{\gamma Z, Z}$}}
\newcommand{\Fzz} {\mbox{$F_3^{\gamma Z, Z}$}}
\newcommand{\Fintz} {\mbox{$F_{2,3}^{\gamma Z}$}}
\newcommand{\Fwkz}  {\mbox{$F_{2,3}^{Z}$}}
\newcommand{\Fzint} {\mbox{$F_3^{\gamma Z}$}}
\newcommand{\Fzwk}  {\mbox{$F_3^{Z}$}}
\newcommand{\Gev}  {\mbox{${\rm GeV}$}}
\newcommand{\Gevv}{\mbox{${\rm GeV}^2$}}
\newcommand{\QQ}  {\mbox{${Q^2}$}}
\newcommand{\gv}{GeV$^2\,$}
\newcommand{\bs}{\bar{s}}
\newcommand{\bc}{\bar{c}}
\newcommand{\bu}{\bar{u}}
\newcommand{\bb}{\bar{b}}
\newcommand{\bU}{\overline{U}}
\newcommand{\bD}{\overline{D}}
\newcommand{\bd}{\bar{d}}
\newcommand{\bq}{\bar{q}}    
\newcommand{\FLc}{$ F_{L}\,$} 
\newcommand{\xg}{$xg(x,Q^2)\,$}
\newcommand{\xgc}{$xg\,$}
\newcommand{\ipb}{pb$^{-1}\,$}               
\newcommand{\TOSS}{x_{{i}/{\PO}}}                                              
\newcommand{\un}[1]{\mbox{\rm #1}}
\newcommand{\pdsi}{$(\partial \sigma_r / \partial \ln y)_{Q^2}\,$}
\newcommand{\pdff}{$(\partial F_{2} / \partial \ln  Q^{2})_x\,$ }
\newcommand{\Fc}{$ F_{2}~$}
\newcommand{\amz}{$\alpha_s(M_Z^2)\,$} 

%
%
\newcommand{\qsq}{\ensuremath{Q^2} }
\newcommand{\gevsq}{\ensuremath{\mathrm{GeV}^2} }
\newcommand{\et}{\ensuremath{E_t^*} }
\newcommand{\rap}{\ensuremath{\eta^*} }
\newcommand{\gp}{\ensuremath{\gamma^*}p }
\newcommand{\dsiget}{\ensuremath{{\rm d}\sigma_{ep}/{\rm d}E_t^*} }
\newcommand{\dsigrap}{\ensuremath{{\rm d}\sigma_{ep}/{\rm d}\eta^*} }
\newcommand{\kd}{\textcolor{black}}
\def\Journal#1#2#3#4{{#1} {\bf #2} (#3) #4}
\def\NCA{\em Nuovo Cimento}
\def\NIM{\em Nucl. Instrum. Methods}
\def\NIMA{{\em Nucl. Instrum. Methods} {\bf A}}
\def\NPB{{\em Nucl. Phys.}   {\bf B}}
\def\PLB{{\em Phys. Lett.}   {\bf B}}
\def\PRL{\em Phys. Rev. Lett.}
\def\PRD{{\em Phys. Rev.}    {\bf D}}
\def\ZPC{{\em Z. Phys.}      {\bf C}}
\def\EJC{{\em Eur. Phys. J.} {\bf C}}
\def\CPC{\em Comp. Phys. Commun.}

\begin{titlepage}

\noindent
\begin{flushleft}
{\tt DESY 13-211    \hfill    ISSN 0418-9833} \\
{\tt December 2013}                  \\
\end{flushleft}

\vspace{1cm}

\begin{center}
\begin{Large}

{\bf Measurement of Inclusive \boldmath${ep}$ Cross Sections at High
\boldmath${Q^2}$ at \boldmath${\sqrt{s}}=225$ and $252$~GeV
  and of the Longitudinal Proton Structure
Function \boldmath${F_L}$ at HERA}
\vspace{1cm}

H1 Collaboration

\end{Large}
\end{center}

\vspace{1cm}

\begin{abstract}
\noindent
Inclusive $ep$ double differential cross sections for neutral current
deep inelastic scattering are measured with the H1 detector at
HERA. The data were taken with a lepton beam energy of $27.6$~GeV and
two proton beam energies of $E_p=460$ and $575$~GeV corresponding to
centre-of-mass energies of $225$ and $252$~GeV, respectively. The measurements
cover the region of $6.5\times 10^{-4} \leq x \leq 0.65$ for $35\leq
Q^2 \leq 800$~GeV$^2$ up to $y=0.85$. The
measurements are used together with previously published H1 data at
$E_p=920$~GeV and lower $Q^2$ data at $E_p=460$, $575$ and $920$~GeV to
extract the longitudinal proton structure function $F_L$ in the region
$1.5\leq Q^2 \leq 800$~GeV$^2$.
\end{abstract}

\vspace{1cm}

\begin{center}
Accepted by EPJC 
\end{center}

\end{titlepage}

%
%

\begin{flushleft}
\end{flushleft}


V.~Andreev$^{22}$,             
A.~Baghdasaryan$^{34}$,        
S.~Baghdasaryan$^{34}$,        
K.~Begzsuren$^{31}$,           
A.~Belousov$^{22}$,            
P.~Belov$^{10}$,               
V.~Boudry$^{25}$,              
G.~Brandt$^{46}$,              
M.~Brinkmann$^{10}$,           
V.~Brisson$^{24}$
% ...


\bigskip{\it
% ...
$ ^{2}$ School of Physics and Astronomy, University of Birmingham,
          Birmingham, UK$^{ b}$ \\
% ...
$ ^{10}$ DESY, Hamburg, Germany \\
% ...
$ ^{22}$ Lebedev Physical Institute, Moscow, Russia \\
% ...
$ ^{24}$ LAL, Universit\'e Paris-Sud, CNRS/IN2P3, Orsay, France \\
 $ ^{25}$ LLR, Ecole Polytechnique, CNRS/IN2P3, Palaiseau, France \\
% ...
$ ^{31}$ Institute of Physics and Technology of the Mongolian
          Academy of Sciences, Ulaanbaatar, Mongolia \\
% ...
$ ^{34}$ Yerevan Physics Institute, Yerevan, Armenia \\
% ...
$ ^{46}$ Department of Physics, Oxford University,
          Oxford, UK$^{ b}$ \\
% ...
$ ^b$ Supported by the UK Science and Technology Facilities Council,
      and formerly by the UK Particle Physics and
      Astronomy Research Council \\
% ...
}

\clearpage
\section{Introduction}
\label{sec:intro}

Deep inelastic scattering (DIS) data provide high precision tests of
perturbative quantum chromodynamics (QCD), and have led to a detailed and comprehensive
understanding of proton structure, see
\cite{Perez:2012um} for a recent review. A measurement of the longitudinal
proton structure function, $F_L$,
provides a unique test of parton dynamics and the
consistency of QCD by allowing a comparison of the gluon density obtained largely from the scaling violations of $F_2$ to an observable directly sensitive to the gluon density.
Previous measurements of $F_L$ have been published by the H1 and ZEUS
collaborations covering the kinematic region of low Bjorken $x$, and
low to medium four-momentum transfer squared, $Q^2$, using data taken
at proton beam energies $E_p=460$, $575$ and $920$~GeV corresponding
to centre-of-mass energies of $\sqrt{s}=225$, $252$ and $319$~GeV respectively
~\cite{h1fl,Collaboration:2010ry,Chekanov:2009na}. \kd{ The new cross
section measurements at $E_p=460$ and $575$~GeV presented here, and recently
published data at $E_p=920$ GeV~\cite{Aaron:2012qi}}
improve the experimental precision on $F_L$ in the
region $35\leq Q^2 \leq 110$~GeV$^2$, and provide the first
measurements of $F_L$ in the region $120\leq Q^2 \leq 800$~GeV$^2$ and
$6.5\times 10^{-4}<x<0.032$.  As the extraction of $F_L$ and $F_2$ is
repeated using all available H1 cross section measurements, the
earlier measurements of $F_L$ and $F_2$
~\cite{h1fl,Collaboration:2010ry} are superseded by the present
analysis.  Furthermore, in the determination of the systematic uncertainties of the
published H1 $F_L$ measurements~\cite{Collaboration:2010ry} an error has been identified in the
procedure of averaging several measurements at fixed $Q^2$ which is corrected here.

The differential cross section for deep inelastic $ep$ scattering can be
described in terms of three proton structure functions
$F_2$, $F_L$ and $xF_3$, which are related to the parton distribution functions
(PDFs) of the proton. The structure functions depend on the
kinematic variables, $x$ and $Q^2$ only, whereas the cross section is additionally
dependent on the inelasticity $y$ related by
$y=Q^2/sx$. The reduced neutral current (NC) differential cross section for $e^+p$
scattering after correcting for
QED radiative effects can be written as
\begin{linenomath}
\begin{equation}
\tilde{\sigma}_{\rm NC}(x,Q^2,y)\equiv
\frac{\rm{d}^2\sigma_{\rm NC}}{{\rm d}x{\rm d}Q^2}\frac{xQ^4}{2\pi\alpha^2}\frac{1}{Y_+}\equiv
\left( F_2 -\frac{y^2}{Y_+}F_L - \frac{Y_-}{Y_+}xF_3 \right) \,,
\label{eq:Rnc}
\end{equation} 
\end{linenomath}
where $Y_{\pm}=1\pm(1-y)^2$ and the fine structure constant is defined as
$\alpha\equiv\alpha(Q^2=0)$.

The cross section for virtual boson ($Z/\gamma^*$) exchange is
related to the $F_2$ and $xF_3$ structure functions in which both the
longitudinal and transverse boson polarisation states contribute. The
$F_L$ term is related to the longitudinally polarised virtual
boson exchange process. 
\kd{This term vanishes at lowest order QCD but has been predicted by Altarelli and Martinelli~\cite{altarelli78} to be non-zero when including higher order QCD terms.}
As can be seen from
equation~\ref{eq:Rnc} the contribution of $F_L$ to the cross
section is significant only at high $y$.  For $Q^2 \lesssim
800$~GeV$^2$ the contribution of $Z$ exchange and the influence of
$xF_3$ is expected to be small.

A direct measurement of $F_L$ is performed by measuring the
differential cross section at different values of $\sqrt{s}$ by
reducing the proton beam energy from $920$~GeV, used for most of
the HERA-II run period, to $E_p=460$ and $575$~GeV. The lepton beam
energy was maintained at $27.6$~GeV.
The two sets of cross section data are combined with recently
published H1 data taken at $E_p=920$~GeV~\cite{Aaron:2012qi}, and cross section
measurements at lower $Q^2$ taken at $E_p=460$,
$575$ and $920$~GeV\cite{Collaboration:2010ry}, to provide a set of
measurements at fixed $x$ and $Q^2$ but at different values of
$y$. This provides an experimental separation between the $F_2$ and
$F_L$ structure functions. Sensitivity to $F_L$ is enhanced by
performing the differential cross sections measurement up to high $y$,
a kinematic region in which the scattered lepton energy is low, and
consequently the background from photoproduction processes is large.
The cross sections are used to extract $F_L$ and the ratio $R$ of the longitudinally
to transversely polarised photon exchange cross sections. 
\kd{ In addition
a direct extraction of the gluon density $xg(x,Q^2)$ is performed
using an approximation at order $\alpha_S$.} 

This paper is organised as follows: in section~\ref{sec:det} the H1
detector, trigger system and data sets are described. The simulation
programs and Monte Carlo models used in the analysis are presented
in section~\ref{sec:mc}. In section~\ref{sec:expt} the analysis
procedure is given in which the event selection and background
suppression methods are discussed followed by an assessment of the
systematic uncertainties of the measurements. The results are presented
in section~\ref{sec:results} and the paper is summarised in
section~\ref{sec:summary}.

\section{H1 Apparatus, Trigger and Data Samples}
\label{sec:det}

\subsection{The H1 Detector}

A detailed description of the H1 detector can be found
elsewhere~\cite{h1detector,h1tracker,h1lar,spacal}. The coordinate
system of H1 is defined such that the positive $z$ axis is in the
direction of the proton beam (forward direction) and the nominal
interaction point is located at $z=0$. The polar angle $\theta$ is
then defined with respect to this axis.  The detector components most
relevant to this analysis are the Liquid Argon (LAr) calorimeter,
which measures the positions and energies of particles over the range
$4^\circ<\theta<154^\circ$, the inner tracking detectors, which
measure the angles and momenta of charged particles over the range
$7^\circ<\theta<165^\circ$, and a lead-fibre calorimeter (SpaCal)
covering the range $153^\circ<\theta<177^\circ$.

The LAr calorimeter consists of an inner electromagnetic section with
lead 
absorbers and an outer hadronic section with steel absorbers.  The
calorimeter is divided into eight wheels along the beam axis, each
consisting of eight stacks arranged in an octagonal formation
around the beam axis. The electromagnetic and the hadronic sections
are highly segmented in the transverse and the longitudinal
directions. Electromagnetic shower energies are measured with a
resolution of $\delta E/E \simeq 0.11/\sqrt{E/{\rm GeV}} \oplus 0.01$
and hadronic energies with $\delta E/E \simeq 0.50/\sqrt{E/{\rm GeV}}
\oplus 0.02$ as
determined using electron and pion test beam
data~\cite{Andrieu:1993tz,Andrieu:1994yn}.

In the central region, $25^{\circ}<\theta<155^{\circ}$, the central
tracking detector (CTD) measures the trajectories of charged particles
in two cylindrical drift chambers (CJC) immersed in a uniform
$1.16\,{\rm T}$ solenoidal magnetic field. The CTD also contains a
further drift chamber (COZ) between the two drift chambers to improve
the $z$ coordinate reconstruction, as well as a multiwire proportional
chamber at inner radii (CIP) mainly used for
triggering~\cite{Becker:2007ms}. The CTD measures charged particle
trajectories with a transverse momentum resolution of
$\sigma(p_T)/p_T\simeq 0.2\% \, p_T/{\rm GeV} \oplus 1.5\%$.  The CJC
also provides a measurement of the specific ionisation energy loss,
${\rm d}E/{\rm d}x$, of charged particles with a relative resolution
of $6.5\%$ for long tracks. The forward tracking detector (FTD) is
used to supplement track reconstruction in the region
$7^{\circ}<\theta<30^{\circ}$~\cite{Laycock:2012xg} and to improve the
hadronic final state (HFS) reconstruction of forward going low transverse
momentum particles.

The CTD tracks are linked to hits in the vertex detectors: the central
silicon tracker (CST)~\cite{h1cst,h1cst2}, the forward silicon tracker
(FST), and the backward silicon tracker (BST). These detectors provide
precise spatial track reconstruction and therefore also improve the
primary vertex reconstruction. The CST consists of two layers
of double-sided silicon strip detectors surrounding the beam pipe
covering an angular range of $30^\circ<\theta<150^\circ$ for tracks
passing through both layers. The FST consists of five double wheels of
single-sided strip detectors~\cite{Glushkov:2007zz} measuring the
transverse coordinates of charged particles. The BST design is very
similar to the FST and consists of six double wheels of strip
detectors~\cite{Kretzschmar:2008zz}. 

In the backward region the SpaCal provides an energy measurement for
electrons \footnote{In this paper ``electron'' refers generically to
  both electrons and positrons. Where distinction is required, the
  terms $e^-$ and $e^+$ are used.} and hadronic particles, and has a
resolution for electromagnetic energy depositions of $\delta E/E
\simeq 0.07/\sqrt{E/{\rm GeV}}\oplus 0.01$, and a hadronic energy
resolution of $\delta E/E \simeq 0.70/\sqrt{E/{\rm GeV}}\oplus 0.01$
as measured using test beam data~\cite{spacal_res}. 

The integrated $ep$ luminosity is determined by measuring the event rate
for the Bethe-Heitler process of QED bremsstrahlung $ep \rightarrow
ep\gamma$. The photons are detected in the photon tagger located at
$z=-103\,{\rm m}$.  An electron tagger is placed at $z=-5.4\,{\rm m}$
adjacent to
the beampipe. It is used to provide information on $ep\rightarrow eX$
events at very low $Q^2$ (photoproduction) where the electron scatters
through a small angle ($\pi - \theta < 5\,{\rm mrad}$).  

At HERA transverse polarisation of the lepton beam arises naturally
through synchrotron radiation via the Sokolov-Ternov
effect~\cite{sokolov-ternov}. Spin rotators installed in the beamline
on either side of the H1 detector allow transversely polarised leptons
to be rotated into longitudinally polarised states and back again.
Two independent polarimeters LPOL~\cite{lpol} and TPOL~\cite{tpol}
monitor the polarisation. Only data where a TPOL or LPOL measurement
is available is used. When both measurements are available they are
averaged~\cite{pubpola}.

\subsection{The Trigger}
The H1 trigger system is a three level trigger with a first level
latency of approximately $2\,\mu{\rm s}$. In the following we describe
only the components relevant to this analysis.
NC events at high $Q^2$ are
triggered mainly using information from the LAr calorimeter to rapidly
identify the scattered lepton.  The calorimeter has a finely segmented
geometry allowing the trigger to select localised energy
deposits in the electromagnetic section of the calorimeter pointing to
the nominal interaction vertex.  For electrons with energy above
$11\,{\rm GeV}$ this LAr electron trigger is determined to be $100\%$ efficient
obtained by using LAr triggers fired by the hadronic final state
particles.  

At high $y$, corresponding to lower electron energies, the backward
going HFS particles can enter the
SpaCal and therefore trigger the event. In addition low energy
scattered electron candidates can be triggered by the Fast Track
Trigger~\cite{FTT,Jung2007a} based on hit information provided by the CJC, and
the LAr Jet Trigger~\cite{jettrigger} using energy depositions in the
LAr calorimeter.  These two
trigger subsystems allow electron identification to be performed at
the third trigger level~\cite{Aaron:2012cj,Sauter}.  This L3 electron trigger
and the SpaCal trigger are used to extend the kinematically accessible
region to high $y$ where scattered leptons have energies as low as
$3\,{\rm GeV}$, the minimum value considered in this analysis. For
electron energies of $3\,{\rm GeV}$, the total trigger efficiency
is found to vary between $91-97\%$ depending on the kinematic region.

\subsection{Data Samples}
\label{sec:datasets}

The data sets used in the measurement of the reduced cross sections
correspond to two short dedicated data taking periods in 2007 in which
the proton beam energy was reduced to $460$~GeV and $575$~GeV, and the
scattered lepton was detected in the LAr calorimeter. The
positron beam was longitudinally polarised with polarisation
$P_e=(N_R-N_L)/(N_R+N_L)$, where $N_R$ ($N_L$) is the number of right
(left) handed leptons in the beam. The integrated luminosity and
longitudinal lepton beam polarisation for each data set are given in
table~\ref{tab:lumi}. The lepton beam
polarisation plays no significant role in this analysis.

\begin{table}[h]
  \begin{center}
    \begin{tabular}{r|c|c}
\hline
 & $E_p=460$~GeV & $E_p=575$~GeV\\
\hline
\multirow{2}{*}{$e^+p$}  
& $\mathcal{L}=11.8\,{\rm pb}^{-1}$ & $\mathcal{L}=5.4\,{\rm pb}^{-1}$ \\
& $P_e=(-42.3\pm 0.8)\%$            &   $P_e=(-41.8\pm 0.8)\%$ \\
\hline
\end{tabular} 
\caption{\sl Integrated luminosities, $\mathcal{L}$, and
  luminosity weighted longitudinal lepton beam polarisation, $P_e$, for
  the data sets presented here.}
\label{tab:lumi}
\end{center}
\end{table}

The extraction of the $F_L$ structure function in section~\ref{sec:fl}
uses the cross section measurements presented here and \kd{  $e^+p$ measurements 
with $P_e=0$ at
$E_p=920$~GeV in which the scattered positron is detected in the LAr
calorimeter (Tables 22 and 26 of~\cite{Aaron:2012qi} scaled by a normalisation factor of $1.018$~\cite{QEDCerratum} which arises from an error in the determination of the integrated luminosity used for this data set). In addition the $F_L$ extraction also
uses cross section measurements from H1 at $E_p=460$, $575$ and $920$~GeV
with the positron detected in the
SpaCal as it is described in~\cite{Collaboration:2010ry}. }The two detectors provide access to
different kinematic regions and the corresponding measurements are
referred to as the LAr and SpaCal data for each of the three values of
$E_p$.

\section{Simulation Programs}
\label{sec:mc}

In order to determine acceptance corrections, DIS processes are
generated at leading order (LO) QCD using the {\sc
  Djangoh\,1.4}~\cite{django} Monte Carlo (MC) simulation program
which is based on {\sc Heracles\,4.6}~\cite{heracles} for the
electroweak interaction and on {\sc Lepto\,6.5.1}~\cite{lepto} for the
hard matrix element calculation. The colour dipole model (CDM) as
implemented in {\sc Ariadne} \cite{cdm} is used to simulate higher
order QCD dynamics.  The {\sc Jetset\,7.410} program~\cite{jetset} is
used to simulate the hadronisation process in the framework of the 
`string-fragmentation' model. The parameters of this model used here
are tuned to describe hadronic $Z$ decay data~\cite{Schael:2004ux}. 
The simulated events are weighted to reproduce the cross sections
predicted by the NLO QCD fit H1PDF\,2012~\cite{Aaron:2012qi}. This fit
includes H1 low $Q^2$ NC data
and high $Q^2$ neutral and charged current (CC) data from
HERA\,I, as well as inclusive NC and CC measurements from H1
at high $Q^2$ based on the full HERA\,II integrated
luminosity at $E_p=920$ GeV~\cite{Aaron:2012qi}. 
In addition the {\sc Compton~22}~\cite{Lendermann:2003rq} MC is used
to simulate 
elastic and quasi-elastic QED Compton processes, and replaces the
Compton processes simulation available in {\sc Djangoh}.

The detector response to events produced by the various generator
programs is simulated in detail using a program based on {\sc
  Geant3}~\cite{Geant}.  The simulation includes a detailed time
dependent modelling of detector noise conditions, beam optics,
polarisation and inefficient channel maps reflecting actual running
conditions throughout the data taking periods.  These simulated
events are then subjected to the same reconstruction, calibration,
alignment and analysis chain as the real data.

\section{Experimental Procedure}
\label{sec:expt}

\subsection{Kinematic Reconstruction}
\label{sec:kine}

Accurate measurements of the event kinematic quantities $Q^2$, $x$ and
$y$ are an essential component of the analysis. Since both the
scattered lepton and the hadronic final state (HFS) are observed in
the detector, several kinematic reconstruction methods are available
allowing for calibration and cross checks.

The primary inputs to the various methods employed are the scattered
lepton's energy $E^{\prime}_e$ and polar scattering angle $\theta_e$,
as well as the quantity $\Sigma=\sum_i(E_i-p_{z,i})$ determined from
the sum over the HFS particles assuming charged particles have the
pion mass, where $E_i$ and $p_{z,i}$ are the energy and longitudinal
momenta respectively~\cite{hfs}. At high $Q^2$ and low $y$ the HFS is
dominated by one or more jets. Therefore the complete HFS can be
approximated by the sum of jet four-momenta corresponding to localised
calorimetric energy sums above threshold. This technique allows a
further suppression of ``hadronic noise'' in the reconstruction
arising from electronic sources in the LAr calorimeter or from
back-scattered low energy particles produced in secondary
interactions.

The most precise kinematic reconstruction method for $y\gtrsim 0.1$ is the $e$-method
which relies solely on $E^{\prime}_e$ and $\theta_e$ to reconstruct
the kinematic variables $Q^2$ and $y$ as
\begin{linenomath}
\begin{equation}
  Q^2_{e} = \frac{ (E^{\prime}_e \sin{\theta_e})^2}{ 1-y_e}\,,
 \hspace*{0.5cm} 
  y_e =1-   \frac{ E^{\prime}_e}{E_e} \sin^2\left(\frac{\theta_e}{2}\right) \,,
\label{eq:emethod}
\end{equation}
\end{linenomath}
and $x$ is determined via the relation $x=Q^2/sy$. This method is used
in the analysis region $y>0.19$ since the resolution of the $e$-method
degrades at low $y$. The method is also susceptible to large QED
radiative corrections at the highest and lowest $y$. A cut on quantity
\mbox{$E-P_z=\Sigma+E^{\prime}_e(1-\cos\theta_e)$} ensures that the radiative
corrections are moderate.

In the $\Sigma$-method \cite{sigma}, $y$ is reconstructed
as $\Sigma/(E-P_z)$ and is
therefore less sensitive to QED radiative effects. The
$e\Sigma$-method \cite{esigma} is an optimum combination of the two
and maintains good resolution throughout the kinematic range of the 
measurement with acceptably small QED radiative corrections. The
kinematic variables are determined using
\begin{linenomath}
\begin{equation}
  Q^2_{e\Sigma} = Q^2_{e} = \frac{ (E^{\prime}_e \sin{\theta_e})^2}{ 1-y_e}\,,
 \hspace*{0.5cm} 
  y_{e\Sigma} = 2 E_e \frac{\Sigma}{[E-P_z ]^2} \,,
\label{eq:esmethod}
\end{equation}
\end{linenomath}
and $x$ is determined as for the $e$-method above. The $e\Sigma$-method is employed to reconstruct the event kinematics
for $y \leq 0.19$ in which $\Sigma$ is determined using hadronic jets
defined using the longitudinally invariant $k_T$ jet
algorithm~\cite{kt-jets1,kt-jets2}.


\subsection{Polar Angle Measurement and Energy Calibration}
\label{sec:calib}

The detector calibration and alignment procedures adopted for this
analysis rely on the methods discussed in detail
in~\cite{Aaron:2012qi} which uses the high statistics
$E_p=920$ GeV data recorded just prior to the $460$ and $575$ GeV
runs. The detector was not moved or opened between these run periods.  The
alignment and calibration constants obtained at $E_p=920$ GeV are verified
using the same methods~\cite{Aaron:2012qi}  for the data presented here.

In this analysis the scattered lepton is detected in the LAr
calorimeter by searching for a compact and isolated electromagnetic
energy deposition. The polar angle of the scattered
lepton, $\theta_e$, is determined using the position of its energy
deposit (cluster) in the LAr calorimeter, and the event
vertex reconstructed with tracks from charged particles. The relative
alignment of the calorimeter and tracking chambers is verified using a
sample of events with a well measured lepton track~\cite{tran} in
which the COZ chamber provides an accurate $z$ reconstruction
of the particle trajectory. The residual difference between the track
and cluster polar angles in data and simulation is found to be less
than $1$~mrad, and this value is used as the systematic uncertainty of
the scattered lepton polar angle.

An {\em in situ} energy calibration of electromagnetic energy
depositions in the LAr calorimeter is performed for both data and
simulation. Briefly, a sample of NC events in which the HFS is well
contained in the detector is used with the Double Angle reconstruction
method~\cite{damethod1,damethod2} to predict the scattered lepton
energy ($E_{DA}$) which is then compared to the measured
electromagnetic energy response allowing local calibration factors to
be determined in a finely segmented grid in $z$ and $\phi$.  The
residual mismatch between $E_{DA}$ and $E_e^{\prime}$ after performing
the calibration step are found to vary within $\simeq 0.3-1\%$
depending on the geometric location of the scattered lepton within the
LAr calorimeter. An additional $0.3\%$ correlated uncertainty is
considered and accounts for a possible bias in the $P_{T,DA}$
reconstruction and is determined by varying $\theta_e$ and a
measurement of the inclusive hadronic polar angle, $\gamma_h$, by the
angular measurement uncertainty. This has been verified by comparing
the residual global shifts between data and MC in the kinematic peak
of the $E_e^{\prime}$ distribution.

At the lowest electron energies the calibration is validated using
QED Compton interactions $ep\rightarrow e\gamma p$ with $E_e^{\prime}$
of $3-8\,{\rm GeV}$ in which the lepton track momentum $P_{\rm track}$
is compared to the measured energy $E_{e}^{\prime}$ of the
cluster. The simulation on average describes the data well in this low
energy region. For energies below $11$~GeV an additional uncorrelated
uncertainty of $0.5\%$ is included to account for a possible 
nonlinearity of the energy scale.

The hadronic response of the detector is calibrated by requiring a
transverse momentum balance between the predicted $P_T$ in the
DA-method ($P_{T,DA}$) and the measured hadronic final state using a
tight selection of well reconstructed events with a single jet.  The
calorimeter calibration constants are then determined in a
minimisation procedure across the detector acceptance separately for
HFS objects inside and outside jets and for electromagnetic and
hadronic contributions to the HFS~\cite{kogler}. The potential bias in
the $P_{T,DA}$ reference scale of $0.3\%$ is also included as a
correlated source of uncertainty.

The mean transverse momentum balance between the hadronic final state
and the scattered lepton both in data and simulation agree to within
$1\%$ precision which is taken as the uncorrelated hadronic scale
uncertainty.  The hadronic SpaCal calibration is performed in a
similar manner and a systematic uncertainty of $5\%$ is
adopted.

\subsection{Measurement Procedure}
\label{sec:ncmeas}

The event selection and analysis of the NC sample follows closely the
procedures discussed in~\cite{Aaron:2012qi}. 
Inelastic $ep$ interactions are required to have a well reconstructed
interaction vertex to suppress beam induced background events. High
\qsq neutral current events are selected by requiring each event to
have a compact and isolated cluster in the electromagnetic part of the
LAr calorimeter.
The scattered lepton candidate is identified as the cluster of highest
transverse momentum and must have an associated CTD track. For high
electron energies the track condition is relaxed as detailed
in~\ref{sec:nominalAnalysis}. The analysis is restricted to the region
$32<Q^2_e<890$~GeV$^2$.

The quantity $E-P_z$ summed over all final state particles (including
the electron) is required by energy-momentum conservation to be
approximately equal to twice the initial electron beam energy.
Restricting $E-P_z$ to be greater than $35$~GeV considerably reduces
the photoproduction background and radiative processes in which 
either the scattered lepton or bremsstrahlung photons escape
undetected in the lepton beam direction. Topological algorithms~\cite{shushkevich} are
employed to suppress non-$ep$ and QED Compton backgrounds $ep\rightarrow
e\gamma p$.

The photoproduction background increases rapidly with decreasing
electron energy (corresponding to high $y$), therefore the analysis is separated into two distinct
regions: the {\em nominal} analysis ($y_e \leq 0.38$), and the {\em high y}
analysis ($0.38<y_e<0.9$).  In the {\em high y} region dedicated techniques are
employed to contend with the large background. The analysis differences in each
kinematic region are described below.

\subsubsection{{\em Nominal} Analysis}
\label{sec:nominalAnalysis}

At low $y\leq 0.38$ the minimum electron energy is kinematically
restricted to be above $18\,{\rm GeV}$. The forward going hadronic final state
particles can undergo interactions with material of the beam pipe
leading sometimes to a bias in the reconstruction of the primary
interaction vertex position. In such cases the vertex position is
calculated using a stand alone reconstruction of the track associated
with the electron cluster~\cite{nikiforov,shushkevich}.  For the {\em
  nominal} analysis the photoproduction contribution is negligible,
and the only sizeable background contribution arises from remaining QED Compton
events which is estimated using simulation. The
electron candidate track verification is supplemented by searching for
hits in the CIP located on the trajectory from the interaction vertex to the
electron cluster. This optimised treatment of the vertex determination
and verification of the electron cluster with the tracker information
improves the reliability of the vertex position determination and
increases the efficiency of the procedure to be larger than $99.5\%$.

For the region $y<0.19$ the hadronic noise has an increasing influence
on $\Sigma$ and on the transverse momentum balance $P_{T,h}/P_{T,e}$ through its
effect on $P_{T,h}$ where $P_{T,h},~P_{T,e}$ are the hadronic and
scattered lepton transverse momenta respectively. The event kinematics
reconstructed with the $e\Sigma$-method in which the HFS is formed
from hadronic jets only, limits the noise contribution and
substantially improves the $P_{T,h}/P_{T,e}$ description by the
simulation.  The jets are found with the longitudinally invariant
$k_T$ jet algorithm~\cite{kt-jets1,kt-jets2} as implemented in
FastJet~\cite{fastjet1,fastjet2} with radius parameter $R=1.0$ and are
required to have transverse momenta $P_{T,{\rm jet}}>2\,{\rm GeV}$. In
figure~\ref{fig:contBulk} the quality of the simulation and its
description of the $E_p=460$~GeV and $E_p=575$~GeV data for $y_e<0.19$
can be seen for the distributions of the $P_{T,h}/P_{T,e}$,
$\theta_{\rm jets}$, and $E-P_z$ where all HFS quantities are obtained
using the vector sum of jet four-momenta.  The simulation provides a
reasonable description of both sets of distributions. The MC
simulation is normalised to the integrated luminosity of the data.

\subsubsection{{\em High y} Analysis}
In the {\em high y} region ($0.38<y_e<0.9$) the analysis is extended
to low energies of the scattered electron, $E_e^{\prime}>3$~GeV.  At
these energies photoproduction background contributions arise from
$\pi^0\rightarrow \gamma \gamma$ decays, from charged hadrons
being misidentified as electron candidates, and from real electrons originating predominantly from
semi-leptonic decays of heavy flavour hadrons. These
contributions increase rapidly with decreasing energy of the
electron candidate. Therefore additional techniques are used to
reduce this background.



The background from $\pi^0\rightarrow \gamma \gamma$ decays leads to
different electromagnetic shower profiles compared to electrons of
similar energy. In addition genuine electrons have a
momentum matched track associated to the cluster. Four cluster shape
variables and the ratio of the candidate electron energy
$E_e^{\prime}$ determined using cluster information, to the momentum of
the associated track $p_e$, are used in a neural network multilayer
perceptron~\cite{Hocker:2007ht} to discriminate signal from
background. Additional information using the specific ionisation
energy loss, ${\rm d}E/{\rm d}x$, of the track is also used to form a
single electron discrimination variable, $D_{ele}$, such that a value of $1$
corresponds to electrons and a value of $0$ corresponds to
hadrons. The neural network is trained using single particle MC
simulations, and validated with samples
of identified electrons and pions from $J/\psi\rightarrow e^+e^-$ and
$K^0_s\rightarrow\pi^+\pi^-$ decays in data and
MC~\cite{Aaron:2012cj,Sauter}. For the region $E_e^{\prime}<10$~GeV
isolated electrons are selected by requiring $D_{ele}>0.80$ which is
estimated to have a pion background rejection of more than $99$\% and a
signal selection efficiency of better than $90$\%~\cite{Aaron:2012cj}. For the region
$E_e^{\prime}>10$~GeV the scattered electron is identified as in the
nominal analysis.

The scattered lepton candidate is required to have positive charge
corresponding to the beam lepton. The remaining background is
estimated from the number of data events with opposite charge. This
background is statistically subtracted from the positively charged
sample. However, a charge asymmetry in photoproduction can arise due
to the different detector response to particles compared to
antiparticles~\cite{Adloff:2000qk,h1fl2010}.  The charge asymmetry has
been determined by measuring the ratio of wrongly charged 
scattered lepton candidates in $e^+p$ to $e^-p$ scattering at
$E_p=920$~GeV data and was found to be $1.03\pm
0.05$~\cite{Aaron:2012qi}. This is cross checked in the $E_p=460$ and
$575$~GeV data using photoproduction events in which the scattered
electron is detected in the electron tagger.  In this sample fake scattered
electron candidates passing all selection criteria are detected in the
LAr calorimeter with both positively and negatively charged tracks
associated to the electromagnetic cluster. The charge asymmetry is obtained by
comparing the two contributions. The results obtained are consistent
with the asymmetry measured in the $E_p=920$~GeV data, however due to
the lower statistical precision of the $E_p=460$ and $575$~GeV data sets,
the uncertainty of the asymmetry is increased to $0.08$. The asymmetry
is taken into account in the subtraction procedure. The efficiency
with which the lepton charge is determined is well described by
simulation within $0.5\%$ and is discussed in section~\ref{sys:sel}.

The control of the background in the most critical region of
$E_e^{\prime}<6$~GeV is demonstrated in figure~\ref{fig:contLowE} for
both data sets.  The MC simulation is normalised to the integrated
luminosity of the data. In all cases the background dominated regions
are well described in shape and overall normalisation, giving
confidence that the background contributions can be reliably estimated
from the wrong charge sample. At low $E-P_z$ a peak is observed
arising from QED initial state radiation (ISR) which is reasonably
well described. The cut $E-P_z>35$~GeV suppresses the influence of
ISR on the measurement. The $D_{ele}$ distribution show two populations
peaking at zero and unity arising from hadrons and real electrons
respectively. The peak at $D_{ele}=1$ for the background indicates
that there are real electrons in the remaining background sample.

The $e$-method has the highest precision in this region of phase space
and is used to reconstruct the event
kinematics. Figure~\ref{fig:contHiY} shows the energy spectrum and the polar angle distribution of the scattered lepton, and the $E-P_z$
spectrum of the {\em high y} sample for the $E_p=460$ and
$575$ GeV data before background subtraction. The background
estimates are shown together with the contribution from the remaining QED
Compton process.  The NC simulation provides a good description of
these distributions.

In the scattered lepton energy spectrum a small discontinuity at
$8$~GeV can be seen. \kd{ This is a consequence of suppressing 
electron candidates with $E<8$~GeV if a second electron candidate is
found with $E>8$~GeV. } This criterion efficiently suppresses background
from the QED Compton process in the region $E_e^{\prime}<8$~GeV.

\subsection{Cross Section Measurement}
\label{sec:xsec}

The simulation is used to correct the selected event samples for
detector acceptance, efficiencies, migrations and QED radiation effects.  The simulation
provides a good description of the data and therefore is expected to
give a reliable determination of the detector acceptance.  The accessible
kinematic ranges of the measurements depend on the resolution of the
reconstructed kinematic variables.  The ranges are determined by
requiring the purity and stability of any measurement bin to be larger
than $30\%$ as determined from signal MC. The purity is defined as the
fraction of events generated and reconstructed in a measurement bin
($N^{g+r}$) from the total number of events reconstructed in the bin
($N^{r}$). The stability is the ratio of the number of events
generated and reconstructed in a bin to the number of events generated
in that bin ($N^{g}$). The purity and stability are typically found to
be above $60\%$. The detector acceptance, $\mathcal{A}=N^{r}/N^{g}$,
corrects the measured signal event yields for detector effects
including resolution smearing and selection efficiency.

The measured differential cross sections $\sigma(x,Q^2)$ are
then determined using the relation
\begin{linenomath}
\begin{eqnarray}
\sigma(x,Q^2) =
\frac{N-B}{\mathcal{L}\cdot\mathcal{A}}\cdot{\mathcal C}
\cdot \left(1+\Delta^{\rm QED}\right)\,,
\end{eqnarray}
\end{linenomath}
where $N$ and $B$ are the selected number of data events and the
estimated number of background events respectively, $\mathcal{L}$ is
the integrated luminosity, $\mathcal{C}$ is the bin centre correction,
and $(1+\Delta^{\rm QED})$ are the QED radiative corrections. These
corrections are defined in~\cite{h19497,h1hiq2} and are calculated to
first order in $\alpha$ using the program {\sc
  Heracles}~\cite{heracles} as implemented in {\sc
  Djangoh}~\cite{django} and verified with the numerical analysis
programs {\sc Hector}~\cite{hector} and {\sc Eprc}~\cite{eprc}.  No
weak radiative corrections are applied to the measurements.

The bin centre correction ${\mathcal C(x,Q^2)}$ is a factor
obtained from NLO QCD expectation, \kd{$\sigma^{th}(x,Q^2)$}, using H1PDF\,2012~\cite{Aaron:2012qi},
 and scales the bin integrated cross section to
a differential cross section at the kinematic point $x,Q^2$
defined as
\begin{linenomath}
\begin{eqnarray}
{\mathcal C}(x,Q^2) =\frac{\sigma^{th}(x,Q^2)}
{\iint_{bin}{\small {\rm d}x^{\prime}{\rm d}Q^{2\prime}}\,\,\,\sigma^{th}(x^{\prime},Q^{2\prime})}\,.
\end{eqnarray}
\end{linenomath}
The cross section measurements are finally corrected for the effects
of lepton beam polarisation using the H1PDF\,2012 fit to yield cross
sections with $P_e=0$. This multiplicative correction does not exceed
$2.5\%$ in the region considered.

In order to optimise the measurement for an extraction of the
structure function $F_L$, the cross sections are measured in $y,Q^2$
bins for $y>0.38$ at $E_p=460$~GeV, and $y>0.304$ at $E_p=575$~GeV. At
$E_p=920$~GeV the $y,Q^2$ binned cross sections are published for
$y>0.19$~\cite{Aaron:2012qi}. This binning is constructed specifically
for a measurement of $F_L$ with fine segmentation in $y$. The lower
limits in $y$ for each proton beam energy are chosen such that they
have the same $x$ for all three values of $E_p$. \kd{ Below these $y$
boundaries for each of the three proton beam energies the cross
sections are measured in $Q^2,x$ bins.} The bin boundaries and bin
centres in the $Q^2-x$ plane are chosen to be the same in the
overlapping region for $E_p=460$, $575$ and $920$~GeV for $35\leq
Q^2\leq 800$~GeV$^2$.

\subsection{Systematic Uncertainties}
\label{sys:sel}
The uncertainties on the measurement lead to systematic errors on the
cross sections, which can be split into bin-to-bin correlated and
uncorrelated parts.  All the correlated systematic errors are found to
be symmetric to a good approximation and are assumed so in the
following. The total systematic error is formed by adding the
individual errors in quadrature. 

The size of each systematic uncertainty source and its region of
applicability are given in table~\ref{tab:syserr}.  Further details
can be found elsewhere~\cite{tran,shushkevich,habib,nikiforov,kogler}
in which several of the sources of uncertainty have been investigated
using the $E_p=920$~GeV LAr data. The results of similar studies
performed using the $E_p=460$~GeV and $575$~GeV LAr data are compared
to these earlier analyses to determine the systematic uncertainties.
The influence of the systematic uncertainties on the cross section
measurements are given in tables~\ref{tab:xsec460}-\ref{tab:xsec575},
and their origin and method of estimation are discussed below.

\begin{table}[ht]\footnotesize
  \begin{center}
    \begin{tabular}{ l|c|r}

\hline
{\bf Source} & {\bf Region} & \multicolumn{1}{c} {\bf Uncertainty} \\ 
\hline
    \multirow{5}{*}{Electron energy scale} 
&        $z_{\rm imp} \leq -150\,{\rm cm}$    &    $0.5\%$ unc. $\oplus$ $0.3\%$ corr. \\ 
& $-150 < z_{\rm imp} \leq  -60\,{\rm cm}$    &    $0.3\%$ unc. $\oplus$ $0.3\%$ corr. \\ 
& $-60  < z_{\rm imp} \leq  +20\,{\rm cm}$    &    $0.5\%$ unc. $\oplus$ $0.3\%$ corr. \\ 
& $+20  < z_{\rm imp} \leq +110\,{\rm cm}$    &    $0.5\%$ unc. $\oplus$ $0.3\%$ corr. \\ 
&        $z_{\rm imp} >    +110\,{\rm cm}$    &    $1.0\%$ unc. $\oplus$ $0.3\%$ corr. \\ 
\hline
Electron scale linearity &  $E_e^{\prime}<11\,{\rm GeV}$ & $0.5\%$ \\
\hline
\multirow{2}{*}{Hadronic energy scale}
& LAr \& Tracks   & $1.0\%$ unc. $\oplus$ $0.3\%$ corr. \\
& SpaCal & $5.0\%$ unc. $\oplus$ $0.3\%$ corr. \\
\hline
Polar angle
&$\theta_e$  & $1\,{\rm mrad}$ corr.\\
\hline
\multirow{2}{*}{Noise} 
& $y<0.19$ & $5\%$ energy not in jets , corr.\\
& $y>0.19$ & $20\%$ corr.\\
\hline
\multirow{2}{*}{Trigger efficiency} 
& {\em high y}  & $0.3 - 2\%$ \\
& {\em nominal} & $0.3\%$ \\
\hline
\multirow{2}{*}{Electron track and vertex efficiency}
&{\em high y}  & $1    \%$ \\
&{\em nominal} & $0.2-1\%$ \\
\hline
Electron charge ID efficiency  &  {\em high $y$}  & $0.5\%$ \\
\hline
\multirow{2}{*}{Electron ID efficiency}
& {\em high y} $z_{\rm imp}<20~(>20)$~cm  & $0.5\%~(1\%)$ \\
& {\em nominal} $z_{\rm imp}<20~(>20)$~cm  & $0.2\%~(1\%)$ \\
\hline
Extra background suppression & $E_e^{\prime}<10\,{\rm GeV}$  &  $D_{ele}>0.80 \pm 0.04$ corr.\\
\hline
{\em High y} background subtraction & {\em high $y$}  &  $1.03 \pm 0.08$ corr.\\
\hline
\multirow{2}{*}{QED radiative corrections  }
& $x<0.1 \, , \, 0.1 \leq x<0.3 \, , \, x\geq0.3$      & $0.3\%$ , $1.0\%$, $2.0\%$ \\
& {\em high y:} $y<0.8$ ($y>0.8$)  & $1\%~(1.5\%)$ \\
\hline
\multirow{2}{*}{Acceptance corrections}
& {\em high y}  & $0.5\%$ \\
& {\em nominal} & $0.2\%$ \\
\hline
Luminosity     &    &    $4\%$ corr.\\

\hline
\end{tabular} 
\caption{ Table of applied systematic uncertainties and regions of
  applicability. Uncertainties which are considered point-to-point
  correlated are labelled corr., and all other sources are considered
  uncorrelated. The effect of these uncertainties  on the cross section
  measurements is given in the tables of
  section~\ref{sec:results} (except for the
  luminosity uncertainty).   }
\label{tab:syserr}
\end{center}
\end{table}

\begin{description}
\item[Electron Energy:] Uncertainties arise from the particular choice
  of calibration samples, and the linearity correction
  uncertainty. These uncertainties are taken from the analysis of the
  $920$ GeV data~\cite{Aaron:2012qi}. The uncertainty varies as a
  function of $z_{\rm imp}$~\cite{Aaron:2012qi}, the $z$ position of
  the scattered electron in the calorimeter, as given in
  table~\ref{tab:syserr}. The correlated part of the uncertainty of
  $0.3\%$ accounts for a possible bias in the $E_{DA}$ reconstruction
  used as a reference scale in the energy calibration procedure. This
  results in a systematic uncertainty which is up to $2-3\%$ at low
  $y$.

\item[Hadronic Calibration:] An uncorrelated uncertainty of $1\%$ is
  used for the hadronic energy measurement. The uncertainty is
  determined by quantifying the agreement between data and simulation
  in the mean of the $P_{T,h}/P_{T,e}$ distribution in different
  kinematic regions.  The correlated part of the uncertainty accounts
  for a possible bias in the $E_{DA}$ reconstruction used as a
  reference scale in the energy calibration. It is determined to be
  $0.3\%$ and results in a correlated systematic error on the cross
  section which is up to $2-3\%$ at low $y$. The resulting correlated
  systematic error is typically below $1\%$ for the cross sections.

\item[Polar Angle:] A correlated $1$~mrad uncertainty on the
  determination of the electron polar angle is considered. This
  contribution leads to a typical uncertainty on the reduced cross
  sections of less than $1\%$.
 
\item[Noise Subtraction:] Energy classified as noise in
  the LAr calorimeter is excluded from the HFS. For $y<0.19$ the
  calorimetric energy not contained within hadronic jets is classified
  as noise. The uncertainty on the subtracted noise is estimated to be
  $5\%$ of the noise contribution as determined from the analysis of
  the HERA\,II $E_p=920$ GeV data~\cite{Aaron:2012qi}. For $y>0.19$
  the noise contribution is restricted to the sum of isolated low energy
  calorimetric depositions. Here the residual noise contribution
  is assigned an uncertainty of $20\%$, to accomodate differences
  between data and simulation.

\item[{\em Nominal} Trigger Efficiency:] The uncertainty on the
  trigger efficiency in the {\em nominal} analysis is determined
  separately for both $E_p=460$ and $575$ GeV data taking
  periods. Three trigger requirements are employed: the global timing,
  the event timing and the calorimeter energy. The inefficiency of
  global timing criteria to suppress out of time beam related
  background was continuously monitored with high precision and found
  to be $0.3\%$ and is corrected for. Finally the event timing trigger
  requirements were also continuously monitored in the data. After
  rejection of local inefficient regions the overall trigger
  efficiency is close to $100\%$ and an uncertainty of $0.3\%$ is
  assigned.

\item[{\em High y} Trigger Efficiency:] 
  At low $E_e^{\prime}$ the LAr electron trigger is supplemented 
  by the SpaCal trigger and by the Level 3 electron trigger
  based on the LAr Jet Trigger and the Fast Track Trigger.
  The same global timing conditions as mentioned above are 
  used in the {\em high y} triggers.
  The SpaCal trigger and the LAr electron trigger
  together with the L3 electron trigger are independent
  since the SpaCal trigger is fired by the backward going hadronic 
  final state particles. The efficiency of each of
  these two groups of triggers is determined using events 
  triggered by the other group as a monitor sample. 
  In the analysis events from both groups of triggers are used.
  The combined efficiency is calculated and is found to vary 
  between $91\%$ and $97\%$ at $E_e^{\prime}=3\,{\rm GeV}$.
  The statistical uncertainty of the combined efficiency together 
  with a $0.3\%$ uncertainty arising from the global timing conditions 
  is adopted as uncorrelated trigger uncertainty.
  It varies from $0.3\%$ at high electron energies
  to $2\%$ at $E_e^{\prime}=3\,{\rm GeV}$.

\item[Electron Track-Vertex Efficiency:] The efficiencies for
  reconstructing a track associated to the scattered lepton and for
  reconstructing the interaction vertex are determined
  simultaneously. The efficiency measurement follows the procedure
  used in the analysis of the HERA\,II $E_p=920$~GeV data and checked
  on the $E_p=460$ and $575$ GeV data. Three algorithms are used to
  determine the interaction vertex. The data and MC efficiencies are
  compared for each contributing algorithm.  The combined efficiency
  in the {\em nominal} analysis is found to be larger than $99.5\%$ in
  the data.  The residual differences between data and simulation
  after correction of simulation by $0.3\%$ define the uncorrelated
  systematic uncertainty which is $0.2\%$ and is considered to be
  uncorrelated.  In the {\em high y} analysis a more stringent
  requirement on the quality of the track associated to the scattered
  lepton is applied.  The efficiency was measured using electrons in
  the region of $E_e^{\prime}>18\,{\rm GeV}$ and checked at low
  $E_e^{\prime}$ using a sample of QED Compton events. It is found to
  be $96\%$ in data with a difference of $1\%$ between data and
  simulation.  This difference was corrected for and a $1\%$
  uncorrelated uncertainty is adopted.

\item[Electron Charge Identification Efficiency:] In the
  {\em high y} analysis the efficiency for correct charge
  identification of the scattered lepton is measured in the region
  $E_e^{\prime}>18\,{\rm GeV}$ which is free from photoproduction
  background. The simulation describes the efficiency of the data
  with an overall difference of $0.5\%$, and no significant time
  dependence is observed.  This is validated using ISR events in which
  the incoming beam positron has reduced energy due to QED radiation,
  yielding a sample of events which is free from photoproduction
  background but has $E_e^{\prime}$ below 12~GeV.  The measured
  cross section is corrected for the overall difference by increasing the
  measured values by $2\times 0.5\%$ with an uncertainty of
  $2\times 0.25\%$. The factor of two accounts for the fact
  that charge misidentification has a dual influence on the
  measurement by causing both a loss of signal events and an increase
  of the subtracted background~\cite{shushkevich}.

\item[Electron Identification:] A calorimetric algorithm based on
  longitudinal and transverse shower shape quantities is used to
  identify electrons in the $E_p=460$ and $575$~GeV data sample. The
  efficiency of this selection can be estimated using a simple track
  based electron finder which searches for an isolated high $p_T$
  track associated to an electromagnetic energy deposition. The
  efficiency is well described by the simulation and the uncertainty
  of $0.2\%~(0.5\%)$ is assigned in the {\em nominal} ({\em high y})
  analysis at $z_{\rm imp} < 20$~cm.  For $z_{\rm imp} > 20$~cm the
  uncertainty is taken to be $1\%$ due to the lack of statistics in
  this region selected by the track based algorithm.

\item[Extra Background Suppression:] The uncertainty on the efficiency
  of the $D_{ele}$ requirement has been studied with
  $J/\psi\rightarrow ee$  decays in
  data and is well described by the simulation. A variation of $\pm
  0.04$ around the nominal $D_{ele}$ cut value accommodates any residual
  difference between data and simulation. This variation leads to a
  cross section uncertainty of up to $2\%$ at highest $y$.

\item[{\em High y} Background Subtraction:] In the {\em high y}
  analysis the photoproduction background asymmetry is measured in the
  $E_p=460$ and $575$ GeV data, and found to be consistent with the
  determination using the $E_p=920$ GeV
  data~\cite{Aaron:2012qi,shushkevich}, albeit with reduced precision.
  Therefore the asymmetry is taken from the analysis of the HERA\,II
  data at $E_p=920$~GeV and the associated uncertainty is increased to
  $0.08$.  The resulting uncertainty on the measured cross sections
  is at most $2.7\%$ at $y=0.85$ and $Q^2 = 35$~GeV$^2$.

\item[QED Radiative Corrections:] An error on the cross
  sections originating from the QED radiative corrections is taken
  into account. This is determined by comparing the predicted
  radiative corrections from the programs {\sc Heracles} (as
  implemented in {\sc Djangoh}), {\sc Hector}, and {\sc
    Eprc}.  The radiative corrections due to the exchange
  of two or more photons between the lepton and the quark lines, which
  are not included in {\sc Djangoh}, vary with the polarity of the
  lepton beam.  This variation, estimated using {\sc Eprc}, is
  found to be small compared to the quoted errors and is
  neglected~\cite{shushkevich}.

\item[Model Uncertainty of Acceptance Correction:] The MC simulation
  is used to determine the acceptance correction to the data and
  relies on a specific choice of PDFs. The assigned uncertainty is
  listed in table~\ref{tab:syserr}.

\item[Luminosity:] The integrated luminosity is measured using the
  Bethe-Heitler process $ep\rightarrow ep\gamma$ with an uncertainty
  of $4\%$, of which $0.5\%$ is from the uncertainty in the
  theoretical calculation of this QED process.

\end{description}


\section{Results}
\label{sec:results}

\subsection{Double Differential Cross Sections}
\label{sec:dxdq2}

The reduced cross sections $\tilde{\sigma}_{\rm NC}(x, Q^2)$ for $P_e=0$ are
measured in the kinematic range $35 \leq Q^2\leq 800\,{\rm GeV}^2$ and
$0.00065\leq x\leq 0.65$ at two different centre-of-mass energies and
are referred to as the LAr data. The data are presented in
tables~\ref{tab:xsec460} --~\ref{tab:xsec575} and shown in
figure~\ref{fig:xsec-xq2}. The figure also includes previously
published H1 data~\cite{Collaboration:2010ry} in the $Q^2$ range of
the new data reported here, and are referred to as the SpaCal
data. The published $920$~GeV $e^+p$ LAr data~\cite{Aaron:2012qi} are scaled
by a normalisation factor of $1.018$~\cite{QEDCerratum}. This
correction factor arises from an error in the {\sc Compton} generator
used in the determination of the integrated luminosity of the HERA-II 920 GeV
data set.  The new LAr data provide additional low $x$ measurements
for $Q^2\geq 35$~GeV$^2$ (from the $E_p=460$ and $575$~GeV data sets). The
data are compared to the H1PDF\,2012 fit~\cite{Aaron:2012qi} which
provides a good description of the data. 

\subsection{Measurement of $\pmb{F}_{\pmb L}$}
\label{sec:fl}

According to equation~\ref{eq:Rnc} it is straightforward to determine
$F_L$ by a linear fit as a function of $y^2/(1+(1-y)^2)$ to the
reduced cross section measured at given values of $x$ and $Q^2$ but at
different centre-of-mass energies. An example of this
procedure is shown in figure~\ref{fig:rosenbluth} for six different
values of $x$ at $Q^2=60$~GeV$^2$. 
This method however does not optimally account
for correlations across all measurements, and therefore an
alternative procedure is applied.

The structure functions $F_L$ and $F_2$ are simultaneously determined
from the cross section measurements at $E_p=460,575$ and $920$~GeV
using a $\chi^2$ minimisation technique as employed
in~\cite{Collaboration:2010ry}. In this approach the values of $F_L$ and
$F_2$ at each measured $x,Q^2$ point are free parameters of the fit.
For $Q^2\leq 800$~GeV$^2$ the influence of the $xF_3$ structure
function is predicted to be small and is neglected. In addition, a set
of nuisance parameters $b_j$ for each correlated systematic error
source $j$ is introduced.  The minimisation is performed using the new
measurements presented here as well as previously published data from
H1~\cite{Collaboration:2010ry,Aaron:2012qi}.  The $\chi^2$ function
for the minimisation is
\begin{linenomath}
\begin{equation} \label{eq:chi2}
\chi^2 
\left(F_{L,i},F_{2,i},b_{j}\right) = \sum_i 
\frac{\textstyle \left[ \left(F_{2,i}- f(y_{i}) F_{L,i}\right) - 
\sum_j \Gamma_{i,j} b_j - \mu_i\right]^2}{\textstyle \Delta^2_i} + \sum_j b^2_j\,,
\end{equation}
\end{linenomath}
where $f(y)=y^2/(1+(1-y)^2)$ and $\mu_i$ is the measured reduced cross
section at an $x,Q^2$ point $i$ with a combined statistical and
uncorrelated systematic uncertainty $\Delta_i = \sqrt{\left(\Delta_{i,
    \rm stat}^2 + \Delta_{i,\rm syst}^2 \right)}$.  The effect of
correlated error sources $j$ on the cross section measurements is
given by the systematic error matrix $\Gamma_{i,j}$.  The correlations
of systematic uncertainties between the SpaCal data sets at different
energies are taken from~\cite{Collaboration:2010ry}. The
systematic uncertainties of the LAr measurements are taken to be $100\%$ correlated among
the $460\,,575$ and $920$ GeV data sets.  There is no correlation
between LAr and SpaCal measurements except for a common integrated
luminosity normalisation of the LAr and SpaCal data at
$E_p=460$ and $575$~GeV. 
For low $y\le 0.35$, the coefficient $f(y)$ is small compared to unity and thus $F_L$ can not be accurately measured. 
To avoid unphysical values for $F_L$ in this kinematic region the $\chi^2$ function is modified by adding an extra prior \cite{Collaboration:2010ry}.
The minimisation of the $\chi^2$ function with respect to these variables leads to a system of linear equations which is solved analytically. 
This technique is identical to the linear fit
discussed above when considering a single $x,~Q^2$ bin and neglecting
correlations between the cross section measurements.

The $\chi^2$ per degree of freedom is found to be $184/210$. 
The systematic sources include
normalisation uncertainties for the SpaCal and LAr data sets for
$E_p=460,\,575$ and $920$~GeV data which are all shifted in the
minimisation procedure by less than
one standard deviation with the exception of the LAr $920$~GeV data
which are re-normalised by $+3.4\%$, or $1.2$ standard
deviations. All other sources of uncertainty including those related
to calibration scales, noise subtractions, background estimates and polar angle
measurements are shifted by typically less than $0.3$ and never more
than $0.8$ standard deviations.

The measured structure functions are given in table~\ref{tab:flf2}
 over the full range in $Q^2$ from $1.5$ to
$800$~GeV$^2$. Only measurements of $F_L$ with a total uncertainty
less than $0.3$ for $Q^2\leq25$~GeV$^2$, or total uncertainty less
than $0.4$ for $Q^2\geq35$~GeV$^2$ are considered. The table also
includes the correlation coefficient $\rho$ between the $F_2$ and
$F_L$ values. In figures~\ref{fig:FLxq2-a} and~\ref{fig:FLxq2-b} the
measured structure functions $F_2$ and $F_L$ are shown in the regions
$2\leq Q^2\leq 25$~GeV$^2$ and $Q^2\geq 35$~GeV$^2$ respectively.  The
new data reported here, in which the scattered electron is recorded in
the LAr calorimeter, provide small additional constraints on the $F_L$
measurement for $1.5\leq Q^2\leq 25$~GeV$^2$ by means of correlations
in the systematic uncertainties.  The SpaCal and LAr data are used
together for $35\leq Q^2 \leq 90$~GeV$^2$ . For 
$Q^2\geq
120$~GeV$^2$ $F_L$ is determined exclusively from the LAr cross
section measurements. Therefore these data supersede the previous measurements of $F_2$ and $F_L$ in \cite{h1fl,Collaboration:2010ry}. 
For precision analyses of H1 data it is recommended to use the published
tables of the reduced differential cross sections given in
tables~\ref{tab:xsec460} to~\ref{tab:xsec575} and the full breakdown
of systematic uncertainties instead of the derived quantities $F_2$ or
$F_L$.


This measurement of $F_L$ and $F_2$ at high $y$ constitutes a model
independent method with no assumptions made on the values of the
structure functions. Within uncertainties the $F_L$ structure function is observed to be
positive everywhere and approximately equal to $20\%$ of
$F_2$. Also shown are the $F_L$ and $F_2$ measurements from the ZEUS
collaboration~\cite{Chekanov:2009na} which agree with the H1 data. The
ZEUS data are moved to the $Q^2$ values of the H1 measurements using
the H1PDF\,2012 NLO QCD fit.  This QCD fit is able to provide a good
description of both measurements of $F_L$ and $F_2$ across the full
$Q^2$ range.

In order to reduce the experimental uncertainties the $F_L$
measurements are combined at each $Q^2$ value. Furthermore the highest
$Q^2$ bins are also averaged to achieve an approximately uniform
experimental precision over the full kinematic range of the
measurement. The $Q^2$ averaging is performed for $Q^2=300$ and
$400$~GeV$^2$, and for the $Q^2=500,~600$ and $800$~GeV$^2$
values.  The resulting data are given in table~\ref{tab:flave} and shown
in figure~\ref{fig:FLq2} where the average $x$ for each $Q^2$ is
provided on the upper scale of the figure. The data are compared to a
suite of QCD predictions at NNLO: HERAPDF1.5~\cite{HERAPDF1.5},
CT10~\cite{CT10}, ABM11~\cite{ABM11}, MSTW2008~\cite{MSTW2008},
JR09~\cite{Gluck:2007ck} and NNPDF2.3~\cite{NNPDF2.3}. In all cases
the perturbative calculations provide a reasonable description of the
data.  

A similar average of $F_L$ measurements over $x$ has already been performed in~\cite{Collaboration:2010ry} for $Q^2<45$~GeV$^2$. A small problem in~\cite{Collaboration:2010ry} has been identified in the averaging procedure which lead to underestimated correlated systematic uncertainties which has been corrected in the measurements reported here. Therefore
the data presented in  tables~\ref{tab:flave} supersedes the corresponding table from~\cite{Collaboration:2010ry}.

The cross section ratio $R$ of longitudinally to transversely polarised
virtual photons is related to the structure functions $F_2$ and $F_L$ as
\begin{equation}
R = \frac{\sigma_L}{\sigma_T} = \frac{F_L}{F_2-F_L} \,.
\label{eq:Rflf2}
\end{equation} 
This ratio has previously been observed to be approximately constant
for $3.5 \leq Q^2 \leq 45$~GeV$^2$~\cite{Collaboration:2010ry}.  

The values of $R$ as a function of $Q^2$ are determined by minimising
the $\chi^2$ function of equation~\ref{eq:chi2} in which $F_L$ is
replaced by
$$F_L=\frac{R}{1+R}F_2$$
assuming the value of $R$ is constant as a function of $x$ for a given
$Q^2$. 
The minimum is found numerically in this case, using the MINUIT package \cite{minuit}.
The asymmetric uncertainties are determined using a MC method
in which the mean squared deviation from the measured value of $R$ is
used to define the asymmetric uncertainties. The resulting value of
$R(Q^2)$ is shown in figure~\ref{fig:Rq2}.
%
%
The measurements are compared to the prediction of the HERAPDF1.5 NNLO for $\sqrt{s}=225$~GeV and $y=0.7$. { The expected small variation of $R$ in the region of $x$ in which the data are sensitive to this quantity is also shown.  

The data are found to be consistent with a constant value across the
entire $Q^2$ range shown. 
The fit is repeated by assuming
that $R$ is constant over the full $Q^2$ range.
 This yields a value of $R=0.23\pm0.04$ with
$\chi^2$/ndf =$314/367$ 
which agrees well with the value obtained
previously~\cite{Collaboration:2010ry} using only data up to
$Q^2=45$~GeV$^2$, and with the ZEUS data.

\kd{
In NLO and NNLO QCD analyses of precision DIS data on $F_2$ and the reduced NC cross sections the gluon density is constrained indirectly via scaling violations.
The Altarelli-Martinelli relation~\cite{altarelli78}, however, would allow for a direct extraction of the gluon density from measurements of $F_L$. This relation cannot be solved analytically for the gluon density, but approximate solutions have been proposed
~\cite{CooperSarkar:1987ds,Zijlstra:1992qd,Boroun:2012}
\begin{equation}
xg(x,Q^2) \approx 1.77 \frac{3\pi}{2\alpha_S(Q^2)} F_L(ax,Q^2) \,,
\label{eq:xgapprox}
\end{equation} 
where $a$ is a numerical factor and is here set to unity.
This relation can be used to demonstrate sensitivity of the direct
measurement of $F_L$} to the gluon density by 
comparing the gluon obtained from the $F_L$ measurements to the
predicted gluon density obtained from a NLO QCD fit to DIS data. 
In figure~\ref{fig:gluon} the gluon density extracted according to equation~\ref{eq:xgapprox}
is compared to the prediction from the gluon density determined in the NLO HERA\-PDF1.5 QCD fit. 
In order to judge on the goodness of the approximation, 
the gluon density as obtained by applying equation~\ref{eq:xgapprox} to the $F_L$ prediction based on  the NLO HERAPDF1.5 QCD fit is also shown.
A reasonable agreement between the gluon density as extracted from the direct measurement of $F_L$ based on the  approximate relation with the gluon derived indirectly from scaling violations
is observed.

\section{Conclusions}
\label{sec:summary}

The unpolarised neutral current inclusive DIS cross section for $ep$
interactions are measured at two centre-of-mass energies of
$\sqrt{s}=225$ and $252$~GeV \kd{ in the region of 
$35<Q^2<800$~GeV$^2$,}
with integrated luminosities of
$11.8$~pb$^{-1}$ and $5.4$~pb$^{-1}$ respectively. The measurements
are performed up to the highest accessible inelasticity of $y=0.85$
where the contribution of the $F_L$ structure function to the reduced
cross section is sizeable. The data are used together with previously
published measurements at $\sqrt{s}=319$ GeV ($E_p=920$~GeV) to
simultaneously extract the $F_L$ and $F_2$ structure functions in a
model independent way. The
new data extend previous measurements of $F_L$ up to 
$Q^2 =800$~GeV$^2$ and supersede previous H1 data. Predictions of
different perturbative QCD calculations at NNLO are compared to
data. Good agreement is observed between the measurements and the
theoretical calculations. The ratio $R$ of the longitudinally to
transversely polarised virtual photon cross section is consistent with
being constant
over the kinematic range of the data, and is determined to be
$=0.23\pm0.04$. 
The $F_L$ measurements are used to perform 
{a gluon density extraction based on a NLO approximation} which is found to agree reasonably well with the gluon determined from scaling violations.

\section*{Acknowledgements}

We are grateful to the HERA machine group whose outstanding efforts
have made this experiment possible.  We thank the engineers and
technicians for their work in constructing and maintaining the H1
detector, our funding agencies for financial support, the DESY
technical staff for continual assistance and the DESY directorate for
support and for the hospitality which they extend to the non DESY
members of the collaboration.  We would like to give credit to all
partners contributing to the EGI computing infrastructure for their
support for the H1 Collaboration. We would also like to thank the
members of the MSTW, CT, ABM, JR, NNPDF and HERAPDF collaborations for
their help in producing theoretical predictions of $F_L$ shown in
figure~\ref{fig:FLq2}.

\newpage


\newpage

\begin{table}[htbp] 
\begin{center} 
\tiny 
\begin{tabular}{|r|c|c|r|r|r|r|r|r|r|r|r|r|r|r|r|} 
\hline 
$Q^2$ & $x$ & $y$ & $\tilde{\sigma}_{\rm NC}$ & 
$\delta_{\rm tot}$ & $\delta_{\rm stat}$ & $\delta_{\rm unc}$ & 
$\delta_{\rm unc}^{E}$ & 
$\delta_{\rm unc}^{h}$& 
$\delta_{\rm cor}$ & 
$\delta_{\rm cor}^{E^+}$ & 
$\delta_{\rm cor}^{\theta^+}$& 
$\delta_{\rm cor}^{h^+}$& 
$\delta_{\rm cor}^{N^+}$& 
$\delta_{\rm cor}^{S^+}$& 
$\delta_{\rm cor}^{D^+}$ \\ 
$(\rm GeV^2)$ & & & & 
$(\%)$ & $(\%)$ & $(\%)$ & $(\%)$ & $(\%)$ & $(\%)$ & 
$(\%)$ & $(\%)$ & $(\%)$ & $(\%)$ & $(\%)$ & $(\%)$ 
\\ \hline 
$35$ & $8.10 \times 10^{-4}$ & $0.850$ & $1.343$ & $6.5$ & $4.5$ & $3.8$ & $0.6$ & $2.8$ & $2.8$ & $-0.3$ & $-0.4$ & $0.2$ & $0.7$ & $2.3$ & $1.2$ \\ 
 
\hline 
$45$ & $1.04 \times 10^{-3}$ & $0.850$ & $1.173$ & $6.3$ & $4.7$ & $3.4$ & $0.4$ & $2.4$ & $2.4$ & $-0.1$ & $-0.5$ & $0.1$ & $0.6$ & $2.1$ & $1.0$ \\ 
 
$45$ & $1.18 \times 10^{-3}$ & $0.750$ & $1.187$ & $5.7$ & $5.1$ & $2.2$ & $0.6$ & $0.8$ & $1.3$ & $0.2$ & $-0.5$ & $0.0$ & $0.3$ & $0.6$ & $1.0$ \\ 
 
\hline 
$60$ & $1.39 \times 10^{-3}$ & $0.850$ & $1.190$ & $6.2$ & $5.0$ & $3.0$ & $0.2$ & $2.0$ & $2.0$ & $-0.1$ & $-0.3$ & $0.1$ & $0.5$ & $1.7$ & $0.8$ \\ 
 
$60$ & $1.58 \times 10^{-3}$ & $0.750$ & $1.117$ & $4.7$ & $4.0$ & $2.0$ & $0.5$ & $0.6$ & $1.4$ & $-0.2$ & $-0.6$ & $0.0$ & $0.2$ & $0.6$ & $1.0$ \\ 
 
\hline 
$90$ & $2.09 \times 10^{-3}$ & $0.850$ & $1.269$ & $6.3$ & $5.3$ & $2.9$ & $0.3$ & $1.8$ & $1.8$ & $-0.2$ & $-0.4$ & $0.1$ & $0.5$ & $1.4$ & $0.9$ \\ 
 
$90$ & $2.36 \times 10^{-3}$ & $0.750$ & $1.193$ & $4.6$ & $4.1$ & $1.9$ & $0.3$ & $0.5$ & $1.1$ & $-0.2$ & $-0.5$ & $0.1$ & $0.2$ & $0.3$ & $0.9$ \\ 
 
$90$ & $2.73 \times 10^{-3}$ & $0.650$ & $1.156$ & $4.2$ & $3.8$ & $1.8$ & $0.4$ & $0.2$ & $0.8$ & $-0.2$ & $-0.6$ & $0.0$ & $0.2$ & $0.3$ & $0.4$ \\ 
 
\hline 
$120$ & $2.78 \times 10^{-3}$ & $0.850$ & $1.249$ & $6.8$ & $6.1$ & $2.7$ & $0.1$ & $1.5$ & $1.6$ & $0.0$ & $-0.4$ & $0.1$ & $0.4$ & $0.9$ & $1.2$ \\ 
 
$120$ & $3.15 \times 10^{-3}$ & $0.750$ & $1.099$ & $5.3$ & $4.8$ & $1.9$ & $0.4$ & $0.4$ & $0.9$ & $-0.2$ & $-0.4$ & $0.0$ & $0.2$ & $0.4$ & $0.6$ \\ 
 
$120$ & $3.63 \times 10^{-3}$ & $0.650$ & $1.052$ & $4.7$ & $4.3$ & $1.9$ & $0.6$ & $0.2$ & $0.6$ & $-0.2$ & $-0.4$ & $0.0$ & $0.2$ & $0.2$ & $0.2$ \\ 
 
$120$ & $4.82 \times 10^{-3}$ & $0.490$ & $1.041$ & $3.3$ & $2.7$ & $1.8$ & $0.6$ & $0.1$ & $0.8$ & $-0.3$ & $-0.7$ & $0.0$ & $0.2$ & $0.1$ & $0.0$ \\ 
 
\hline 
$150$ & $3.47 \times 10^{-3}$ & $0.850$ & $1.230$ & $7.8$ & $7.1$ & $2.6$ & $0.4$ & $1.4$ & $1.9$ & $-0.2$ & $-0.3$ & $0.1$ & $0.4$ & $0.8$ & $1.6$ \\ 
 
$150$ & $3.94 \times 10^{-3}$ & $0.750$ & $1.024$ & $6.1$ & $5.8$ & $1.9$ & $0.4$ & $0.3$ & $0.8$ & $-0.2$ & $-0.4$ & $0.0$ & $0.2$ & $0.2$ & $0.6$ \\ 
 
$150$ & $4.54 \times 10^{-3}$ & $0.650$ & $1.010$ & $5.4$ & $5.0$ & $2.0$ & $0.9$ & $0.1$ & $0.5$ & $-0.2$ & $-0.4$ & $0.0$ & $0.2$ & $0.2$ & $0.1$ \\ 
 
$150$ & $6.03 \times 10^{-3}$ & $0.490$ & $1.060$ & $3.2$ & $2.5$ & $1.8$ & $0.5$ & $0.1$ & $0.7$ & $-0.3$ & $-0.6$ & $0.0$ & $0.2$ & $0.1$ & $0.0$ \\ 
 
$150$ & $8.00 \times 10^{-3}$ & $0.369$ & $0.9774$ & $3.0$ & $2.6$ & $1.4$ & $0.6$ & $0.0$ & $0.9$ & $-0.4$ & $-0.8$ & $0.0$ & $0.1$ & $0.0$ & $0.0$ \\ 
 
$150$ & $1.30 \times 10^{-2}$ & $0.227$ & $0.8384$ & $3.8$ & $3.3$ & $1.5$ & $1.2$ & $0.0$ & $1.1$ & $-0.8$ & $-0.7$ & $0.0$ & $0.0$ & $0.0$ & $0.0$ \\ 
 
$150$ & $2.00 \times 10^{-2}$ & $0.148$ & $0.7006$ & $5.2$ & $4.5$ & $2.2$ & $1.7$ & $1.0$ & $1.7$ & $-1.0$ & $-1.0$ & $-0.4$ & $-0.8$ & $0.0$ & $0.0$ \\ 
 
\hline 
$200$ & $4.63 \times 10^{-3}$ & $0.850$ & $1.117$ & $9.6$ & $9.1$ & $2.5$ & $0.3$ & $1.1$ & $2.1$ & $-0.1$ & $-0.4$ & $0.0$ & $0.3$ & $0.6$ & $1.9$ \\ 
 
$200$ & $5.25 \times 10^{-3}$ & $0.750$ & $1.011$ & $8.1$ & $7.7$ & $1.9$ & $0.3$ & $0.3$ & $1.1$ & $0.2$ & $-0.4$ & $0.0$ & $0.2$ & $0.1$ & $1.0$ \\ 
 
$200$ & $6.06 \times 10^{-3}$ & $0.650$ & $0.9997$ & $6.8$ & $6.5$ & $2.0$ & $0.9$ & $0.2$ & $0.6$ & $-0.1$ & $-0.5$ & $0.1$ & $0.2$ & $0.2$ & $0.0$ \\ 
 
$200$ & $8.04 \times 10^{-3}$ & $0.490$ & $0.9567$ & $3.8$ & $3.4$ & $1.7$ & $0.3$ & $0.1$ & $0.6$ & $-0.3$ & $-0.6$ & $0.0$ & $0.2$ & $0.1$ & $0.0$ \\ 
 
$200$ & $1.30 \times 10^{-2}$ & $0.303$ & $0.8430$ & $3.4$ & $3.1$ & $1.0$ & $0.6$ & $0.0$ & $0.8$ & $-0.4$ & $-0.7$ & $0.0$ & $0.0$ & $0.0$ & $0.0$ \\ 
 
$200$ & $2.00 \times 10^{-2}$ & $0.197$ & $0.6517$ & $4.1$ & $3.5$ & $1.8$ & $1.6$ & $0.0$ & $1.2$ & $-1.0$ & $-0.7$ & $0.0$ & $0.0$ & $0.0$ & $0.0$ \\ 
 
$200$ & $3.20 \times 10^{-2}$ & $0.123$ & $0.5275$ & $4.2$ & $4.0$ & $0.9$ & $0.1$ & $0.1$ & $0.6$ & $-0.1$ & $-0.5$ & $-0.1$ & $0.3$ & $0.0$ & $0.0$ \\ 
 
$200$ & $5.00 \times 10^{-2}$ & $0.079$ & $0.5297$ & $4.3$ & $4.1$ & $1.2$ & $0.9$ & $0.2$ & $0.7$ & $-0.5$ & $-0.4$ & $-0.1$ & $0.2$ & $0.0$ & $0.0$ \\ 
 
$200$ & $8.00 \times 10^{-2}$ & $0.049$ & $0.4587$ & $5.0$ & $4.7$ & $1.3$ & $0.9$ & $0.3$ & $1.0$ & $-0.6$ & $-0.7$ & $-0.2$ & $0.2$ & $0.0$ & $0.0$ \\ 
 
$200$ & $1.30 \times 10^{-1}$ & $0.030$ & $0.3610$ & $5.6$ & $5.1$ & $1.9$ & $1.5$ & $0.1$ & $1.6$ & $-0.9$ & $-0.7$ & $-0.1$ & $1.0$ & $0.0$ & $0.0$ \\ 
 
$200$ & $1.80 \times 10^{-1}$ & $0.022$ & $0.3201$ & $6.8$ & $5.8$ & $2.3$ & $1.1$ & $1.4$ & $2.6$ & $-0.7$ & $-0.9$ & $-0.4$ & $-2.3$ & $0.0$ & $0.0$ \\ 
 
$200$ & $4.00 \times 10^{-1}$ & $0.010$ & $0.1694$ & $13.1$ & $8.2$ & $4.5$ & $0.3$ & $4.0$ & $9.2$ & $0.2$ & $-1.1$ & $-0.6$ & $-9.1$ & $0.0$ & $0.0$ \\ 
 
\hline 
$250$ & $5.79 \times 10^{-3}$ & $0.850$ & $1.049$ & $10.9$ & $10.4$ & $2.5$ & $0.3$ & $0.9$ & $2.2$ & $-0.1$ & $-0.2$ & $0.0$ & $0.3$ & $0.6$ & $2.0$ \\ 
 
$250$ & $6.56 \times 10^{-3}$ & $0.750$ & $1.036$ & $9.1$ & $8.8$ & $1.9$ & $0.2$ & $0.3$ & $1.3$ & $-0.2$ & $-0.4$ & $0.1$ & $0.2$ & $0.2$ & $1.1$ \\ 
 
$250$ & $7.57 \times 10^{-3}$ & $0.650$ & $0.9480$ & $8.0$ & $7.7$ & $2.0$ & $0.9$ & $0.1$ & $0.5$ & $-0.2$ & $-0.4$ & $0.0$ & $0.2$ & $0.2$ & $0.0$ \\ 
 
$250$ & $1.00 \times 10^{-2}$ & $0.490$ & $0.8829$ & $4.3$ & $3.9$ & $1.7$ & $0.3$ & $0.1$ & $0.6$ & $-0.3$ & $-0.5$ & $0.0$ & $0.2$ & $0.1$ & $0.0$ \\ 
 
$250$ & $1.30 \times 10^{-2}$ & $0.379$ & $0.8281$ & $4.0$ & $3.7$ & $1.3$ & $0.4$ & $0.0$ & $0.6$ & $-0.4$ & $-0.5$ & $0.0$ & $0.1$ & $0.0$ & $0.0$ \\ 
 
$250$ & $2.00 \times 10^{-2}$ & $0.246$ & $0.6799$ & $4.0$ & $3.8$ & $1.0$ & $0.6$ & $0.0$ & $0.8$ & $-0.5$ & $-0.7$ & $0.0$ & $0.0$ & $0.0$ & $0.0$ \\ 
 
$250$ & $3.20 \times 10^{-2}$ & $0.154$ & $0.5817$ & $4.4$ & $4.1$ & $1.3$ & $1.0$ & $0.2$ & $0.9$ & $0.4$ & $-0.6$ & $0.1$ & $0.6$ & $0.0$ & $0.0$ \\ 
 
$250$ & $5.00 \times 10^{-2}$ & $0.098$ & $0.5025$ & $4.4$ & $4.1$ & $1.3$ & $0.9$ & $0.1$ & $0.9$ & $0.3$ & $-0.6$ & $0.0$ & $0.7$ & $0.0$ & $0.0$ \\ 
 
$250$ & $8.00 \times 10^{-2}$ & $0.062$ & $0.4429$ & $4.7$ & $4.4$ & $1.3$ & $1.0$ & $0.1$ & $1.0$ & $0.5$ & $-0.6$ & $-0.1$ & $0.7$ & $0.0$ & $0.0$ \\ 
 
$250$ & $1.30 \times 10^{-1}$ & $0.038$ & $0.3750$ & $4.9$ & $4.4$ & $1.4$ & $0.8$ & $0.1$ & $1.6$ & $0.3$ & $-0.5$ & $-0.1$ & $1.5$ & $0.0$ & $0.0$ \\ 
 
$250$ & $1.80 \times 10^{-1}$ & $0.027$ & $0.3582$ & $5.1$ & $4.5$ & $2.1$ & $1.3$ & $0.9$ & $1.3$ & $0.6$ & $-0.7$ & $-0.3$ & $-0.8$ & $0.0$ & $0.0$ \\ 
 
$250$ & $4.00 \times 10^{-1}$ & $0.012$ & $0.1675$ & $12.6$ & $6.6$ & $4.9$ & $2.7$ & $3.6$ & $9.5$ & $1.6$ & $-1.0$ & $-0.6$ & $-9.3$ & $0.0$ & $0.0$ \\ 
 
\hline 
$300$ & $6.95 \times 10^{-3}$ & $0.850$ & $0.8700$ & $13.8$ & $13.3$ & $2.4$ & $0.2$ & $0.8$ & $2.5$ & $-0.2$ & $-0.2$ & $0.0$ & $0.3$ & $0.8$ & $2.3$ \\ 
 
$300$ & $7.88 \times 10^{-3}$ & $0.750$ & $0.8274$ & $11.1$ & $10.9$ & $2.0$ & $0.2$ & $0.3$ & $1.0$ & $-0.2$ & $-0.3$ & $0.0$ & $0.2$ & $0.3$ & $0.9$ \\ 
 
$300$ & $9.09 \times 10^{-3}$ & $0.650$ & $0.8411$ & $9.8$ & $9.6$ & $1.9$ & $0.3$ & $0.1$ & $0.5$ & $0.1$ & $-0.4$ & $0.0$ & $0.2$ & $0.2$ & $0.0$ \\ 
 
$300$ & $1.21 \times 10^{-2}$ & $0.490$ & $0.9058$ & $4.8$ & $4.5$ & $1.7$ & $0.3$ & $0.1$ & $0.5$ & $-0.3$ & $-0.4$ & $0.0$ & $0.1$ & $0.0$ & $0.0$ \\ 
 
$300$ & $2.00 \times 10^{-2}$ & $0.295$ & $0.7296$ & $4.4$ & $4.2$ & $1.0$ & $0.6$ & $0.0$ & $0.8$ & $-0.6$ & $-0.5$ & $0.0$ & $0.0$ & $0.0$ & $0.0$ \\ 
 
$300$ & $3.20 \times 10^{-2}$ & $0.185$ & $0.6231$ & $4.7$ & $4.5$ & $0.9$ & $0.3$ & $0.2$ & $0.7$ & $0.3$ & $-0.3$ & $0.1$ & $0.5$ & $0.0$ & $0.0$ \\ 
 
$300$ & $5.00 \times 10^{-2}$ & $0.118$ & $0.5210$ & $4.9$ & $4.7$ & $1.1$ & $0.6$ & $0.2$ & $0.9$ & $0.4$ & $-0.5$ & $0.1$ & $0.6$ & $0.0$ & $0.0$ \\ 
 
$300$ & $8.00 \times 10^{-2}$ & $0.074$ & $0.4584$ & $5.2$ & $4.9$ & $1.4$ & $1.0$ & $0.2$ & $1.0$ & $0.6$ & $-0.6$ & $-0.1$ & $0.5$ & $0.0$ & $0.0$ \\ 
 
$300$ & $1.30 \times 10^{-1}$ & $0.045$ & $0.3695$ & $5.5$ & $5.1$ & $1.5$ & $1.0$ & $0.2$ & $1.6$ & $0.5$ & $-0.6$ & $0.0$ & $1.4$ & $0.0$ & $0.0$ \\ 
 
$300$ & $1.80 \times 10^{-1}$ & $0.033$ & $0.3330$ & $5.8$ & $5.2$ & $2.2$ & $1.5$ & $0.9$ & $1.2$ & $0.9$ & $-0.8$ & $-0.3$ & $-0.3$ & $0.0$ & $0.0$ \\ 
 
$300$ & $4.00 \times 10^{-1}$ & $0.015$ & $0.1567$ & $13.0$ & $7.7$ & $5.2$ & $3.1$ & $3.6$ & $9.1$ & $2.1$ & $-1.3$ & $-0.6$ & $-8.8$ & $0.0$ & $0.0$ \\ 
 
\hline 

\end{tabular} 
\end{center} 
\caption[RESULT] 
{\sl \label{tab:xsec460} The NC 
$e^+p$ reduced cross section $\tilde{\sigma}_{\rm NC}(x,Q^2)$ 
for $E_p=460$ GeV and $P_e=0$ 
with 
total $(\delta_{tot})$, 
statistical 
$(\delta_{\rm stat})$, 
total uncorrelated systematic $(\delta_{\rm unc})$ 
errors and two of its contributions from the 
 electron energy error ($\delta_{unc}^{E}$)  
and the hadronic energy error  
($\delta_{\rm unc}^{h}$). 
The effect of the other uncorrelated 
systematic errors is included in $\delta_{\rm unc}$. 
In addition the correlated systematic  
$(\delta_{\rm cor})$ and its contributions from a 
positive variation of one  
standard deviation of the 
electron energy error ($\delta_{cor}^{E^+}$), of 
the polar electron angle error 
($\delta_{\rm cor}^{\theta^+}$), of the hadronic 
energy error ($\delta_{\rm cor}^{h^+}$), of the error 
due to noise subtraction ($\delta_{\rm cor}^{N^+}$), 
of the error 
due to background subtraction charge asymmetry 
 ($\delta_{\rm cor}^{S^+}$) 
and of the error 
due to variation of the cut value on the electron discriminator $D_{ele}$  
 ($\delta_{\rm cor}^{D^+}$) are given. 
The overall normalisation uncertainty of $4\%$ is 
not included in the errors. 
}
\end{table} 
\begin{table}[htbp] 
\begin{center} 
\tiny 
\begin{tabular}{|r|c|c|r|r|r|r|r|r|r|r|r|r|r|r|r|} 
\hline 
$Q^2$ & $x$ & $y$ & $\tilde{\sigma}_{\rm NC}$ & 
$\delta_{\rm tot}$ & $\delta_{\rm stat}$ & $\delta_{\rm unc}$ & 
$\delta_{\rm unc}^{E}$ & 
$\delta_{\rm unc}^{h}$& 
$\delta_{\rm cor}$ & 
$\delta_{\rm cor}^{E^+}$ & 
$\delta_{\rm cor}^{\theta^+}$& 
$\delta_{\rm cor}^{h^+}$& 
$\delta_{\rm cor}^{N^+}$& 
$\delta_{\rm cor}^{S^+}$& 
$\delta_{\rm cor}^{D^+}$ \\ 
$(\rm GeV^2)$ & & & & 
$(\%)$ & $(\%)$ & $(\%)$ & $(\%)$ & $(\%)$ & $(\%)$ & 
$(\%)$ & $(\%)$ & $(\%)$ & $(\%)$ & $(\%)$ & $(\%)$ 
\\ \hline 
$400$ & $9.27 \times 10^{-3}$ & $0.850$ & $1.025$ & $13.8$ & $13.3$ & $2.4$ & $0.6$ & $0.6$ & $2.5$ & $-0.2$ & $-0.3$ & $0.0$ & $0.2$ & $0.6$ & $2.4$ \\ 
 
$400$ & $1.05 \times 10^{-2}$ & $0.750$ & $1.074$ & $10.4$ & $10.1$ & $2.2$ & $1.0$ & $0.3$ & $0.7$ & $0.2$ & $-0.3$ & $0.1$ & $0.2$ & $0.1$ & $0.6$ \\ 
 
$400$ & $1.21 \times 10^{-2}$ & $0.650$ & $0.9263$ & $10.0$ & $9.8$ & $1.9$ & $0.2$ & $0.1$ & $0.4$ & $-0.2$ & $-0.3$ & $0.0$ & $0.2$ & $0.2$ & $0.0$ \\ 
 
$400$ & $1.61 \times 10^{-2}$ & $0.490$ & $0.8145$ & $5.7$ & $5.4$ & $1.7$ & $0.2$ & $0.0$ & $0.5$ & $-0.2$ & $-0.5$ & $0.0$ & $0.1$ & $0.0$ & $0.0$ \\ 
 
$400$ & $3.20 \times 10^{-2}$ & $0.246$ & $0.6305$ & $5.2$ & $5.0$ & $1.0$ & $0.6$ & $0.0$ & $0.8$ & $-0.6$ & $-0.6$ & $0.0$ & $0.0$ & $0.0$ & $0.0$ \\ 
 
$400$ & $5.00 \times 10^{-2}$ & $0.157$ & $0.5686$ & $5.4$ & $5.2$ & $1.0$ & $0.6$ & $0.2$ & $0.9$ & $0.5$ & $-0.5$ & $-0.1$ & $0.4$ & $0.0$ & $0.0$ \\ 
 
$400$ & $8.00 \times 10^{-2}$ & $0.098$ & $0.4493$ & $5.8$ & $5.7$ & $1.0$ & $0.4$ & $0.1$ & $0.7$ & $0.4$ & $-0.4$ & $0.1$ & $0.5$ & $0.0$ & $0.0$ \\ 
 
$400$ & $1.30 \times 10^{-1}$ & $0.061$ & $0.4300$ & $5.6$ & $5.3$ & $1.2$ & $0.4$ & $0.1$ & $1.2$ & $0.4$ & $-0.4$ & $-0.1$ & $1.1$ & $0.0$ & $0.0$ \\ 
 
$400$ & $1.80 \times 10^{-1}$ & $0.044$ & $0.3375$ & $6.2$ & $5.8$ & $1.7$ & $0.8$ & $0.7$ & $0.9$ & $0.7$ & $-0.6$ & $-0.2$ & $-0.2$ & $0.0$ & $0.0$ \\ 
 
$400$ & $4.00 \times 10^{-1}$ & $0.020$ & $0.1494$ & $13.1$ & $8.7$ & $4.6$ & $1.9$ & $3.7$ & $8.6$ & $1.9$ & $-0.9$ & $-0.7$ & $-8.3$ & $0.0$ & $0.0$ \\ 
 
\hline 
$500$ & $1.16 \times 10^{-2}$ & $0.850$ & $1.002$ & $15.0$ & $14.6$ & $2.4$ & $0.1$ & $0.4$ & $2.2$ & $0.2$ & $-0.2$ & $0.0$ & $0.2$ & $0.0$ & $2.2$ \\ 
 
$500$ & $1.31 \times 10^{-2}$ & $0.750$ & $0.7577$ & $13.8$ & $13.6$ & $2.1$ & $0.6$ & $0.2$ & $0.5$ & $-0.3$ & $-0.3$ & $0.0$ & $0.2$ & $0.1$ & $0.1$ \\ 
 
$500$ & $1.51 \times 10^{-2}$ & $0.650$ & $0.6938$ & $12.4$ & $12.2$ & $1.9$ & $0.3$ & $0.1$ & $0.4$ & $-0.2$ & $-0.4$ & $0.1$ & $0.2$ & $0.0$ & $0.0$ \\ 
 
$500$ & $2.01 \times 10^{-2}$ & $0.490$ & $0.7395$ & $6.7$ & $6.5$ & $1.7$ & $0.1$ & $0.0$ & $0.4$ & $-0.1$ & $-0.4$ & $0.0$ & $0.1$ & $0.0$ & $0.0$ \\ 
 
$500$ & $3.20 \times 10^{-2}$ & $0.308$ & $0.6559$ & $6.1$ & $6.0$ & $1.0$ & $0.3$ & $0.0$ & $0.6$ & $-0.3$ & $-0.5$ & $0.0$ & $0.0$ & $0.0$ & $0.0$ \\ 
 
$500$ & $5.00 \times 10^{-2}$ & $0.197$ & $0.6106$ & $6.1$ & $5.9$ & $1.3$ & $0.9$ & $0.0$ & $1.1$ & $-0.9$ & $-0.6$ & $0.0$ & $0.0$ & $0.0$ & $0.0$ \\ 
 
$500$ & $8.00 \times 10^{-2}$ & $0.123$ & $0.4712$ & $6.5$ & $6.3$ & $1.0$ & $0.5$ & $0.0$ & $0.8$ & $0.5$ & $-0.4$ & $0.0$ & $0.4$ & $0.0$ & $0.0$ \\ 
 
$500$ & $1.30 \times 10^{-1}$ & $0.076$ & $0.4112$ & $7.7$ & $7.5$ & $1.4$ & $0.6$ & $0.1$ & $1.2$ & $0.6$ & $-0.5$ & $0.0$ & $0.9$ & $0.0$ & $0.0$ \\ 
 
$500$ & $1.80 \times 10^{-1}$ & $0.055$ & $0.3045$ & $8.7$ & $8.4$ & $1.5$ & $0.6$ & $0.1$ & $1.4$ & $0.6$ & $-0.4$ & $-0.1$ & $1.1$ & $0.0$ & $0.0$ \\ 
 
$500$ & $2.50 \times 10^{-1}$ & $0.039$ & $0.2759$ & $8.5$ & $8.3$ & $1.9$ & $0.9$ & $0.8$ & $1.2$ & $0.9$ & $-0.6$ & $-0.2$ & $-0.3$ & $0.0$ & $0.0$ \\ 
 
$500$ & $4.00 \times 10^{-1}$ & $0.025$ & $0.1311$ & $13.7$ & $11.8$ & $4.1$ & $1.9$ & $3.1$ & $5.8$ & $1.8$ & $-0.7$ & $-0.7$ & $-5.3$ & $0.0$ & $0.0$ \\ 
 
$500$ & $6.50 \times 10^{-1}$ & $0.015$ & $0.01698$ & $27.9$ & $23.0$ & $7.0$ & $2.8$ & $5.9$ & $14.3$ & $2.8$ & $-1.4$ & $-1.0$ & $-13.9$ & $0.0$ & $0.0$ \\ 
 
\hline 
$650$ & $1.51 \times 10^{-2}$ & $0.850$ & $0.8058$ & $19.6$ & $19.4$ & $2.7$ & $0.5$ & $0.5$ & $1.1$ & $-0.1$ & $0.2$ & $0.1$ & $0.3$ & $0.7$ & $0.7$ \\ 
 
$650$ & $1.71 \times 10^{-2}$ & $0.750$ & $0.9192$ & $14.0$ & $13.9$ & $2.0$ & $0.1$ & $0.2$ & $0.4$ & $0.1$ & $-0.3$ & $0.0$ & $0.2$ & $0.2$ & $0.0$ \\ 
 
$650$ & $1.97 \times 10^{-2}$ & $0.650$ & $0.9125$ & $12.1$ & $12.0$ & $1.9$ & $0.0$ & $0.1$ & $0.4$ & $-0.1$ & $-0.4$ & $0.0$ & $0.1$ & $0.1$ & $0.0$ \\ 
 
$650$ & $2.61 \times 10^{-2}$ & $0.490$ & $0.6085$ & $8.0$ & $7.8$ & $1.8$ & $0.5$ & $0.0$ & $0.5$ & $-0.2$ & $-0.4$ & $0.0$ & $0.1$ & $0.0$ & $0.0$ \\ 
 
$650$ & $5.00 \times 10^{-2}$ & $0.256$ & $0.4952$ & $7.9$ & $7.8$ & $1.1$ & $0.6$ & $0.0$ & $0.8$ & $-0.6$ & $-0.6$ & $0.0$ & $0.0$ & $0.0$ & $0.0$ \\ 
 
$650$ & $8.00 \times 10^{-2}$ & $0.160$ & $0.4515$ & $7.9$ & $7.8$ & $1.0$ & $0.4$ & $0.2$ & $0.8$ & $0.5$ & $-0.4$ & $0.1$ & $0.5$ & $0.0$ & $0.0$ \\ 
 
$650$ & $1.30 \times 10^{-1}$ & $0.098$ & $0.3732$ & $9.5$ & $9.3$ & $1.4$ & $0.6$ & $0.2$ & $0.8$ & $0.6$ & $-0.4$ & $-0.1$ & $0.3$ & $0.0$ & $0.0$ \\ 
 
$650$ & $1.80 \times 10^{-1}$ & $0.071$ & $0.3397$ & $9.7$ & $9.5$ & $1.5$ & $0.5$ & $0.2$ & $1.1$ & $0.6$ & $-0.5$ & $0.0$ & $0.9$ & $0.0$ & $0.0$ \\ 
 
$650$ & $2.50 \times 10^{-1}$ & $0.051$ & $0.2520$ & $10.3$ & $10.1$ & $1.7$ & $0.7$ & $0.5$ & $0.9$ & $0.7$ & $-0.3$ & $0.1$ & $0.5$ & $0.0$ & $0.0$ \\ 
 
$650$ & $4.00 \times 10^{-1}$ & $0.032$ & $0.1915$ & $12.8$ & $11.2$ & $3.9$ & $1.9$ & $2.7$ & $4.8$ & $1.9$ & $-0.8$ & $-0.7$ & $-4.3$ & $0.0$ & $0.0$ \\ 
 
$650$ & $6.50 \times 10^{-1}$ & $0.020$ & $0.02382$ & $27.6$ & $22.4$ & $7.8$ & $3.7$ & $6.4$ & $14.0$ & $3.5$ & $-1.1$ & $-1.2$ & $-13.5$ & $0.0$ & $0.0$ \\ 
 
\hline 
$800$ & $1.85 \times 10^{-2}$ & $0.850$ & $0.2872$ & $37.1$ & $36.9$ & $3.9$ & $1.1$ & $0.4$ & $0.5$ & $0.2$ & $-0.3$ & $0.0$ & $0.2$ & $0.3$ & $0.1$ \\ 
 
$800$ & $2.10 \times 10^{-2}$ & $0.750$ & $0.6634$ & $19.2$ & $19.0$ & $2.3$ & $0.2$ & $0.2$ & $0.3$ & $-0.1$ & $-0.2$ & $0.1$ & $0.2$ & $0.0$ & $0.0$ \\ 
 
$800$ & $2.42 \times 10^{-2}$ & $0.650$ & $0.6620$ & $16.0$ & $15.9$ & $2.1$ & $0.4$ & $0.1$ & $0.4$ & $-0.3$ & $-0.2$ & $0.0$ & $0.1$ & $0.0$ & $0.0$ \\ 
 
$800$ & $3.21 \times 10^{-2}$ & $0.490$ & $0.6172$ & $8.8$ & $8.6$ & $1.8$ & $0.6$ & $0.0$ & $0.4$ & $-0.3$ & $-0.3$ & $0.0$ & $0.1$ & $0.0$ & $0.0$ \\ 
 
$800$ & $5.00 \times 10^{-2}$ & $0.315$ & $0.4847$ & $9.1$ & $9.0$ & $1.2$ & $0.7$ & $0.0$ & $0.6$ & $-0.5$ & $-0.3$ & $0.0$ & $0.0$ & $0.0$ & $0.0$ \\ 
 
$800$ & $8.00 \times 10^{-2}$ & $0.197$ & $0.4527$ & $9.3$ & $9.1$ & $1.3$ & $0.8$ & $0.0$ & $0.7$ & $-0.5$ & $-0.6$ & $0.0$ & $0.0$ & $0.0$ & $0.0$ \\ 
 
$800$ & $1.30 \times 10^{-1}$ & $0.121$ & $0.3868$ & $10.8$ & $10.6$ & $1.4$ & $0.5$ & $0.3$ & $0.9$ & $0.7$ & $-0.4$ & $-0.2$ & $0.4$ & $0.0$ & $0.0$ \\ 
 
$800$ & $1.80 \times 10^{-1}$ & $0.087$ & $0.3642$ & $11.0$ & $10.9$ & $1.5$ & $0.2$ & $0.1$ & $0.8$ & $0.4$ & $-0.3$ & $0.0$ & $0.6$ & $0.0$ & $0.0$ \\ 
 
$800$ & $2.50 \times 10^{-1}$ & $0.063$ & $0.2749$ & $11.7$ & $11.6$ & $1.7$ & $0.6$ & $0.6$ & $1.0$ & $0.8$ & $-0.4$ & $-0.2$ & $0.5$ & $0.0$ & $0.0$ \\ 
 
$800$ & $4.00 \times 10^{-1}$ & $0.039$ & $0.1262$ & $16.7$ & $15.8$ & $3.5$ & $1.5$ & $2.3$ & $3.9$ & $1.5$ & $-0.4$ & $-0.6$ & $-3.5$ & $0.0$ & $0.0$ \\ 
 
$800$ & $6.50 \times 10^{-1}$ & $0.024$ & $0.01953$ & $31.8$ & $28.9$ & $7.1$ & $2.9$ & $5.9$ & $11.4$ & $2.8$ & $-0.7$ & $-1.0$ & $-11.0$ & $0.0$ & $0.0$ \\ 
 
\hline 
\end{tabular} 
\end{center} 
\captcont{\sl continued.}
\end{table} 

\begin{table}[htbp] 
\begin{center} 
\tiny 
\begin{tabular}{|r|c|c|r|r|r|r|r|r|r|r|r|r|r|r|r|} 
\hline 
$Q^2$ & $x$ & $y$ & $\tilde{\sigma}_{\rm NC}$ & 
$\delta_{\rm tot}$ & $\delta_{\rm stat}$ & $\delta_{\rm unc}$ & 
$\delta_{\rm unc}^{E}$ & 
$\delta_{\rm unc}^{h}$& 
$\delta_{\rm cor}$ & 
$\delta_{\rm cor}^{E^+}$ & 
$\delta_{\rm cor}^{\theta^+}$& 
$\delta_{\rm cor}^{h^+}$& 
$\delta_{\rm cor}^{N^+}$& 
$\delta_{\rm cor}^{S^+}$& 
$\delta_{\rm cor}^{D^+}$ \\ 
$(\rm GeV^2)$ & & & & 
$(\%)$ & $(\%)$ & $(\%)$ & $(\%)$ & $(\%)$ & $(\%)$ & 
$(\%)$ & $(\%)$ & $(\%)$ & $(\%)$ & $(\%)$ & $(\%)$ 
\\ \hline 
$35$ & $6.50 \times 10^{-4}$ & $0.848$ & $1.303$ & $8.6$ & $7.1$ & $3.8$ & $0.5$ & $2.7$ & $3.1$ & $-0.2$ & $-0.5$ & $0.2$ & $0.6$ & $2.7$ & $1.2$ \\ 
 
\hline 
$45$ & $8.40 \times 10^{-4}$ & $0.848$ & $1.413$ & $7.2$ & $6.0$ & $3.4$ & $0.4$ & $2.2$ & $2.0$ & $-0.2$ & $-0.4$ & $0.1$ & $0.5$ & $1.6$ & $1.1$ \\ 
 
$45$ & $9.30 \times 10^{-4}$ & $0.760$ & $1.235$ & $8.2$ & $7.7$ & $2.6$ & $0.5$ & $0.7$ & $1.4$ & $-0.2$ & $-0.6$ & $0.1$ & $0.3$ & $0.7$ & $0.9$ \\ 
 
\hline 
$60$ & $1.11 \times 10^{-3}$ & $0.848$ & $1.259$ & $8.0$ & $7.1$ & $3.2$ & $0.3$ & $2.0$ & $1.8$ & $-0.1$ & $-0.4$ & $0.1$ & $0.5$ & $1.5$ & $0.8$ \\ 
 
$60$ & $1.24 \times 10^{-3}$ & $0.760$ & $1.411$ & $6.4$ & $5.8$ & $2.3$ & $0.7$ & $0.6$ & $1.3$ & $-0.4$ & $-0.5$ & $0.0$ & $0.2$ & $0.6$ & $1.0$ \\ 
 
$60$ & $1.39 \times 10^{-3}$ & $0.680$ & $1.268$ & $7.5$ & $7.2$ & $2.0$ & $0.5$ & $0.2$ & $1.1$ & $-0.2$ & $-0.6$ & $0.0$ & $0.2$ & $0.5$ & $0.7$ \\ 
 
\hline 
$90$ & $1.67 \times 10^{-3}$ & $0.848$ & $1.310$ & $8.6$ & $7.9$ & $2.9$ & $0.3$ & $1.7$ & $1.7$ & $-0.2$ & $-0.4$ & $0.1$ & $0.4$ & $1.4$ & $0.8$ \\ 
 
$90$ & $1.86 \times 10^{-3}$ & $0.760$ & $1.326$ & $6.9$ & $6.5$ & $2.1$ & $0.3$ & $0.5$ & $1.1$ & $-0.1$ & $-0.5$ & $0.0$ & $0.3$ & $0.3$ & $0.9$ \\ 
 
$90$ & $2.09 \times 10^{-3}$ & $0.680$ & $1.316$ & $6.2$ & $5.8$ & $1.9$ & $0.5$ & $0.2$ & $1.0$ & $-0.2$ & $-0.6$ & $0.0$ & $0.2$ & $0.2$ & $0.7$ \\ 
 
$90$ & $2.36 \times 10^{-3}$ & $0.600$ & $1.342$ & $6.4$ & $6.0$ & $2.0$ & $0.8$ & $0.1$ & $0.8$ & $-0.2$ & $-0.8$ & $0.0$ & $0.2$ & $0.1$ & $0.0$ \\ 
 
\hline 
$120$ & $2.23 \times 10^{-3}$ & $0.848$ & $1.374$ & $9.0$ & $8.4$ & $2.7$ & $0.3$ & $1.5$ & $1.4$ & $-0.2$ & $-0.4$ & $0.1$ & $0.4$ & $0.6$ & $1.1$ \\ 
 
$120$ & $2.49 \times 10^{-3}$ & $0.760$ & $1.173$ & $8.0$ & $7.7$ & $2.0$ & $0.6$ & $0.4$ & $0.8$ & $-0.3$ & $-0.3$ & $0.0$ & $0.2$ & $0.3$ & $0.6$ \\ 
 
$120$ & $2.78 \times 10^{-3}$ & $0.680$ & $1.161$ & $7.2$ & $6.9$ & $1.9$ & $0.4$ & $0.2$ & $0.6$ & $0.1$ & $-0.4$ & $0.0$ & $0.2$ & $0.2$ & $0.3$ \\ 
 
$120$ & $3.15 \times 10^{-3}$ & $0.600$ & $1.115$ & $6.8$ & $6.5$ & $1.8$ & $0.4$ & $0.1$ & $0.7$ & $-0.3$ & $-0.6$ & $0.0$ & $0.2$ & $0.3$ & $0.0$ \\ 
 
$120$ & $3.63 \times 10^{-3}$ & $0.520$ & $1.185$ & $6.0$ & $5.7$ & $1.8$ & $0.5$ & $0.1$ & $0.8$ & $-0.4$ & $-0.7$ & $0.0$ & $0.2$ & $0.1$ & $0.0$ \\ 
 
$120$ & $4.82 \times 10^{-3}$ & $0.392$ & $1.074$ & $5.5$ & $5.1$ & $1.6$ & $0.5$ & $0.0$ & $1.0$ & $-0.3$ & $-0.9$ & $0.0$ & $0.2$ & $0.0$ & $0.0$ \\ 
 
\hline 
$150$ & $2.79 \times 10^{-3}$ & $0.848$ & $1.291$ & $10.8$ & $10.3$ & $2.6$ & $0.3$ & $1.2$ & $1.8$ & $-0.1$ & $-0.3$ & $0.1$ & $0.4$ & $0.8$ & $1.5$ \\ 
 
$150$ & $3.11 \times 10^{-3}$ & $0.760$ & $1.171$ & $9.8$ & $9.5$ & $2.0$ & $0.3$ & $0.4$ & $1.0$ & $-0.2$ & $-0.5$ & $0.1$ & $0.3$ & $0.5$ & $0.6$ \\ 
 
$150$ & $3.47 \times 10^{-3}$ & $0.680$ & $1.324$ & $7.9$ & $7.6$ & $2.3$ & $1.3$ & $0.2$ & $0.6$ & $-0.1$ & $-0.5$ & $0.0$ & $0.2$ & $0.3$ & $0.2$ \\ 
 
$150$ & $3.94 \times 10^{-3}$ & $0.600$ & $1.244$ & $7.2$ & $6.9$ & $1.8$ & $0.2$ & $0.1$ & $0.6$ & $-0.2$ & $-0.5$ & $0.0$ & $0.1$ & $0.2$ & $0.0$ \\ 
 
$150$ & $4.54 \times 10^{-3}$ & $0.520$ & $1.041$ & $7.1$ & $6.8$ & $1.8$ & $0.3$ & $0.1$ & $0.7$ & $-0.3$ & $-0.6$ & $0.0$ & $0.2$ & $0.2$ & $0.0$ \\ 
 
$150$ & $6.03 \times 10^{-3}$ & $0.392$ & $1.020$ & $4.0$ & $3.6$ & $1.4$ & $0.7$ & $0.0$ & $0.8$ & $-0.5$ & $-0.6$ & $0.0$ & $0.1$ & $0.0$ & $0.0$ \\ 
 
$150$ & $8.00 \times 10^{-3}$ & $0.295$ & $0.9700$ & $4.2$ & $3.9$ & $1.1$ & $0.8$ & $0.0$ & $0.9$ & $-0.5$ & $-0.8$ & $0.0$ & $0.0$ & $0.0$ & $0.0$ \\ 
 
$150$ & $1.30 \times 10^{-2}$ & $0.182$ & $0.8609$ & $5.7$ & $5.2$ & $1.8$ & $1.4$ & $0.9$ & $1.2$ & $-0.9$ & $-0.6$ & $-0.2$ & $-0.5$ & $0.0$ & $0.0$ \\ 
 
$150$ & $2.00 \times 10^{-2}$ & $0.118$ & $0.7980$ & $7.8$ & $6.9$ & $2.8$ & $2.5$ & $1.0$ & $2.0$ & $-1.6$ & $-1.0$ & $-0.3$ & $-0.8$ & $0.0$ & $0.0$ \\ 
 
\hline 
$200$ & $3.72 \times 10^{-3}$ & $0.848$ & $1.296$ & $13.3$ & $12.9$ & $2.6$ & $0.1$ & $1.2$ & $2.0$ & $-0.2$ & $-0.3$ & $0.1$ & $0.3$ & $0.9$ & $1.7$ \\ 
 
$200$ & $4.15 \times 10^{-3}$ & $0.760$ & $1.288$ & $11.8$ & $11.5$ & $2.1$ & $0.2$ & $0.3$ & $1.2$ & $0.2$ & $-0.5$ & $0.0$ & $0.2$ & $0.1$ & $1.0$ \\ 
 
$200$ & $4.63 \times 10^{-3}$ & $0.680$ & $1.051$ & $11.3$ & $11.1$ & $2.2$ & $1.1$ & $0.2$ & $0.5$ & $0.2$ & $-0.3$ & $0.0$ & $0.2$ & $0.3$ & $0.2$ \\ 
 
$200$ & $5.25 \times 10^{-3}$ & $0.600$ & $1.169$ & $9.1$ & $8.9$ & $1.8$ & $0.2$ & $0.1$ & $0.5$ & $-0.2$ & $-0.4$ & $0.0$ & $0.2$ & $0.1$ & $0.0$ \\ 
 
$200$ & $6.06 \times 10^{-3}$ & $0.520$ & $1.110$ & $8.3$ & $8.1$ & $1.8$ & $0.2$ & $0.1$ & $0.5$ & $-0.2$ & $-0.4$ & $0.0$ & $0.2$ & $0.1$ & $0.0$ \\ 
 
$200$ & $8.04 \times 10^{-3}$ & $0.392$ & $0.9625$ & $4.8$ & $4.6$ & $1.3$ & $0.4$ & $0.0$ & $0.8$ & $-0.3$ & $-0.7$ & $0.0$ & $0.1$ & $0.1$ & $0.0$ \\ 
 
$200$ & $1.30 \times 10^{-2}$ & $0.242$ & $0.8743$ & $4.7$ & $4.4$ & $1.3$ & $1.0$ & $0.0$ & $1.0$ & $-0.6$ & $-0.7$ & $0.0$ & $0.0$ & $0.0$ & $0.0$ \\ 
 
$200$ & $2.00 \times 10^{-2}$ & $0.157$ & $0.7573$ & $5.1$ & $4.9$ & $0.9$ & $0.3$ & $0.0$ & $0.8$ & $0.2$ & $-0.6$ & $-0.1$ & $0.5$ & $0.0$ & $0.0$ \\ 
 
$200$ & $3.20 \times 10^{-2}$ & $0.098$ & $0.6151$ & $5.6$ & $5.5$ & $1.0$ & $0.5$ & $0.0$ & $0.6$ & $-0.3$ & $-0.5$ & $-0.1$ & $0.3$ & $0.0$ & $0.0$ \\ 
 
$200$ & $5.00 \times 10^{-2}$ & $0.063$ & $0.5041$ & $6.5$ & $6.4$ & $1.2$ & $0.6$ & $0.5$ & $0.8$ & $-0.4$ & $-0.6$ & $-0.2$ & $-0.1$ & $0.0$ & $0.0$ \\ 
 
$200$ & $8.00 \times 10^{-2}$ & $0.039$ & $0.4211$ & $7.7$ & $7.3$ & $1.7$ & $1.4$ & $0.2$ & $1.5$ & $-0.9$ & $-0.6$ & $-0.1$ & $1.1$ & $0.0$ & $0.0$ \\ 
 
$200$ & $1.30 \times 10^{-1}$ & $0.024$ & $0.3857$ & $7.6$ & $7.2$ & $1.9$ & $1.5$ & $0.2$ & $1.5$ & $-0.7$ & $-0.9$ & $-0.2$ & $0.9$ & $0.0$ & $0.0$ \\ 
 
$200$ & $1.80 \times 10^{-1}$ & $0.018$ & $0.3034$ & $10.4$ & $9.3$ & $2.6$ & $0.6$ & $2.1$ & $3.7$ & $-0.5$ & $-0.8$ & $-0.3$ & $-3.6$ & $0.0$ & $0.0$ \\ 
 
$200$ & $4.00 \times 10^{-1}$ & $0.008$ & $0.1910$ & $13.2$ & $11.0$ & $3.7$ & $0.7$ & $3.1$ & $6.3$ & $-0.5$ & $-1.2$ & $-0.4$ & $-6.1$ & $0.0$ & $0.0$ \\ 
 
\hline 
$250$ & $4.64 \times 10^{-3}$ & $0.848$ & $0.8545$ & $19.8$ & $19.5$ & $2.5$ & $0.2$ & $1.0$ & $2.5$ & $0.1$ & $-0.3$ & $0.1$ & $0.3$ & $1.4$ & $2.0$ \\ 
 
$250$ & $5.18 \times 10^{-3}$ & $0.760$ & $1.080$ & $14.3$ & $14.1$ & $2.2$ & $0.2$ & $0.3$ & $1.5$ & $0.1$ & $-0.4$ & $0.1$ & $0.2$ & $0.0$ & $1.4$ \\ 
 
$250$ & $5.79 \times 10^{-3}$ & $0.680$ & $0.9481$ & $13.9$ & $13.7$ & $2.3$ & $1.2$ & $0.1$ & $0.5$ & $-0.2$ & $-0.3$ & $0.0$ & $0.2$ & $0.2$ & $0.0$ \\ 
 
$250$ & $6.56 \times 10^{-3}$ & $0.600$ & $0.9475$ & $11.6$ & $11.5$ & $1.8$ & $0.2$ & $0.1$ & $0.5$ & $-0.1$ & $-0.4$ & $0.0$ & $0.2$ & $0.1$ & $0.0$ \\ 
 
$250$ & $7.57 \times 10^{-3}$ & $0.520$ & $1.018$ & $9.8$ & $9.6$ & $1.8$ & $0.2$ & $0.1$ & $0.6$ & $-0.2$ & $-0.5$ & $0.0$ & $0.2$ & $0.1$ & $0.0$ \\ 
 
$250$ & $1.00 \times 10^{-2}$ & $0.392$ & $0.9523$ & $5.3$ & $5.1$ & $1.2$ & $0.4$ & $0.0$ & $0.6$ & $-0.3$ & $-0.5$ & $0.0$ & $0.1$ & $0.0$ & $0.0$ \\ 
 
$250$ & $1.30 \times 10^{-2}$ & $0.303$ & $0.8513$ & $5.3$ & $5.1$ & $1.1$ & $0.6$ & $0.0$ & $0.8$ & $-0.6$ & $-0.5$ & $0.0$ & $0.0$ & $0.0$ & $0.0$ \\ 
 
$250$ & $2.00 \times 10^{-2}$ & $0.197$ & $0.7707$ & $5.5$ & $5.2$ & $1.5$ & $1.2$ & $0.0$ & $1.1$ & $-0.9$ & $-0.6$ & $0.0$ & $0.0$ & $0.0$ & $0.0$ \\ 
 
$250$ & $3.20 \times 10^{-2}$ & $0.123$ & $0.6210$ & $5.9$ & $5.7$ & $1.4$ & $1.0$ & $0.3$ & $1.0$ & $0.4$ & $-0.6$ & $0.0$ & $0.6$ & $0.0$ & $0.0$ \\ 
 
$250$ & $5.00 \times 10^{-2}$ & $0.079$ & $0.5412$ & $6.1$ & $5.9$ & $1.4$ & $1.0$ & $0.1$ & $0.9$ & $0.5$ & $-0.6$ & $0.0$ & $0.6$ & $0.0$ & $0.0$ \\ 
 
$250$ & $8.00 \times 10^{-2}$ & $0.049$ & $0.4602$ & $6.7$ & $6.4$ & $1.3$ & $0.9$ & $0.1$ & $1.3$ & $0.3$ & $-0.5$ & $-0.1$ & $1.2$ & $0.0$ & $0.0$ \\ 
 
$250$ & $1.30 \times 10^{-1}$ & $0.030$ & $0.3906$ & $6.6$ & $6.2$ & $1.4$ & $0.7$ & $0.2$ & $1.8$ & $0.3$ & $-0.5$ & $0.0$ & $1.8$ & $0.0$ & $0.0$ \\ 
 
$250$ & $1.80 \times 10^{-1}$ & $0.022$ & $0.3514$ & $7.8$ & $6.9$ & $2.5$ & $1.5$ & $1.5$ & $2.5$ & $0.7$ & $-0.6$ & $-0.4$ & $-2.3$ & $0.0$ & $0.0$ \\ 
 
$250$ & $4.00 \times 10^{-1}$ & $0.010$ & $0.1556$ & $13.0$ & $10.1$ & $4.2$ & $2.5$ & $2.7$ & $7.1$ & $1.5$ & $-1.0$ & $-0.5$ & $-6.8$ & $0.0$ & $0.0$ \\ 
 
\hline 

\end{tabular} 
\end{center} 
\caption[RESULT] 
{\sl \label{tab:xsec575} The NC 
$e^+p$ reduced cross section $\tilde{\sigma}_{\rm NC}(x,Q^2)$ 
for $E_p=575$ GeV and $P_e=0$ 
with 
total $(\delta_{tot})$, 
statistical 
$(\delta_{\rm stat})$, 
total uncorrelated systematic $(\delta_{\rm unc})$ 
errors and two of its contributions from the 
 electron energy error ($\delta_{unc}^{E}$)  
and the hadronic energy error  
($\delta_{\rm unc}^{h}$). 
The effect of the other uncorrelated 
systematic errors is included in $\delta_{\rm unc}$. 
In addition the correlated systematic  
$(\delta_{\rm cor})$ and its contributions from a 
positive variation of one  
standard deviation of the 
electron energy error ($\delta_{cor}^{E^+}$), of 
the polar electron angle error 
($\delta_{\rm cor}^{\theta^+}$), of the hadronic 
energy error ($\delta_{\rm cor}^{h^+}$), of the error 
due to noise subtraction ($\delta_{\rm cor}^{N^+}$), 
of the error 
due to background subtraction charge asymmetry 
 ($\delta_{\rm cor}^{S^+}$) 
and of the error 
due to variation of the cut value on the electron discriminator $D_{ele}$  
 ($\delta_{\rm cor}^{D^+}$) are given. 
The overall normalisation uncertainty of $4\%$ is 
not included in the errors. 
}
\end{table} 
\begin{table}[htbp] 
\begin{center} 
\tiny 
\begin{tabular}{|r|c|c|r|r|r|r|r|r|r|r|r|r|r|r|r|} 
\hline 
$Q^2$ & $x$ & $y$ & $\tilde{\sigma}_{\rm NC}$ & 
$\delta_{\rm tot}$ & $\delta_{\rm stat}$ & $\delta_{\rm unc}$ & 
$\delta_{\rm unc}^{E}$ & 
$\delta_{\rm unc}^{h}$& 
$\delta_{\rm cor}$ & 
$\delta_{\rm cor}^{E^+}$ & 
$\delta_{\rm cor}^{\theta^+}$& 
$\delta_{\rm cor}^{h^+}$& 
$\delta_{\rm cor}^{N^+}$& 
$\delta_{\rm cor}^{S^+}$& 
$\delta_{\rm cor}^{D^+}$ \\ 
$(\rm GeV^2)$ & & & & 
$(\%)$ & $(\%)$ & $(\%)$ & $(\%)$ & $(\%)$ & $(\%)$ & 
$(\%)$ & $(\%)$ & $(\%)$ & $(\%)$ & $(\%)$ & $(\%)$ 
\\ \hline 
$300$ & $5.57 \times 10^{-3}$ & $0.848$ & $1.208$ & $16.0$ & $15.7$ & $2.5$ & $0.6$ & $0.8$ & $2.1$ & $-0.3$ & $-0.3$ & $0.1$ & $0.3$ & $0.2$ & $2.1$ \\ 
 
$300$ & $6.22 \times 10^{-3}$ & $0.760$ & $0.8707$ & $18.1$ & $18.0$ & $2.1$ & $0.2$ & $0.3$ & $1.3$ & $-0.2$ & $-0.4$ & $0.1$ & $0.2$ & $0.4$ & $1.1$ \\ 
 
$300$ & $6.95 \times 10^{-3}$ & $0.680$ & $0.9694$ & $15.0$ & $14.8$ & $2.1$ & $0.8$ & $0.1$ & $0.5$ & $-0.1$ & $-0.5$ & $0.0$ & $0.2$ & $0.0$ & $0.0$ \\ 
 
$300$ & $7.88 \times 10^{-3}$ & $0.600$ & $1.035$ & $12.9$ & $12.7$ & $1.9$ & $0.1$ & $0.1$ & $0.5$ & $0.1$ & $-0.5$ & $0.0$ & $0.2$ & $0.0$ & $0.0$ \\ 
 
$300$ & $9.09 \times 10^{-3}$ & $0.520$ & $0.8632$ & $12.1$ & $12.0$ & $1.8$ & $0.3$ & $0.1$ & $0.6$ & $-0.3$ & $-0.5$ & $0.0$ & $0.1$ & $0.0$ & $0.0$ \\ 
 
$300$ & $1.21 \times 10^{-2}$ & $0.392$ & $0.9079$ & $6.2$ & $6.0$ & $1.3$ & $0.4$ & $0.0$ & $0.6$ & $-0.4$ & $-0.5$ & $0.0$ & $0.1$ & $0.0$ & $0.0$ \\ 
 
$300$ & $2.00 \times 10^{-2}$ & $0.236$ & $0.6653$ & $6.5$ & $6.3$ & $1.0$ & $0.6$ & $0.0$ & $0.9$ & $-0.6$ & $-0.7$ & $0.0$ & $0.0$ & $0.0$ & $0.0$ \\ 
 
$300$ & $3.20 \times 10^{-2}$ & $0.148$ & $0.6171$ & $6.7$ & $6.6$ & $1.0$ & $0.5$ & $0.1$ & $0.8$ & $0.3$ & $-0.4$ & $0.0$ & $0.6$ & $0.0$ & $0.0$ \\ 
 
$300$ & $5.00 \times 10^{-2}$ & $0.094$ & $0.5364$ & $6.9$ & $6.8$ & $1.2$ & $0.8$ & $0.0$ & $0.9$ & $0.5$ & $-0.6$ & $-0.1$ & $0.5$ & $0.0$ & $0.0$ \\ 
 
$300$ & $8.00 \times 10^{-2}$ & $0.059$ & $0.4802$ & $7.4$ & $7.2$ & $1.3$ & $0.9$ & $0.0$ & $1.2$ & $0.5$ & $-0.6$ & $-0.1$ & $0.9$ & $0.0$ & $0.0$ \\ 
 
$300$ & $1.30 \times 10^{-1}$ & $0.036$ & $0.3762$ & $7.6$ & $7.2$ & $1.5$ & $1.0$ & $0.1$ & $1.9$ & $0.4$ & $-0.6$ & $-0.1$ & $1.7$ & $0.0$ & $0.0$ \\ 
 
$300$ & $1.80 \times 10^{-1}$ & $0.026$ & $0.3190$ & $8.7$ & $8.1$ & $2.5$ & $1.8$ & $1.1$ & $2.0$ & $1.1$ & $-0.9$ & $-0.3$ & $-1.4$ & $0.0$ & $0.0$ \\ 
 
$300$ & $4.00 \times 10^{-1}$ & $0.012$ & $0.1469$ & $15.5$ & $12.0$ & $4.8$ & $2.8$ & $3.4$ & $8.6$ & $1.7$ & $-1.1$ & $-0.6$ & $-8.4$ & $0.0$ & $0.0$ \\ 
 
\hline 
$400$ & $7.43 \times 10^{-3}$ & $0.848$ & $0.8123$ & $23.1$ & $22.9$ & $2.6$ & $0.3$ & $0.7$ & $2.3$ & $-0.3$ & $-0.3$ & $0.1$ & $0.3$ & $0.6$ & $2.2$ \\ 
 
$400$ & $8.29 \times 10^{-3}$ & $0.760$ & $0.5949$ & $23.1$ & $23.0$ & $2.3$ & $0.6$ & $0.2$ & $0.8$ & $-0.1$ & $-0.3$ & $0.0$ & $0.2$ & $0.0$ & $0.8$ \\ 
 
$400$ & $9.27 \times 10^{-3}$ & $0.680$ & $1.013$ & $16.1$ & $16.0$ & $2.0$ & $0.4$ & $0.2$ & $0.4$ & $0.2$ & $-0.2$ & $0.0$ & $0.2$ & $0.2$ & $0.0$ \\ 
 
$400$ & $1.05 \times 10^{-2}$ & $0.600$ & $0.8806$ & $15.6$ & $15.5$ & $1.9$ & $0.3$ & $0.1$ & $0.5$ & $-0.2$ & $-0.4$ & $0.0$ & $0.2$ & $0.0$ & $0.0$ \\ 
 
$400$ & $1.21 \times 10^{-2}$ & $0.520$ & $0.9991$ & $13.0$ & $12.9$ & $1.9$ & $0.1$ & $0.1$ & $0.5$ & $0.2$ & $-0.4$ & $0.0$ & $0.1$ & $0.0$ & $0.0$ \\ 
 
$400$ & $1.61 \times 10^{-2}$ & $0.392$ & $0.8791$ & $7.1$ & $7.0$ & $1.2$ & $0.3$ & $0.0$ & $0.6$ & $-0.3$ & $-0.5$ & $0.0$ & $0.1$ & $0.0$ & $0.0$ \\ 
 
$400$ & $3.20 \times 10^{-2}$ & $0.197$ & $0.6501$ & $7.3$ & $7.2$ & $1.1$ & $0.8$ & $0.0$ & $1.0$ & $-0.8$ & $-0.6$ & $0.0$ & $0.0$ & $0.0$ & $0.0$ \\ 
 
$400$ & $5.00 \times 10^{-2}$ & $0.126$ & $0.5099$ & $8.0$ & $7.9$ & $1.1$ & $0.6$ & $0.1$ & $1.0$ & $0.6$ & $-0.6$ & $0.0$ & $0.5$ & $0.0$ & $0.0$ \\ 
 
$400$ & $8.00 \times 10^{-2}$ & $0.079$ & $0.4452$ & $8.6$ & $8.5$ & $1.1$ & $0.6$ & $0.1$ & $1.1$ & $0.6$ & $-0.6$ & $-0.1$ & $0.7$ & $0.0$ & $0.0$ \\ 
 
$400$ & $1.30 \times 10^{-1}$ & $0.049$ & $0.3769$ & $8.5$ & $8.2$ & $1.3$ & $0.4$ & $0.2$ & $1.7$ & $0.4$ & $-0.4$ & $0.0$ & $1.6$ & $0.0$ & $0.0$ \\ 
 
$400$ & $1.80 \times 10^{-1}$ & $0.035$ & $0.3421$ & $8.9$ & $8.6$ & $2.0$ & $1.0$ & $1.1$ & $1.6$ & $0.9$ & $-0.6$ & $-0.3$ & $-1.1$ & $0.0$ & $0.0$ \\ 
 
$400$ & $4.00 \times 10^{-1}$ & $0.016$ & $0.1488$ & $16.6$ & $13.4$ & $4.6$ & $2.0$ & $3.5$ & $8.6$ & $1.9$ & $-0.9$ & $-0.7$ & $-8.4$ & $0.0$ & $0.0$ \\ 
 
\hline 
$500$ & $9.29 \times 10^{-3}$ & $0.848$ & $0.7285$ & $27.8$ & $27.6$ & $2.7$ & $0.3$ & $0.5$ & $2.1$ & $0.0$ & $-0.1$ & $0.1$ & $0.2$ & $0.4$ & $2.1$ \\ 
 
$500$ & $1.04 \times 10^{-2}$ & $0.760$ & $0.7348$ & $22.8$ & $22.7$ & $2.3$ & $1.0$ & $0.3$ & $0.5$ & $-0.2$ & $-0.4$ & $0.1$ & $0.2$ & $0.0$ & $0.1$ \\ 
 
$500$ & $1.16 \times 10^{-2}$ & $0.680$ & $1.177$ & $16.2$ & $16.1$ & $2.0$ & $0.1$ & $0.1$ & $0.4$ & $-0.1$ & $-0.3$ & $0.0$ & $0.2$ & $0.0$ & $0.0$ \\ 
 
$500$ & $1.31 \times 10^{-2}$ & $0.600$ & $0.8538$ & $17.7$ & $17.6$ & $1.9$ & $0.2$ & $0.1$ & $0.5$ & $-0.2$ & $-0.4$ & $0.0$ & $0.1$ & $0.2$ & $0.0$ \\ 
 
$500$ & $1.51 \times 10^{-2}$ & $0.520$ & $1.040$ & $14.2$ & $14.1$ & $1.9$ & $0.4$ & $0.1$ & $0.4$ & $-0.3$ & $-0.3$ & $0.0$ & $0.1$ & $0.0$ & $0.0$ \\ 
 
$500$ & $2.01 \times 10^{-2}$ & $0.392$ & $0.7340$ & $9.1$ & $9.0$ & $1.3$ & $0.4$ & $0.0$ & $0.6$ & $-0.4$ & $-0.4$ & $0.0$ & $0.1$ & $0.0$ & $0.0$ \\ 
 
$500$ & $3.20 \times 10^{-2}$ & $0.246$ & $0.6891$ & $8.4$ & $8.3$ & $1.1$ & $0.7$ & $0.0$ & $0.8$ & $-0.7$ & $-0.4$ & $0.0$ & $0.0$ & $0.0$ & $0.0$ \\ 
 
$500$ & $5.00 \times 10^{-2}$ & $0.157$ & $0.5602$ & $8.9$ & $8.8$ & $1.0$ & $0.3$ & $0.4$ & $0.8$ & $0.3$ & $-0.3$ & $0.1$ & $0.7$ & $0.0$ & $0.0$ \\ 
 
$500$ & $8.00 \times 10^{-2}$ & $0.098$ & $0.4454$ & $9.8$ & $9.7$ & $1.0$ & $0.5$ & $0.2$ & $0.7$ & $0.5$ & $-0.3$ & $-0.1$ & $0.3$ & $0.0$ & $0.0$ \\ 
 
$500$ & $1.30 \times 10^{-1}$ & $0.061$ & $0.3831$ & $11.5$ & $11.4$ & $1.3$ & $0.5$ & $0.2$ & $1.4$ & $0.5$ & $-0.4$ & $0.1$ & $1.3$ & $0.0$ & $0.0$ \\ 
 
$500$ & $1.80 \times 10^{-1}$ & $0.044$ & $0.3467$ & $11.6$ & $11.3$ & $1.5$ & $0.6$ & $0.1$ & $1.6$ & $0.6$ & $-0.5$ & $-0.1$ & $1.4$ & $0.0$ & $0.0$ \\ 
 
$500$ & $2.50 \times 10^{-1}$ & $0.032$ & $0.2290$ & $13.8$ & $13.5$ & $2.0$ & $0.8$ & $1.2$ & $1.7$ & $0.8$ & $-0.4$ & $-0.3$ & $-1.4$ & $0.0$ & $0.0$ \\ 
 
$500$ & $4.00 \times 10^{-1}$ & $0.020$ & $0.1687$ & $18.1$ & $16.3$ & $4.2$ & $1.8$ & $3.2$ & $6.7$ & $1.8$ & $-0.6$ & $-0.7$ & $-6.3$ & $0.0$ & $0.0$ \\ 
 
$500$ & $6.50 \times 10^{-1}$ & $0.012$ & $0.02022$ & $31.5$ & $28.9$ & $5.5$ & $2.3$ & $4.3$ & $11.2$ & $2.3$ & $-0.9$ & $-0.7$ & $-10.9$ & $0.0$ & $0.0$ \\ 
 
\hline 
$650$ & $1.21 \times 10^{-2}$ & $0.848$ & $0.4914$ & $38.7$ & $38.5$ & $3.0$ & $0.4$ & $0.4$ & $1.0$ & $0.2$ & $0.0$ & $0.0$ & $0.2$ & $0.5$ & $0.8$ \\ 
 
$650$ & $1.35 \times 10^{-2}$ & $0.760$ & $0.6986$ & $28.3$ & $28.2$ & $2.2$ & $0.3$ & $0.2$ & $0.6$ & $-0.2$ & $-0.2$ & $0.1$ & $0.2$ & $0.4$ & $0.0$ \\ 
 
$650$ & $1.51 \times 10^{-2}$ & $0.680$ & $0.6789$ & $25.2$ & $25.1$ & $2.1$ & $0.2$ & $0.1$ & $0.5$ & $0.1$ & $-0.3$ & $0.0$ & $0.2$ & $0.3$ & $0.0$ \\ 
 
$650$ & $1.71 \times 10^{-2}$ & $0.600$ & $0.6957$ & $21.4$ & $21.3$ & $2.0$ & $0.3$ & $0.1$ & $0.5$ & $-0.2$ & $-0.4$ & $0.0$ & $0.1$ & $0.0$ & $0.0$ \\ 
 
$650$ & $1.97 \times 10^{-2}$ & $0.520$ & $0.4817$ & $22.7$ & $22.6$ & $2.0$ & $0.6$ & $0.1$ & $0.6$ & $-0.4$ & $-0.4$ & $0.0$ & $0.1$ & $0.0$ & $0.0$ \\ 
 
$650$ & $2.61 \times 10^{-2}$ & $0.392$ & $0.6348$ & $10.8$ & $10.7$ & $1.3$ & $0.4$ & $0.0$ & $0.5$ & $-0.2$ & $-0.5$ & $0.0$ & $0.1$ & $0.0$ & $0.0$ \\ 
 
$650$ & $5.00 \times 10^{-2}$ & $0.205$ & $0.4685$ & $11.7$ & $11.6$ & $1.3$ & $0.9$ & $0.0$ & $1.0$ & $-0.8$ & $-0.6$ & $0.0$ & $0.0$ & $0.0$ & $0.0$ \\ 
 
$650$ & $8.00 \times 10^{-2}$ & $0.128$ & $0.4525$ & $11.4$ & $11.3$ & $1.1$ & $0.5$ & $0.0$ & $0.9$ & $0.6$ & $-0.4$ & $-0.1$ & $0.5$ & $0.0$ & $0.0$ \\ 
 
$650$ & $1.30 \times 10^{-1}$ & $0.079$ & $0.3975$ & $13.4$ & $13.3$ & $1.4$ & $0.6$ & $0.0$ & $1.2$ & $0.6$ & $-0.5$ & $0.0$ & $0.9$ & $0.0$ & $0.0$ \\ 
 
$650$ & $1.80 \times 10^{-1}$ & $0.057$ & $0.3285$ & $14.0$ & $13.9$ & $1.4$ & $0.3$ & $0.2$ & $1.4$ & $0.4$ & $-0.2$ & $0.0$ & $1.4$ & $0.0$ & $0.0$ \\ 
 
$650$ & $2.50 \times 10^{-1}$ & $0.041$ & $0.2401$ & $15.5$ & $15.3$ & $2.0$ & $1.0$ & $0.9$ & $1.3$ & $1.0$ & $-0.6$ & $-0.3$ & $-0.6$ & $0.0$ & $0.0$ \\ 
 
$650$ & $4.00 \times 10^{-1}$ & $0.026$ & $0.1563$ & $20.4$ & $18.9$ & $4.5$ & $2.1$ & $3.3$ & $6.2$ & $2.2$ & $-0.9$ & $-0.9$ & $-5.7$ & $0.0$ & $0.0$ \\ 
 
$650$ & $6.50 \times 10^{-1}$ & $0.016$ & $0.02266$ & $35.8$ & $33.3$ & $6.0$ & $2.3$ & $4.9$ & $11.4$ & $2.2$ & $-0.9$ & $-0.8$ & $-11.1$ & $0.0$ & $0.0$ \\ 
 
\hline 
$800$ & $1.49 \times 10^{-2}$ & $0.848$ & $0.6679$ & $31.9$ & $31.8$ & $3.1$ & $0.9$ & $0.4$ & $0.5$ & $-0.4$ & $-0.2$ & $0.1$ & $0.2$ & $0.0$ & $0.1$ \\ 
 
$800$ & $1.66 \times 10^{-2}$ & $0.760$ & $0.4843$ & $38.5$ & $38.4$ & $2.7$ & $0.6$ & $0.2$ & $0.7$ & $0.6$ & $-0.3$ & $0.1$ & $0.3$ & $0.0$ & $0.0$ \\ 
 
$800$ & $1.85 \times 10^{-2}$ & $0.680$ & $0.6761$ & $27.1$ & $27.0$ & $2.2$ & $0.2$ & $0.1$ & $0.4$ & $-0.2$ & $-0.3$ & $0.0$ & $0.1$ & $0.0$ & $0.0$ \\ 
 
$800$ & $2.10 \times 10^{-2}$ & $0.600$ & $0.6604$ & $24.5$ & $24.4$ & $2.1$ & $0.1$ & $0.1$ & $0.5$ & $-0.1$ & $-0.5$ & $0.0$ & $0.1$ & $0.0$ & $0.0$ \\ 
 
$800$ & $2.42 \times 10^{-2}$ & $0.520$ & $0.6435$ & $21.9$ & $21.8$ & $1.9$ & $0.1$ & $0.0$ & $0.6$ & $0.0$ & $-0.6$ & $0.0$ & $0.1$ & $0.0$ & $0.0$ \\ 
 
$800$ & $3.21 \times 10^{-2}$ & $0.392$ & $0.4923$ & $13.6$ & $13.5$ & $1.4$ & $0.5$ & $0.0$ & $0.4$ & $-0.3$ & $-0.3$ & $0.0$ & $0.1$ & $0.0$ & $0.0$ \\ 
 
$800$ & $5.00 \times 10^{-2}$ & $0.252$ & $0.5837$ & $12.0$ & $11.9$ & $1.2$ & $0.7$ & $0.0$ & $0.8$ & $-0.6$ & $-0.6$ & $0.0$ & $0.0$ & $0.0$ & $0.0$ \\ 
 
$800$ & $8.00 \times 10^{-2}$ & $0.157$ & $0.5522$ & $12.0$ & $12.0$ & $1.1$ & $0.4$ & $0.1$ & $0.8$ & $0.6$ & $-0.4$ & $0.1$ & $0.3$ & $0.0$ & $0.0$ \\ 
 
$800$ & $1.30 \times 10^{-1}$ & $0.097$ & $0.2926$ & $18.3$ & $18.3$ & $1.3$ & $0.1$ & $0.1$ & $0.7$ & $0.4$ & $-0.2$ & $-0.1$ & $0.6$ & $0.0$ & $0.0$ \\ 
 
$800$ & $1.80 \times 10^{-1}$ & $0.070$ & $0.2636$ & $18.7$ & $18.6$ & $1.5$ & $0.5$ & $0.1$ & $1.5$ & $0.7$ & $-0.5$ & $0.0$ & $1.2$ & $0.0$ & $0.0$ \\ 
 
$800$ & $2.50 \times 10^{-1}$ & $0.050$ & $0.1811$ & $20.5$ & $20.4$ & $1.9$ & $0.8$ & $0.8$ & $1.3$ & $1.1$ & $-0.5$ & $-0.2$ & $-0.5$ & $0.0$ & $0.0$ \\ 
 
$800$ & $4.00 \times 10^{-1}$ & $0.032$ & $0.1614$ & $22.2$ & $21.3$ & $3.8$ & $1.5$ & $2.7$ & $4.9$ & $1.5$ & $-0.4$ & $-0.5$ & $-4.6$ & $0.0$ & $0.0$ \\ 
 
$800$ & $6.50 \times 10^{-1}$ & $0.019$ & $0.02134$ & $43.1$ & $40.8$ & $6.7$ & $2.6$ & $5.4$ & $12.1$ & $2.7$ & $-0.6$ & $-1.1$ & $-11.8$ & $0.0$ & $0.0$ \\ 
 
\hline 
\end{tabular} 
\end{center} 
\captcont{\sl continued.}
\end{table}

\begin{table}[htbp] 
\begin{center} 
\tiny 
\begin{tabular}{|r|c|c|r|r|r|r|r|r|r|r|r|r|} 
\hline 
$Q^2$ & $x$ &
$F_{L}$ &
$\Delta_{\rm stat}F_{L}$ &
$\Delta_{\rm uncor}F_{L}$ &
$\Delta_{\rm cor}F_{L}$ &
$\Delta_{\rm tot}F_{L}$ &
$F_{2}$ &
$\Delta_{\rm stat}F_{2}$ &
$\Delta_{\rm uncor}F_{2}$ &
$\Delta_{\rm cor}F_{2}$ &
$\Delta_{\rm tot}F_{2}$ &
$\rho$ \\ 
(GeV$^2$) & & & & & & & & & & & & \\
\hline
$   1.5 $&$    0.279 \times 10^{-4} $&$ 0.088 $&$ 0.113 $&$ 0.186 $&$ 0.053 $&$ 0.224 $&$ 0.732 $&$ 0.066 $&$ 0.096 $&$ 0.028 $&$ 0.120 $&$ 0.882$\\
$   2.0 $&$    0.372 \times 10^{-4} $&$ 0.110 $&$ 0.069 $&$ 0.131 $&$ 0.062 $&$ 0.160 $&$ 0.843 $&$ 0.028 $&$ 0.051 $&$ 0.032 $&$ 0.066 $&$ 0.855$\\
$   2.0 $&$    0.415 \times 10^{-4} $&$ 0.437 $&$ 0.110 $&$ 0.181 $&$ 0.071 $&$ 0.223 $&$ 0.904 $&$ 0.039 $&$ 0.060 $&$ 0.030 $&$ 0.078 $&$ 0.852$\\
$   2.0 $&$    0.464 \times 10^{-4} $&$ 0.043 $&$ 0.052 $&$ 0.104 $&$ 0.033 $&$ 0.121 $&$ 0.740 $&$ 0.033 $&$ 0.052 $&$ 0.009 $&$ 0.062 $&$ 0.822$\\
$   2.5 $&$    0.465 \times 10^{-4} $&$ 0.013 $&$ 0.057 $&$ 0.120 $&$ 0.046 $&$ 0.141 $&$ 0.846 $&$ 0.022 $&$ 0.045 $&$ 0.016 $&$ 0.053 $&$ 0.856$\\
$   2.5 $&$    0.519 \times 10^{-4} $&$ 0.103 $&$ 0.062 $&$ 0.129 $&$ 0.042 $&$ 0.149 $&$ 0.897 $&$ 0.023 $&$ 0.045 $&$ 0.016 $&$ 0.053 $&$ 0.860$\\
$   2.5 $&$    0.580 \times 10^{-4} $&$ 0.174 $&$ 0.047 $&$ 0.090 $&$ 0.058 $&$ 0.117 $&$ 0.889 $&$ 0.021 $&$ 0.034 $&$ 0.028 $&$ 0.049 $&$ 0.821$\\
$   2.5 $&$    0.658 \times 10^{-4} $&$ 0.169 $&$ 0.043 $&$ 0.099 $&$ 0.063 $&$ 0.125 $&$ 0.865 $&$ 0.019 $&$ 0.035 $&$ 0.031 $&$ 0.050 $&$ 0.840$\\
$   2.5 $&$    0.759 \times 10^{-4} $&$ 0.413 $&$ 0.096 $&$ 0.155 $&$ 0.079 $&$ 0.198 $&$ 0.877 $&$ 0.024 $&$ 0.035 $&$ 0.026 $&$ 0.050 $&$ 0.783$\\
$   3.5 $&$    0.651 \times 10^{-4} $&$ 0.130 $&$ 0.065 $&$ 0.135 $&$ 0.052 $&$ 0.158 $&$ 0.973 $&$ 0.025 $&$ 0.050 $&$ 0.022 $&$ 0.060 $&$ 0.846$\\
$   3.5 $&$    0.727 \times 10^{-4} $&$ 0.199 $&$ 0.061 $&$ 0.133 $&$ 0.044 $&$ 0.152 $&$ 0.989 $&$ 0.024 $&$ 0.047 $&$ 0.021 $&$ 0.057 $&$ 0.850$\\
$   3.5 $&$    0.812 \times 10^{-4} $&$ 0.253 $&$ 0.044 $&$ 0.094 $&$ 0.041 $&$ 0.112 $&$ 0.981 $&$ 0.019 $&$ 0.036 $&$ 0.016 $&$ 0.044 $&$ 0.811$\\
$   3.5 $&$    0.921 \times 10^{-4} $&$ 0.230 $&$ 0.037 $&$ 0.099 $&$ 0.037 $&$ 0.112 $&$ 0.968 $&$ 0.015 $&$ 0.033 $&$ 0.014 $&$ 0.039 $&$ 0.816$\\
$   3.5 $&$    0.106 \times 10^{-3} $&$ 0.155 $&$ 0.049 $&$ 0.123 $&$ 0.046 $&$ 0.141 $&$ 0.934 $&$ 0.015 $&$ 0.032 $&$ 0.010 $&$ 0.037 $&$ 0.797$\\
$   3.5 $&$    0.141 \times 10^{-3} $&$ 0.665 $&$ 0.112 $&$ 0.221 $&$ 0.123 $&$ 0.276 $&$ 0.937 $&$ 0.011 $&$ 0.028 $&$ 0.012 $&$ 0.032 $&$ 0.735$\\
$   5.0 $&$    0.931 \times 10^{-4} $&$ 0.411 $&$ 0.081 $&$ 0.162 $&$ 0.068 $&$ 0.193 $&$ 1.149 $&$ 0.031 $&$ 0.060 $&$ 0.031 $&$ 0.075 $&$ 0.846$\\
$   5.0 $&$    0.104 \times 10^{-3} $&$ 0.344 $&$ 0.065 $&$ 0.142 $&$ 0.044 $&$ 0.163 $&$ 1.072 $&$ 0.027 $&$ 0.052 $&$ 0.024 $&$ 0.063 $&$ 0.859$\\
$   5.0 $&$    0.116 \times 10^{-3} $&$ 0.258 $&$ 0.048 $&$ 0.108 $&$ 0.049 $&$ 0.128 $&$ 1.127 $&$ 0.021 $&$ 0.042 $&$ 0.018 $&$ 0.050 $&$ 0.828$\\
$   5.0 $&$    0.131 \times 10^{-3} $&$ 0.306 $&$ 0.037 $&$ 0.109 $&$ 0.041 $&$ 0.122 $&$ 1.082 $&$ 0.016 $&$ 0.037 $&$ 0.017 $&$ 0.044 $&$ 0.830$\\
$   5.0 $&$    0.152 \times 10^{-3} $&$ 0.224 $&$ 0.044 $&$ 0.134 $&$ 0.045 $&$ 0.148 $&$ 1.060 $&$ 0.014 $&$ 0.034 $&$ 0.015 $&$ 0.040 $&$ 0.834$\\
$   5.0 $&$    0.201 \times 10^{-3} $&$ 0.533 $&$ 0.057 $&$ 0.203 $&$ 0.084 $&$ 0.227 $&$ 1.018 $&$ 0.008 $&$ 0.028 $&$ 0.012 $&$ 0.032 $&$ 0.809$\\
$   6.5 $&$    0.121 \times 10^{-3} $&$ 0.435 $&$ 0.096 $&$ 0.179 $&$ 0.077 $&$ 0.218 $&$ 1.215 $&$ 0.037 $&$ 0.066 $&$ 0.027 $&$ 0.080 $&$ 0.853$\\
$   6.5 $&$    0.135 \times 10^{-3} $&$ 0.199 $&$ 0.071 $&$ 0.151 $&$ 0.042 $&$ 0.172 $&$ 1.103 $&$ 0.030 $&$ 0.055 $&$ 0.020 $&$ 0.066 $&$ 0.862$\\
$   6.5 $&$    0.151 \times 10^{-3} $&$ 0.137 $&$ 0.051 $&$ 0.114 $&$ 0.054 $&$ 0.136 $&$ 1.135 $&$ 0.023 $&$ 0.044 $&$ 0.023 $&$ 0.055 $&$ 0.844$\\
$   6.5 $&$    0.171 \times 10^{-3} $&$ 0.357 $&$ 0.040 $&$ 0.119 $&$ 0.044 $&$ 0.133 $&$ 1.158 $&$ 0.017 $&$ 0.041 $&$ 0.020 $&$ 0.048 $&$ 0.844$\\
$   6.5 $&$    0.197 \times 10^{-3} $&$ 0.318 $&$ 0.044 $&$ 0.145 $&$ 0.053 $&$ 0.161 $&$ 1.147 $&$ 0.014 $&$ 0.038 $&$ 0.019 $&$ 0.044 $&$ 0.855$\\
$   6.5 $&$    0.262 \times 10^{-3} $&$ 0.188 $&$ 0.046 $&$ 0.205 $&$ 0.090 $&$ 0.229 $&$ 1.044 $&$ 0.007 $&$ 0.029 $&$ 0.017 $&$ 0.034 $&$ 0.842$\\
$   8.5 $&$    0.158 \times 10^{-3} $&$ 0.499 $&$ 0.109 $&$ 0.195 $&$ 0.095 $&$ 0.243 $&$ 1.352 $&$ 0.044 $&$ 0.074 $&$ 0.033 $&$ 0.092 $&$ 0.845$\\
$   8.5 $&$    0.177 \times 10^{-3} $&$ 0.489 $&$ 0.089 $&$ 0.184 $&$ 0.051 $&$ 0.210 $&$ 1.335 $&$ 0.038 $&$ 0.067 $&$ 0.022 $&$ 0.080 $&$ 0.862$\\
$   8.5 $&$    0.197 \times 10^{-3} $&$ 0.271 $&$ 0.057 $&$ 0.123 $&$ 0.058 $&$ 0.147 $&$ 1.196 $&$ 0.027 $&$ 0.048 $&$ 0.021 $&$ 0.059 $&$ 0.841$\\
$   8.5 $&$    0.224 \times 10^{-3} $&$ 0.242 $&$ 0.045 $&$ 0.125 $&$ 0.042 $&$ 0.139 $&$ 1.158 $&$ 0.019 $&$ 0.043 $&$ 0.017 $&$ 0.050 $&$ 0.849$\\
$   8.5 $&$    0.258 \times 10^{-3} $&$-0.123 $&$ 0.045 $&$ 0.140 $&$ 0.051 $&$ 0.156 $&$ 1.038 $&$ 0.015 $&$ 0.036 $&$ 0.016 $&$ 0.042 $&$ 0.853$\\
$   8.5 $&$    0.342 \times 10^{-3} $&$ 0.167 $&$ 0.045 $&$ 0.216 $&$ 0.089 $&$ 0.238 $&$ 1.095 $&$ 0.007 $&$ 0.030 $&$ 0.017 $&$ 0.035 $&$ 0.846$\\
$   12  $&$    0.223 \times 10^{-3} $&$ 0.094 $&$ 0.101 $&$ 0.159 $&$ 0.084 $&$ 0.206 $&$ 1.314 $&$ 0.039 $&$ 0.041 $&$ 0.044 $&$ 0.072 $&$ 0.855$\\
$   12  $&$    0.249 \times 10^{-3} $&$ 0.544 $&$ 0.098 $&$ 0.155 $&$ 0.058 $&$ 0.193 $&$ 1.389 $&$ 0.035 $&$ 0.035 $&$ 0.028 $&$ 0.057 $&$ 0.835$\\
$   12  $&$    0.278 \times 10^{-3} $&$ 0.281 $&$ 0.059 $&$ 0.098 $&$ 0.047 $&$ 0.124 $&$ 1.310 $&$ 0.024 $&$ 0.024 $&$ 0.019 $&$ 0.039 $&$ 0.757$\\
$   12  $&$    0.316 \times 10^{-3} $&$ 0.248 $&$ 0.050 $&$ 0.100 $&$ 0.038 $&$ 0.118 $&$ 1.258 $&$ 0.019 $&$ 0.022 $&$ 0.015 $&$ 0.033 $&$ 0.733$\\
$   12  $&$    0.364 \times 10^{-3} $&$ 0.435 $&$ 0.055 $&$ 0.121 $&$ 0.041 $&$ 0.139 $&$ 1.268 $&$ 0.016 $&$ 0.022 $&$ 0.013 $&$ 0.030 $&$ 0.728$\\
$   12  $&$    0.483 \times 10^{-3} $&$ 0.414 $&$ 0.050 $&$ 0.162 $&$ 0.064 $&$ 0.181 $&$ 1.189 $&$ 0.007 $&$ 0.016 $&$ 0.012 $&$ 0.021 $&$ 0.651$\\
$   15  $&$    0.279 \times 10^{-3} $&$ 0.510 $&$ 0.109 $&$ 0.183 $&$ 0.085 $&$ 0.230 $&$ 1.485 $&$ 0.040 $&$ 0.047 $&$ 0.049 $&$ 0.079 $&$ 0.854$\\
$   15  $&$    0.312 \times 10^{-3} $&$ 0.148 $&$ 0.088 $&$ 0.150 $&$ 0.052 $&$ 0.181 $&$ 1.370 $&$ 0.032 $&$ 0.035 $&$ 0.027 $&$ 0.054 $&$ 0.834$\\
$   15  $&$    0.348 \times 10^{-3} $&$ 0.188 $&$ 0.061 $&$ 0.099 $&$ 0.039 $&$ 0.122 $&$ 1.329 $&$ 0.023 $&$ 0.023 $&$ 0.017 $&$ 0.036 $&$ 0.748$\\
$   15  $&$    0.395 \times 10^{-3} $&$ 0.419 $&$ 0.051 $&$ 0.100 $&$ 0.036 $&$ 0.118 $&$ 1.321 $&$ 0.017 $&$ 0.021 $&$ 0.015 $&$ 0.031 $&$ 0.710$\\
$   15  $&$    0.455 \times 10^{-3} $&$ 0.257 $&$ 0.062 $&$ 0.117 $&$ 0.045 $&$ 0.140 $&$ 1.269 $&$ 0.015 $&$ 0.018 $&$ 0.013 $&$ 0.027 $&$ 0.693$\\
$   15  $&$    0.604 \times 10^{-3} $&$ 0.066 $&$ 0.054 $&$ 0.157 $&$ 0.066 $&$ 0.179 $&$ 1.180 $&$ 0.007 $&$ 0.014 $&$ 0.012 $&$ 0.019 $&$ 0.620$\\
$   20  $&$    0.372 \times 10^{-3} $&$ 0.216 $&$ 0.116 $&$ 0.197 $&$ 0.065 $&$ 0.238 $&$ 1.452 $&$ 0.041 $&$ 0.051 $&$ 0.033 $&$ 0.073 $&$ 0.877$\\
$   20  $&$    0.415 \times 10^{-3} $&$ 0.322 $&$ 0.092 $&$ 0.158 $&$ 0.044 $&$ 0.188 $&$ 1.424 $&$ 0.032 $&$ 0.037 $&$ 0.021 $&$ 0.054 $&$ 0.837$\\
$   20  $&$    0.464 \times 10^{-3} $&$ 0.412 $&$ 0.070 $&$ 0.108 $&$ 0.037 $&$ 0.134 $&$ 1.396 $&$ 0.024 $&$ 0.025 $&$ 0.015 $&$ 0.037 $&$ 0.752$\\
$   20  $&$    0.526 \times 10^{-3} $&$ 0.358 $&$ 0.052 $&$ 0.103 $&$ 0.037 $&$ 0.121 $&$ 1.354 $&$ 0.018 $&$ 0.021 $&$ 0.015 $&$ 0.032 $&$ 0.708$\\
$   20  $&$    0.607 \times 10^{-3} $&$ 0.304 $&$ 0.062 $&$ 0.119 $&$ 0.041 $&$ 0.140 $&$ 1.295 $&$ 0.015 $&$ 0.019 $&$ 0.013 $&$ 0.027 $&$ 0.693$\\
$   20  $&$    0.805 \times 10^{-3} $&$ 0.212 $&$ 0.060 $&$ 0.163 $&$ 0.068 $&$ 0.186 $&$ 1.222 $&$ 0.007 $&$ 0.014 $&$ 0.012 $&$ 0.019 $&$ 0.608$\\
$   25  $&$    0.493 \times 10^{-3} $&$ 0.363 $&$ 0.072 $&$ 0.157 $&$ 0.043 $&$ 0.178 $&$ 1.484 $&$ 0.022 $&$ 0.040 $&$ 0.024 $&$ 0.052 $&$ 0.851$\\
$   25  $&$    0.616 \times 10^{-3} $&$ 0.284 $&$ 0.043 $&$ 0.089 $&$ 0.031 $&$ 0.103 $&$ 1.382 $&$ 0.013 $&$ 0.021 $&$ 0.014 $&$ 0.028 $&$ 0.698$\\
$   25  $&$    0.759 \times 10^{-3} $&$ 0.296 $&$ 0.065 $&$ 0.124 $&$ 0.042 $&$ 0.146 $&$ 1.330 $&$ 0.015 $&$ 0.020 $&$ 0.013 $&$ 0.028 $&$ 0.700$\\
$   25  $&$    0.101 \times 10^{-2} $&$ 0.168 $&$ 0.064 $&$ 0.167 $&$ 0.068 $&$ 0.191 $&$ 1.236 $&$ 0.007 $&$ 0.014 $&$ 0.012 $&$ 0.020 $&$ 0.616$\\
\hline 
\end{tabular} 
\end{center} 
\caption{\sl
   \label{tab:flf2} The proton structure functions $F_L$ and
   $F_2$ measured at the given values of $Q^2$ and $x$ without model
   assumptions.  $\Delta_{\rm stat}F_{L}$, $\Delta_{\rm uncor}F_{L}$,
   $\Delta_{\rm cor}F_{L}$ and $\Delta_{\rm tot}F_{L}$ are the
   statistical, uncorrelated systematic, correlated systematic, and
   total uncertainty on $F_L$ respectively.  $\Delta_{\rm stat}F_{2}$,
   $\Delta_{\rm uncor}F_{2}$, $\Delta_{\rm cor}F_{2}$ and $\Delta_{\rm
     tot}F_{2}$ are the statistical, uncorrelated systematic and total
   uncertainty on $F_2$, respectively.  The correlation
   coefficient between the $F_L$ and $F_2$ values ,$\rho$, is also given.}

\end{table}

\begin{table}[htbp] 
\begin{center} 
\tiny 
\begin{tabular}{|r|c|c|r|r|r|r|r|r|r|r|r|r|} 
\hline 
$Q^2$ & $x$ &
$F_{L}$ &
$\Delta_{\rm stat}F_{L}$ &
$\Delta_{\rm uncor}F_{L}$ &
$\Delta_{\rm cor}F_{L}$ &
$\Delta_{\rm tot}F_{L}$ &
$F_{2}$ &
$\Delta_{\rm stat}F_{2}$ &
$\Delta_{\rm uncor}F_{2}$ &
$\Delta_{\rm cor}F_{2}$ &
$\Delta_{\rm tot}F_{2}$ &
$\rho$ \\ 
(GeV$^2$) & & & & & & & & & & & & \\
\hline
$   35 $&$    0.651 \times 10^{-3} $&$ 0.453 $&$ 0.124 $&$ 0.214 $&$ 0.091 $&$ 0.264 $&$ 1.612 $&$ 0.043 $&$ 0.058 $&$ 0.030 $&$ 0.078 $&$ 0.889$\\
$   35 $&$    0.727 \times 10^{-3} $&$ 0.041 $&$ 0.144 $&$ 0.232 $&$ 0.065 $&$ 0.281 $&$ 1.419 $&$ 0.038 $&$ 0.048 $&$ 0.020 $&$ 0.065 $&$ 0.884$\\
$   35 $&$    0.812 \times 10^{-3} $&$ 0.106 $&$ 0.075 $&$ 0.107 $&$ 0.054 $&$ 0.142 $&$ 1.411 $&$ 0.026 $&$ 0.027 $&$ 0.019 $&$ 0.042 $&$ 0.753$\\
$   35 $&$    0.921 \times 10^{-3} $&$ 0.436 $&$ 0.080 $&$ 0.125 $&$ 0.040 $&$ 0.153 $&$ 1.405 $&$ 0.022 $&$ 0.024 $&$ 0.014 $&$ 0.035 $&$ 0.727$\\
$   35 $&$    0.106 \times 10^{-2} $&$ 0.196 $&$ 0.072 $&$ 0.130 $&$ 0.042 $&$ 0.155 $&$ 1.325 $&$ 0.017 $&$ 0.021 $&$ 0.012 $&$ 0.030 $&$ 0.698$\\
$   35 $&$    0.141 \times 10^{-2} $&$ 0.057 $&$ 0.067 $&$ 0.170 $&$ 0.065 $&$ 0.194 $&$ 1.226 $&$ 0.008 $&$ 0.015 $&$ 0.011 $&$ 0.021 $&$ 0.639$\\
$   45 $&$    0.837 \times 10^{-3} $&$ 0.179 $&$ 0.117 $&$ 0.188 $&$ 0.061 $&$ 0.230 $&$ 1.518 $&$ 0.042 $&$ 0.054 $&$ 0.022 $&$ 0.072 $&$ 0.875$\\
$   45 $&$    0.934 \times 10^{-3} $&$ 0.516 $&$ 0.167 $&$ 0.238 $&$ 0.058 $&$ 0.296 $&$ 1.517 $&$ 0.043 $&$ 0.052 $&$ 0.022 $&$ 0.071 $&$ 0.869$\\
$   45 $&$    0.104 \times 10^{-2} $&$ 0.366 $&$ 0.084 $&$ 0.107 $&$ 0.054 $&$ 0.146 $&$ 1.430 $&$ 0.029 $&$ 0.027 $&$ 0.016 $&$ 0.042 $&$ 0.731$\\
$   45 $&$    0.118 \times 10^{-2} $&$ 0.396 $&$ 0.108 $&$ 0.118 $&$ 0.042 $&$ 0.165 $&$ 1.395 $&$ 0.025 $&$ 0.025 $&$ 0.014 $&$ 0.038 $&$ 0.732$\\
$   45 $&$    0.137 \times 10^{-2} $&$ 0.255 $&$ 0.100 $&$ 0.151 $&$ 0.047 $&$ 0.187 $&$ 1.350 $&$ 0.021 $&$ 0.023 $&$ 0.013 $&$ 0.034 $&$ 0.729$\\
$   45 $&$    0.181 \times 10^{-2} $&$ 0.099 $&$ 0.075 $&$ 0.175 $&$ 0.065 $&$ 0.202 $&$ 1.210 $&$ 0.009 $&$ 0.016 $&$ 0.011 $&$ 0.021 $&$ 0.659$\\
$   60 $&$    0.112 \times 10^{-2} $&$ 0.282 $&$ 0.146 $&$ 0.179 $&$ 0.051 $&$ 0.237 $&$ 1.446 $&$ 0.058 $&$ 0.058 $&$ 0.021 $&$ 0.084 $&$ 0.851$\\
$   60 $&$    0.125 \times 10^{-2} $&$ 0.279 $&$ 0.165 $&$ 0.198 $&$ 0.055 $&$ 0.263 $&$ 1.548 $&$ 0.048 $&$ 0.050 $&$ 0.018 $&$ 0.072 $&$ 0.844$\\
$   60 $&$    0.139 \times 10^{-2} $&$ 0.383 $&$ 0.095 $&$ 0.105 $&$ 0.049 $&$ 0.150 $&$ 1.450 $&$ 0.033 $&$ 0.030 $&$ 0.016 $&$ 0.047 $&$ 0.731$\\
$   60 $&$    0.158 \times 10^{-2} $&$ 0.464 $&$ 0.102 $&$ 0.101 $&$ 0.047 $&$ 0.151 $&$ 1.369 $&$ 0.027 $&$ 0.024 $&$ 0.014 $&$ 0.039 $&$ 0.711$\\
$   60 $&$    0.182 \times 10^{-2} $&$ 0.159 $&$ 0.230 $&$ 0.320 $&$ 0.047 $&$ 0.397 $&$ 1.288 $&$ 0.028 $&$ 0.033 $&$ 0.012 $&$ 0.045 $&$ 0.818$\\
$   60 $&$    0.242 \times 10^{-2} $&$-0.044 $&$ 0.094 $&$ 0.185 $&$ 0.069 $&$ 0.218 $&$ 1.186 $&$ 0.011 $&$ 0.016 $&$ 0.011 $&$ 0.023 $&$ 0.683$\\
$   90 $&$    0.187 \times 10^{-2} $&$ 0.041 $&$ 0.222 $&$ 0.207 $&$ 0.045 $&$ 0.307 $&$ 1.330 $&$ 0.095 $&$ 0.077 $&$ 0.017 $&$ 0.123 $&$ 0.862$\\
$   90 $&$    0.209 \times 10^{-2} $&$ 0.060 $&$ 0.109 $&$ 0.119 $&$ 0.041 $&$ 0.166 $&$ 1.313 $&$ 0.051 $&$ 0.045 $&$ 0.014 $&$ 0.069 $&$ 0.801$\\
$   90 $&$    0.237 \times 10^{-2} $&$ 0.007 $&$ 0.109 $&$ 0.101 $&$ 0.040 $&$ 0.154 $&$ 1.218 $&$ 0.037 $&$ 0.029 $&$ 0.012 $&$ 0.048 $&$ 0.769$\\
$   90 $&$    0.273 \times 10^{-2} $&$ 0.447 $&$ 0.143 $&$ 0.135 $&$ 0.048 $&$ 0.202 $&$ 1.325 $&$ 0.031 $&$ 0.027 $&$ 0.013 $&$ 0.043 $&$ 0.717$\\
$   90 $&$    0.362 \times 10^{-2} $&$ 0.163 $&$ 0.167 $&$ 0.233 $&$ 0.058 $&$ 0.293 $&$ 1.145 $&$ 0.015 $&$ 0.017 $&$ 0.011 $&$ 0.025 $&$ 0.699$\\
$  120 $&$    0.220 \times 10^{-2} $&$ 0.070 $&$ 0.067 $&$ 0.241 $&$ 0.041 $&$ 0.253 $&$ 1.400 $&$ 0.019 $&$ 0.060 $&$ 0.027 $&$ 0.069 $&$ 0.908$\\
$  120 $&$    0.250 \times 10^{-2} $&$ 0.450 $&$ 0.096 $&$ 0.252 $&$ 0.037 $&$ 0.272 $&$ 1.414 $&$ 0.022 $&$ 0.051 $&$ 0.025 $&$ 0.061 $&$ 0.875$\\
$  120 $&$    0.280 \times 10^{-2} $&$ 0.136 $&$ 0.094 $&$ 0.103 $&$ 0.036 $&$ 0.144 $&$ 1.299 $&$ 0.022 $&$ 0.025 $&$ 0.021 $&$ 0.039 $&$ 0.711$\\
$  120 $&$    0.320 \times 10^{-2} $&$ 0.073 $&$ 0.175 $&$ 0.342 $&$ 0.032 $&$ 0.385 $&$ 1.129 $&$ 0.103 $&$ 0.146 $&$ 0.018 $&$ 0.179 $&$ 0.963$\\
$  120 $&$    0.360 \times 10^{-2} $&$ 0.480 $&$ 0.178 $&$ 0.234 $&$ 0.039 $&$ 0.296 $&$ 1.245 $&$ 0.062 $&$ 0.063 $&$ 0.012 $&$ 0.089 $&$ 0.886$\\
$  120 $&$    0.480 \times 10^{-2} $&$ 0.152 $&$ 0.166 $&$ 0.212 $&$ 0.071 $&$ 0.278 $&$ 1.069 $&$ 0.024 $&$ 0.024 $&$ 0.010 $&$ 0.035 $&$ 0.762$\\
$  150 $&$    0.280 \times 10^{-2} $&$ 0.123 $&$ 0.083 $&$ 0.270 $&$ 0.046 $&$ 0.286 $&$ 1.357 $&$ 0.024 $&$ 0.065 $&$ 0.025 $&$ 0.074 $&$ 0.931$\\
$  150 $&$    0.310 \times 10^{-2} $&$ 0.306 $&$ 0.099 $&$ 0.290 $&$ 0.035 $&$ 0.308 $&$ 1.330 $&$ 0.022 $&$ 0.056 $&$ 0.023 $&$ 0.065 $&$ 0.904$\\
$  150 $&$    0.350 \times 10^{-2} $&$ 0.038 $&$ 0.085 $&$ 0.127 $&$ 0.038 $&$ 0.157 $&$ 1.274 $&$ 0.017 $&$ 0.027 $&$ 0.021 $&$ 0.038 $&$ 0.728$\\
$  150 $&$    0.390 \times 10^{-2} $&$ 0.401 $&$ 0.095 $&$ 0.124 $&$ 0.029 $&$ 0.159 $&$ 1.266 $&$ 0.015 $&$ 0.022 $&$ 0.020 $&$ 0.033 $&$ 0.675$\\
$  150 $&$    0.450 \times 10^{-2} $&$ 0.554 $&$ 0.114 $&$ 0.153 $&$ 0.038 $&$ 0.195 $&$ 1.209 $&$ 0.014 $&$ 0.023 $&$ 0.020 $&$ 0.034 $&$ 0.668$\\
$  150 $&$    0.600 \times 10^{-2} $&$ 0.137 $&$ 0.118 $&$ 0.194 $&$ 0.060 $&$ 0.235 $&$ 1.069 $&$ 0.009 $&$ 0.020 $&$ 0.018 $&$ 0.028 $&$ 0.677$\\
$  200 $&$    0.370 \times 10^{-2} $&$-0.039 $&$ 0.110 $&$ 0.305 $&$ 0.044 $&$ 0.327 $&$ 1.231 $&$ 0.033 $&$ 0.072 $&$ 0.023 $&$ 0.083 $&$ 0.944$\\
$  200 $&$    0.410 \times 10^{-2} $&$-0.157 $&$ 0.141 $&$ 0.321 $&$ 0.033 $&$ 0.352 $&$ 1.160 $&$ 0.032 $&$ 0.059 $&$ 0.020 $&$ 0.070 $&$ 0.929$\\
$  200 $&$    0.460 \times 10^{-2} $&$ 0.146 $&$ 0.115 $&$ 0.131 $&$ 0.038 $&$ 0.179 $&$ 1.188 $&$ 0.023 $&$ 0.025 $&$ 0.019 $&$ 0.039 $&$ 0.758$\\
$  200 $&$    0.520 \times 10^{-2} $&$ 0.184 $&$ 0.126 $&$ 0.138 $&$ 0.033 $&$ 0.189 $&$ 1.136 $&$ 0.019 $&$ 0.022 $&$ 0.019 $&$ 0.035 $&$ 0.719$\\
$  200 $&$    0.610 \times 10^{-2} $&$ 0.253 $&$ 0.141 $&$ 0.169 $&$ 0.033 $&$ 0.223 $&$ 1.107 $&$ 0.016 $&$ 0.021 $&$ 0.018 $&$ 0.032 $&$ 0.690$\\
$  200 $&$    0.800 \times 10^{-2} $&$ 0.228 $&$ 0.126 $&$ 0.203 $&$ 0.057 $&$ 0.246 $&$ 0.995 $&$ 0.008 $&$ 0.018 $&$ 0.017 $&$ 0.026 $&$ 0.654$\\
$  250 $&$    0.460 \times 10^{-2} $&$ 0.620 $&$ 0.136 $&$ 0.365 $&$ 0.052 $&$ 0.393 $&$ 1.340 $&$ 0.041 $&$ 0.086 $&$ 0.025 $&$ 0.099 $&$ 0.950$\\
$  250 $&$    0.520 \times 10^{-2} $&$ 0.214 $&$ 0.159 $&$ 0.349 $&$ 0.039 $&$ 0.385 $&$ 1.186 $&$ 0.037 $&$ 0.065 $&$ 0.020 $&$ 0.077 $&$ 0.931$\\
$  250 $&$    0.580 \times 10^{-2} $&$ 0.243 $&$ 0.130 $&$ 0.142 $&$ 0.038 $&$ 0.196 $&$ 1.176 $&$ 0.026 $&$ 0.027 $&$ 0.019 $&$ 0.041 $&$ 0.770$\\
$  250 $&$    0.660 \times 10^{-2} $&$ 0.163 $&$ 0.145 $&$ 0.146 $&$ 0.030 $&$ 0.208 $&$ 1.087 $&$ 0.022 $&$ 0.022 $&$ 0.016 $&$ 0.035 $&$ 0.742$\\
$  250 $&$    0.760 \times 10^{-2} $&$ 0.117 $&$ 0.159 $&$ 0.173 $&$ 0.031 $&$ 0.237 $&$ 0.998 $&$ 0.018 $&$ 0.020 $&$ 0.016 $&$ 0.032 $&$ 0.714$\\
$  250 $&$    0.100 \times 10^{-1} $&$ 0.105 $&$ 0.139 $&$ 0.197 $&$ 0.050 $&$ 0.246 $&$ 0.914 $&$ 0.008 $&$ 0.015 $&$ 0.015 $&$ 0.023 $&$ 0.650$\\
$  250 $&$    0.130 \times 10^{-1} $&$ 0.140 $&$ 0.228 $&$ 0.280 $&$ 0.095 $&$ 0.374 $&$ 0.842 $&$ 0.008 $&$ 0.014 $&$ 0.014 $&$ 0.021 $&$ 0.650$\\
$  300 $&$    0.560 \times 10^{-2} $&$-0.038 $&$ 0.161 $&$ 0.316 $&$ 0.041 $&$ 0.357 $&$ 1.138 $&$ 0.048 $&$ 0.075 $&$ 0.021 $&$ 0.091 $&$ 0.942$\\
$  300 $&$    0.690 \times 10^{-2} $&$ 0.345 $&$ 0.151 $&$ 0.149 $&$ 0.039 $&$ 0.216 $&$ 1.118 $&$ 0.030 $&$ 0.028 $&$ 0.019 $&$ 0.045 $&$ 0.781$\\
$  300 $&$    0.790 \times 10^{-2} $&$ 0.377 $&$ 0.168 $&$ 0.148 $&$ 0.027 $&$ 0.225 $&$ 1.058 $&$ 0.025 $&$ 0.022 $&$ 0.016 $&$ 0.037 $&$ 0.752$\\
$  300 $&$    0.910 \times 10^{-2} $&$ 0.349 $&$ 0.193 $&$ 0.176 $&$ 0.028 $&$ 0.263 $&$ 0.967 $&$ 0.021 $&$ 0.019 $&$ 0.015 $&$ 0.033 $&$ 0.734$\\
$  300 $&$    0.121 \times 10^{-1} $&$-0.324 $&$ 0.157 $&$ 0.200 $&$ 0.047 $&$ 0.258 $&$ 0.839 $&$ 0.009 $&$ 0.014 $&$ 0.014 $&$ 0.022 $&$ 0.663$\\
$  400 $&$    0.930 \times 10^{-2} $&$-0.093 $&$ 0.164 $&$ 0.135 $&$ 0.033 $&$ 0.215 $&$ 0.950 $&$ 0.033 $&$ 0.025 $&$ 0.015 $&$ 0.044 $&$ 0.790$\\
$  400 $&$    0.105 \times 10^{-1} $&$-0.199 $&$ 0.180 $&$ 0.131 $&$ 0.020 $&$ 0.223 $&$ 0.923 $&$ 0.028 $&$ 0.019 $&$ 0.014 $&$ 0.037 $&$ 0.760$\\
$  400 $&$    0.121 \times 10^{-1} $&$-0.051 $&$ 0.207 $&$ 0.179 $&$ 0.027 $&$ 0.275 $&$ 0.913 $&$ 0.023 $&$ 0.021 $&$ 0.015 $&$ 0.034 $&$ 0.736$\\
$  400 $&$    0.161 \times 10^{-1} $&$-0.180 $&$ 0.182 $&$ 0.202 $&$ 0.043 $&$ 0.276 $&$ 0.788 $&$ 0.011 $&$ 0.013 $&$ 0.012 $&$ 0.021 $&$ 0.680$\\
$  500 $&$    0.116 \times 10^{-1} $&$-0.255 $&$ 0.184 $&$ 0.119 $&$ 0.028 $&$ 0.221 $&$ 0.868 $&$ 0.037 $&$ 0.022 $&$ 0.014 $&$ 0.046 $&$ 0.790$\\
$  500 $&$    0.131 \times 10^{-1} $&$ 0.340 $&$ 0.207 $&$ 0.143 $&$ 0.019 $&$ 0.252 $&$ 0.946 $&$ 0.033 $&$ 0.022 $&$ 0.015 $&$ 0.042 $&$ 0.755$\\
$  500 $&$    0.152 \times 10^{-1} $&$ 0.279 $&$ 0.244 $&$ 0.149 $&$ 0.021 $&$ 0.287 $&$ 0.860 $&$ 0.028 $&$ 0.017 $&$ 0.012 $&$ 0.035 $&$ 0.742$\\
$  500 $&$    0.201 \times 10^{-1} $&$ 0.192 $&$ 0.214 $&$ 0.211 $&$ 0.043 $&$ 0.304 $&$ 0.770 $&$ 0.013 $&$ 0.014 $&$ 0.013 $&$ 0.023 $&$ 0.691$\\
$  650 $&$    0.151 \times 10^{-1} $&$ 0.229 $&$ 0.219 $&$ 0.145 $&$ 0.018 $&$ 0.263 $&$ 0.917 $&$ 0.043 $&$ 0.026 $&$ 0.013 $&$ 0.052 $&$ 0.804$\\
$  650 $&$    0.171 \times 10^{-1} $&$-0.229 $&$ 0.209 $&$ 0.132 $&$ 0.016 $&$ 0.248 $&$ 0.743 $&$ 0.033 $&$ 0.021 $&$ 0.012 $&$ 0.041 $&$ 0.769$\\
$  650 $&$    0.197 \times 10^{-1} $&$-0.204 $&$ 0.254 $&$ 0.148 $&$ 0.019 $&$ 0.294 $&$ 0.735 $&$ 0.030 $&$ 0.017 $&$ 0.011 $&$ 0.036 $&$ 0.750$\\
$  650 $&$    0.261 \times 10^{-1} $&$ 0.651 $&$ 0.244 $&$ 0.201 $&$ 0.036 $&$ 0.318 $&$ 0.739 $&$ 0.016 $&$ 0.014 $&$ 0.012 $&$ 0.024 $&$ 0.698$\\
$  800 $&$    0.185 \times 10^{-1} $&$ 0.625 $&$ 0.228 $&$ 0.158 $&$ 0.014 $&$ 0.278 $&$ 0.821 $&$ 0.045 $&$ 0.028 $&$ 0.013 $&$ 0.054 $&$ 0.812$\\
$  800 $&$    0.210 \times 10^{-1} $&$ 0.205 $&$ 0.230 $&$ 0.167 $&$ 0.015 $&$ 0.285 $&$ 0.762 $&$ 0.036 $&$ 0.026 $&$ 0.012 $&$ 0.046 $&$ 0.774$\\
$  800 $&$    0.242 \times 10^{-1} $&$ 0.123 $&$ 0.281 $&$ 0.148 $&$ 0.016 $&$ 0.318 $&$ 0.698 $&$ 0.033 $&$ 0.017 $&$ 0.010 $&$ 0.039 $&$ 0.753$\\
$  800 $&$    0.322 \times 10^{-1} $&$ 0.276 $&$ 0.253 $&$ 0.202 $&$ 0.031 $&$ 0.325 $&$ 0.642 $&$ 0.016 $&$ 0.014 $&$ 0.010 $&$ 0.023 $&$ 0.714$\\
\hline 
\end{tabular} 
\end{center} 
\captcont{\sl continued.}
\end{table}

\begin{table}[htbp] 
\begin{center} 
\tiny 
\begin{tabular}{|r|c|c|r|r|r|r|} 
\hline 
$Q^2$ & $x$ &
$F_{L}$ &
$\Delta_{\rm stat}$ &
$\Delta_{\rm uncor}$ &
$\Delta_{\rm cor}$ &
$\Delta_{\rm tot}$ \\ 
(GeV$^2$) & & & & & & \\
\hline
$   1.5 $ &$ 0.279 \times 10^{-4}$ & $ 0.088 $ &$ 0.113 $ &$ 0.186 $ &$ 0.053 $ &$0.224$  \\
$   2.0 $ &$ 0.427 \times 10^{-4}$ & $ 0.127 $ &$ 0.039 $ &$ 0.074 $ &$ 0.044 $ &$0.095$  \\
$   2.5 $ &$ 0.588 \times 10^{-4}$ & $ 0.156 $ &$ 0.025 $ &$ 0.050 $ &$ 0.053 $ &$0.077$  \\
$   3.5 $ &$ 0.877 \times 10^{-4}$ & $ 0.227 $ &$ 0.021 $ &$ 0.049 $ &$ 0.040 $ &$0.067$  \\
$   5.0 $ &$ 0.129 \times 10^{-3}$ & $ 0.314 $ &$ 0.022 $ &$ 0.055 $ &$ 0.045 $ &$0.074$  \\
$   6.5 $ &$ 0.169 \times 10^{-3}$ & $ 0.264 $ &$ 0.023 $ &$ 0.058 $ &$ 0.050 $ &$0.080$  \\
$   8.5 $ &$ 0.224 \times 10^{-3}$ & $ 0.216 $ &$ 0.025 $ &$ 0.062 $ &$ 0.051 $ &$0.084$  \\
$   12  $ &$ 0.319 \times 10^{-3}$ & $ 0.324 $ &$ 0.026 $ &$ 0.051 $ &$ 0.044 $ &$0.072$  \\
$   15  $ &$ 0.402 \times 10^{-3}$ & $ 0.266 $ &$ 0.027 $ &$ 0.051 $ &$ 0.042 $ &$0.071$  \\
$   20  $ &$ 0.540 \times 10^{-3}$ & $ 0.327 $ &$ 0.029 $ &$ 0.053 $ &$ 0.040 $ &$0.072$  \\
$   25  $ &$ 0.687 \times 10^{-3}$ & $ 0.282 $ &$ 0.029 $ &$ 0.061 $ &$ 0.037 $ &$0.077$  \\
$   35  $ &$ 0.958 \times 10^{-3}$ & $ 0.213 $ &$ 0.035 $ &$ 0.059 $ &$ 0.040 $ &$0.080$  \\
$   45  $ &$ 0.121 \times 10^{-2}$ & $ 0.303 $ &$ 0.043 $ &$ 0.060 $ &$ 0.044 $ &$0.086$  \\
$   60  $ &$ 0.157 \times 10^{-2}$ & $ 0.315 $ &$ 0.051 $ &$ 0.060 $ &$ 0.044 $ &$0.090$  \\
$   90  $ &$ 0.243 \times 10^{-2}$ & $ 0.125 $ &$ 0.061 $ &$ 0.062 $ &$ 0.039 $ &$0.095$  \\
$  120  $ &$ 0.303 \times 10^{-2}$ & $ 0.198 $ &$ 0.054 $ &$ 0.077 $ &$ 0.029 $ &$0.098$  \\
$  150  $ &$ 0.402 \times 10^{-2}$ & $ 0.264 $ &$ 0.044 $ &$ 0.068 $ &$ 0.035 $ &$0.088$  \\
$  200  $ &$ 0.541 \times 10^{-2}$ & $ 0.150 $ &$ 0.056 $ &$ 0.073 $ &$ 0.034 $ &$0.099$  \\
$  250  $ &$ 0.736 \times 10^{-2}$ & $ 0.196 $ &$ 0.061 $ &$ 0.075 $ &$ 0.033 $ &$0.102$  \\
$  346  $ &$ 0.986 \times 10^{-2}$ & $ 0.039 $ &$ 0.059 $ &$ 0.057 $ &$ 0.029 $ &$0.087$  \\
$  636  $ &$ 0.184 \times 10^{-1}$ & $ 0.152 $ &$ 0.066 $ &$ 0.045 $ &$ 0.020 $ &$0.082$  \\

\hline 
\end{tabular} 
\caption{\sl The proton structure function $F_L(x,Q^2)$ 
obtained by averaging $F_L$ data from table~\ref{tab:flf2}
at the given values of $Q^2$ and $x$.
$\Delta_{\rm stat}$, $\Delta_{\rm uncor}$,
$\Delta_{\rm cor}$ and $\Delta_{\rm tot}$ 
are the statistical, uncorrelated systematic, correlated systematic, and total uncertainty on $F_L$,
respectively.}
\label{tab:flave}
\end{center} 
\end{table}

\newpage
\begin{figure}[htbp]
\begin{center}
\subfigure[\label{fig:contBulk460}]{\includegraphics[width=\columnwidth]{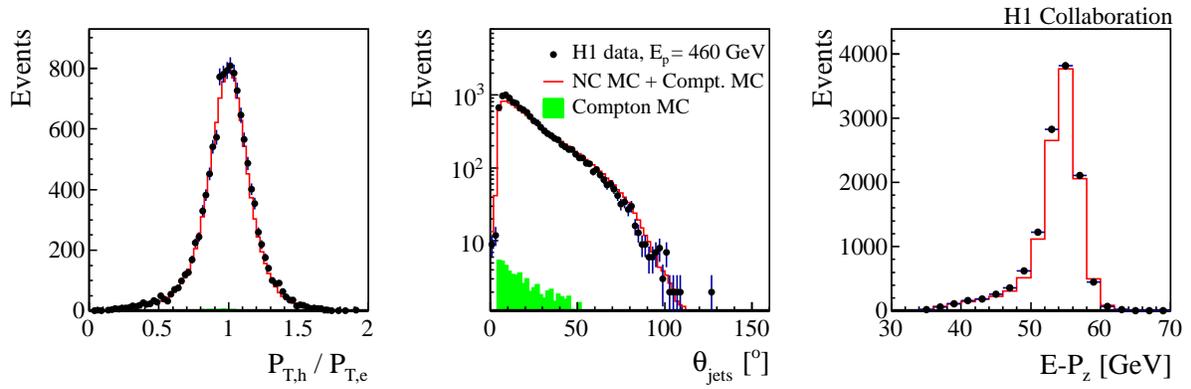} }
\subfigure[\label{fig:contBulk575}]{\includegraphics[width=\columnwidth]{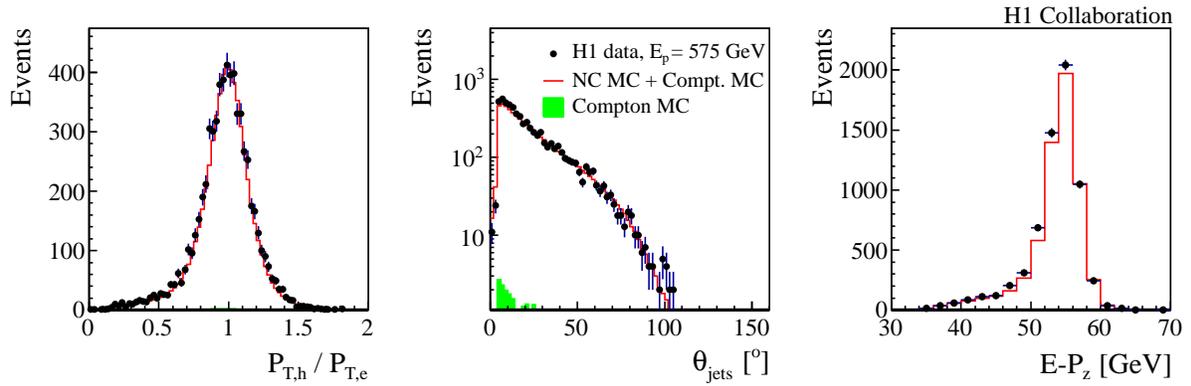} }
\end{center}
\caption{\sl Distributions of $P_{\rm T,h}/P_{\rm T,e}$, $\theta_{\rm
    jets}$ and $E-P_z$ for (a) $E_p=460$~GeV and (b) $E_p=575$~GeV for
  $y<0.19$ data (solid points) and simulation and estimated background
  (histograms)
  normalised to the integrated luminosity of the data. The estimated
  QED Compton background contribution is shown as shaded
  histogram.}
\label{fig:contBulk}
\end{figure}

\begin{figure}[htbp]
\begin{center}
\subfigure[\label{fig:contLoE460}]{\includegraphics[width=\columnwidth]{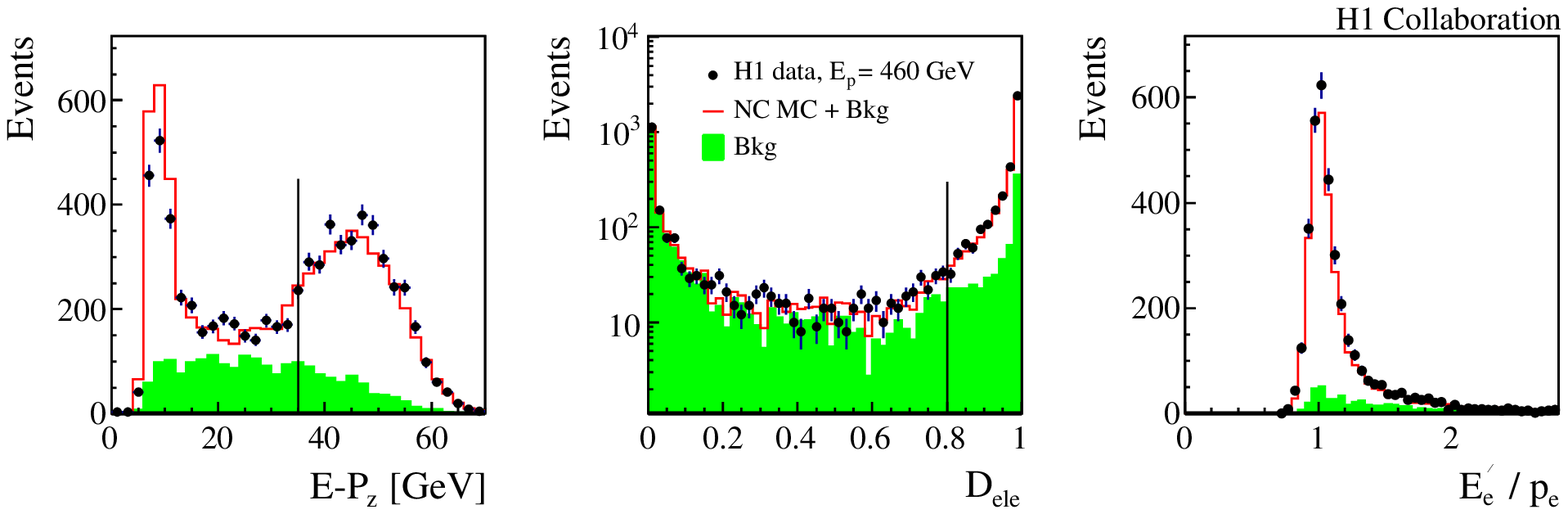} }
\subfigure[\label{fig:contLoE575}]{\includegraphics[width=\columnwidth]{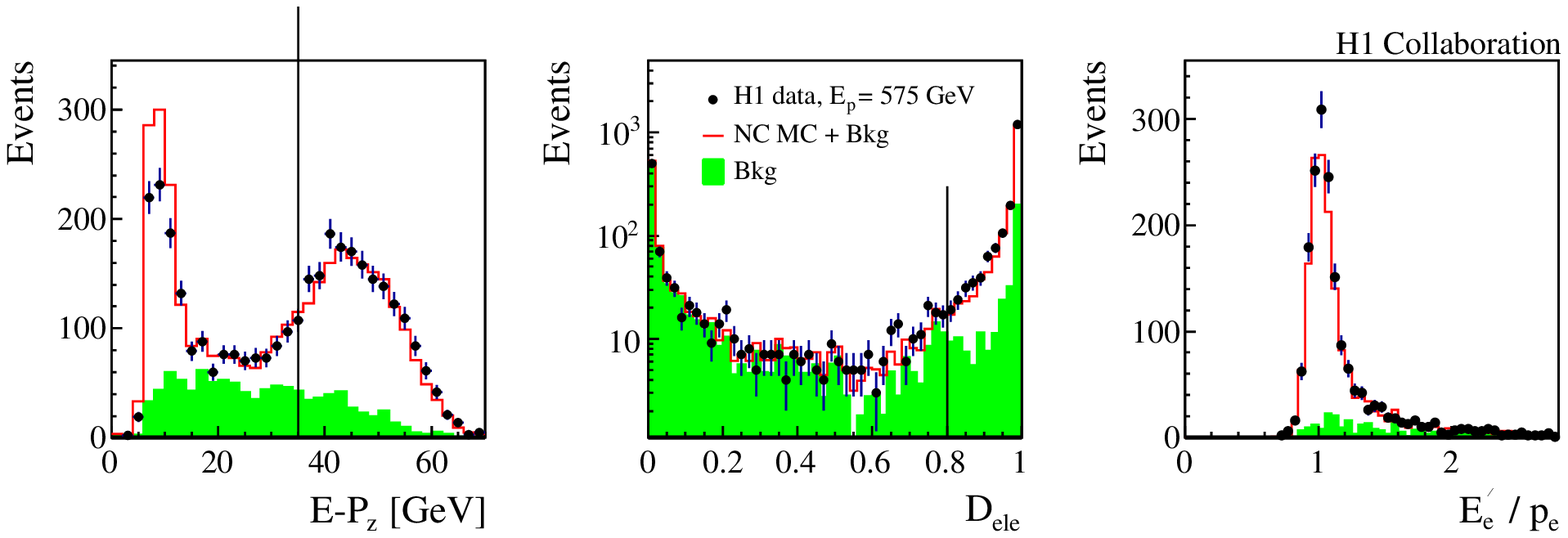} }
\end{center}
\caption{\sl Distributions of $E-P_z$, $D_{ele}$, and $E^\prime_e/p_e$
  for the sample of events with $E^\prime_e<6$~GeV. The selection
  requirements on $E-P_z$ and $D_{ele}$ are shown as vertical lines with all
  other selection criteria applied. The
  distributions are shown for (a) $E_p=460$~GeV and (b) $E_p=575$~GeV
  for data (solid points) and simulation and estimated background (histograms) normalised to
  the integrated luminosity of the data. The estimated background is
  shown as shaded histogram and includes the photoproduction
  contribution estimated using wrong charge scattered lepton
  candidates as well as the QED Compton contribution.}
\label{fig:contLowE}
\end{figure}

\begin{figure}[htbp]
\begin{center}
\subfigure[\label{fig:contHiY460}]{\includegraphics[width=\columnwidth]{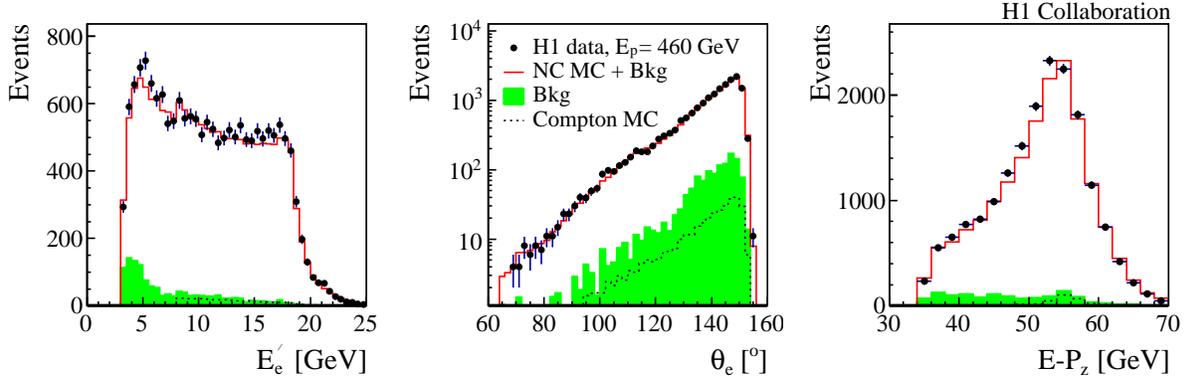} }
\subfigure[\label{fig:contHiY575}]{\includegraphics[width=\columnwidth]{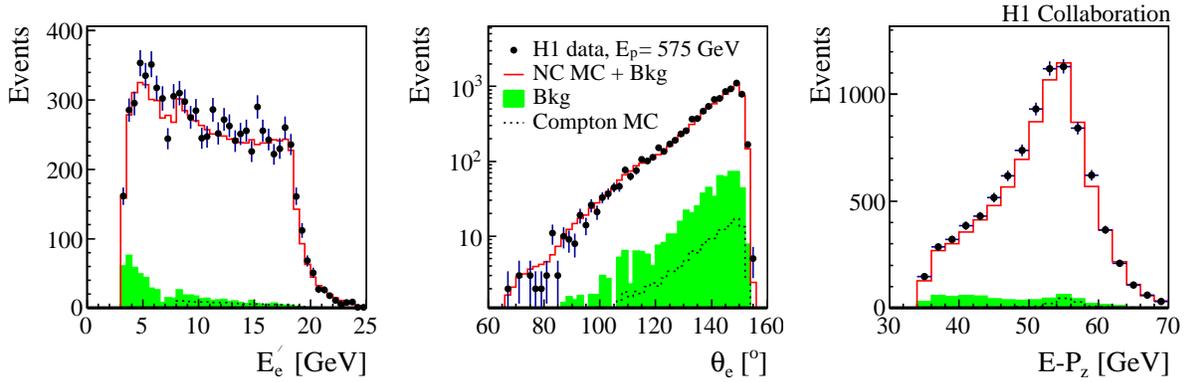} }
\end{center}
\caption{\sl Distributions of $E^\prime_e$, $\theta_e$ and $E-P_z$ for
  (a) $E_p=460$~GeV and (b) $E_p=575$~GeV for $high~y$ data (solid
  points) and simulation and estimated background (histograms) normalised to the integrated
  luminosity of the data. The estimated background is shown as
  shaded histogram and includes the photoproduction contribution estimated using wrong charge
  scattered lepton candidates and the QED Compton contribution (dashed
  line).}
\label{fig:contHiY}
\end{figure}

\begin{figure}[htbp]
\begin{center}
\includegraphics[width=\columnwidth]{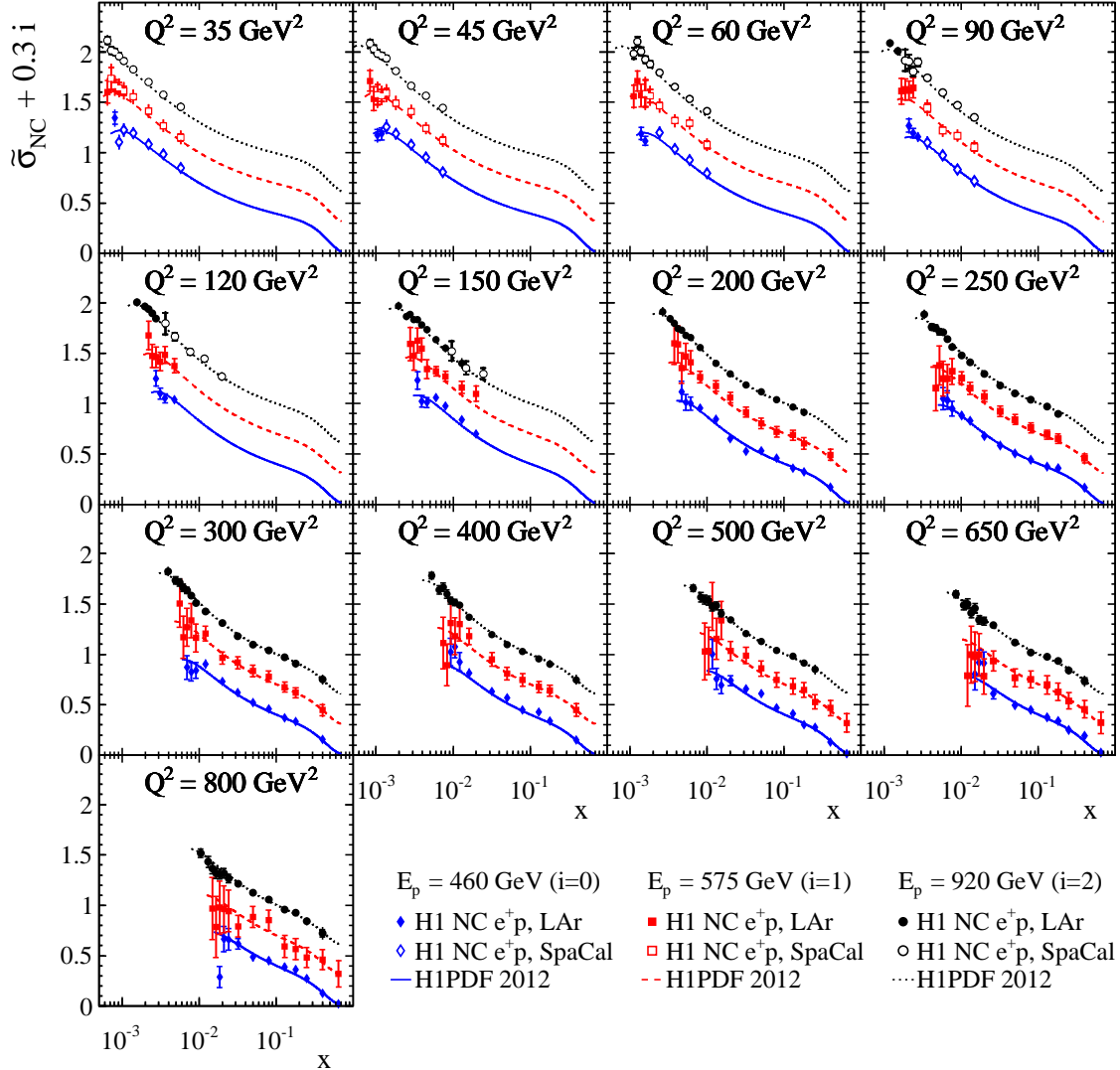}
\end{center}
\caption{\sl The reduced cross section $\tilde{\sigma}_{NC}(x,Q^2)+0.3i$
  measured at three proton beam energies $E_p=460$~GeV (diamonds, i=0),
  $575$~GeV (squares, i=1) and $920$~GeV (circles, i=2). 
  The previously published H1 SpaCal data are shown by the open symbols.
  The solid symbols are the H1 LAr data. The new measurements
  reported here correspond to the filled diamonds
  and squares. The inner
  error bars represent the statistical errors, the full error bars include
  the statistical and systematic uncertainties added in quadrature,
  excluding the normalisation uncertainty. The curves represent the
  prediction from the H1PDF2012 NLO QCD fit.}
\label{fig:xsec-xq2}
\end{figure}

\begin{figure}[htbp]
\begin{center}
\includegraphics[width=\columnwidth]{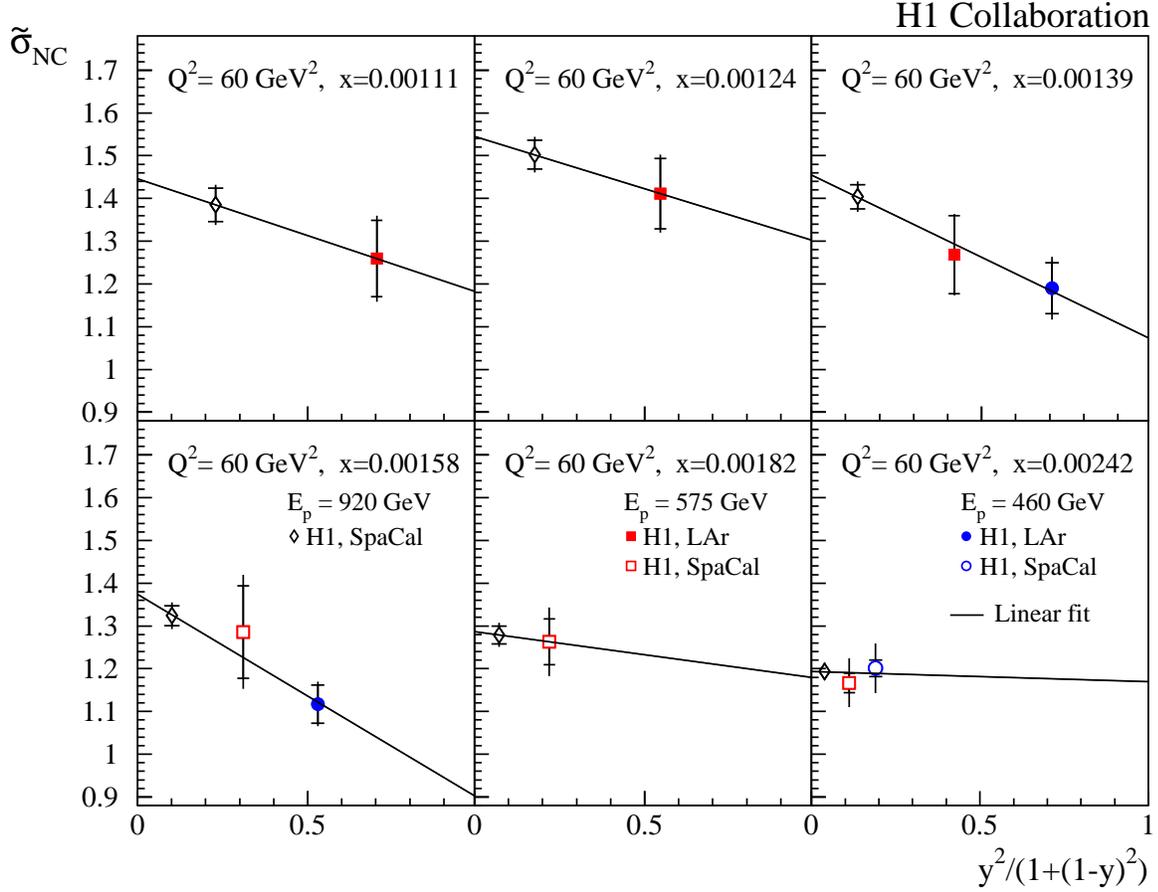}
\end{center}
\caption{\sl The reduced cross section $\tilde{\sigma}_{NC}(x,Q^2)$ as
  a function of $y^2/(1 + (1-y)^2)$ for six values of $x$ at $Q^2 =
  60$~GeV$^2$, measured for proton beam energies of $E_p = 920,~575$
  and $460$~GeV. The inner error bars denote the statistical error,
  the outer error bars show statistical and systematic uncertainties
  added in quadrature. The luminosity uncertainty is not included in
  the error bars. The negative slopes of the linear fits (solid line)
  which were performed using total errors, illustrate the
  non-vanishing values of the structure function $F_L(x,Q^2)$.}
\label{fig:rosenbluth}
\end{figure}

\begin{figure}[htbp]
\begin{center}
\includegraphics[width=\columnwidth]{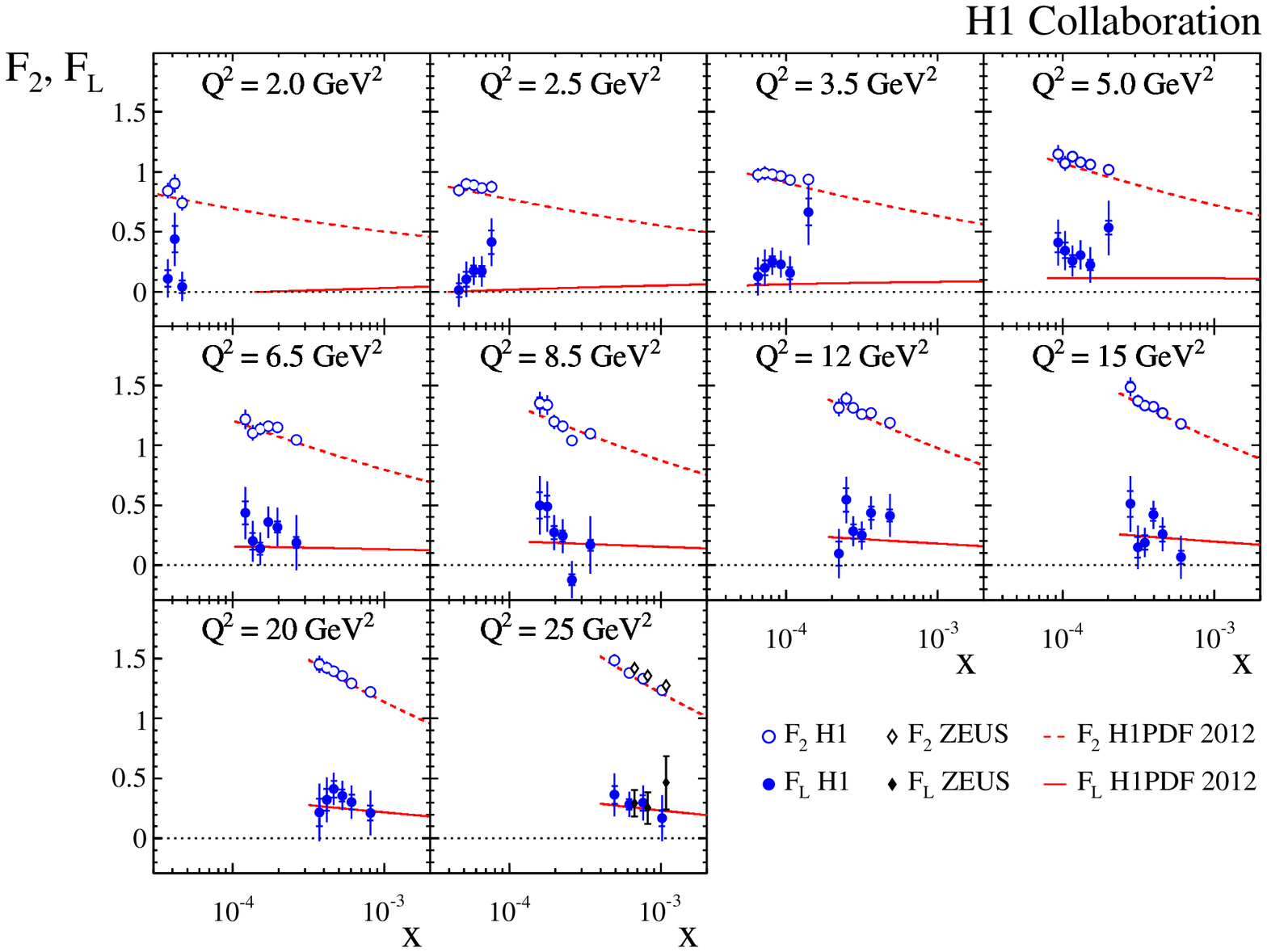}
\end{center}
\caption{\sl The proton structure functions $F_L(x,Q^2)$ (solid
  symbols) and $F_2(x,Q^2)$ (open symbols) measured by H1 (circles)
  and ZEUS (diamonds) in the region $2\leq Q^2 \leq 25$~GeV$^2$. \kd{ Only
  the $F_2(x,Q^2)$ measurements obtained in the determinations of $F_L$ by
  H1 and ZEUS are shown.} The inner error bars represent the
  statistical uncertainties, the full error bars include the
  statistical and systematic uncertainties added in quadrature,
  including all correlated and uncorrelated uncertainties. The curves
  represent the prediction from the H1PDF2012 NLO QCD fit.}
\label{fig:FLxq2-a}
\end{figure}


\begin{figure}[htbp]
\begin{center}
\includegraphics[width=\columnwidth]{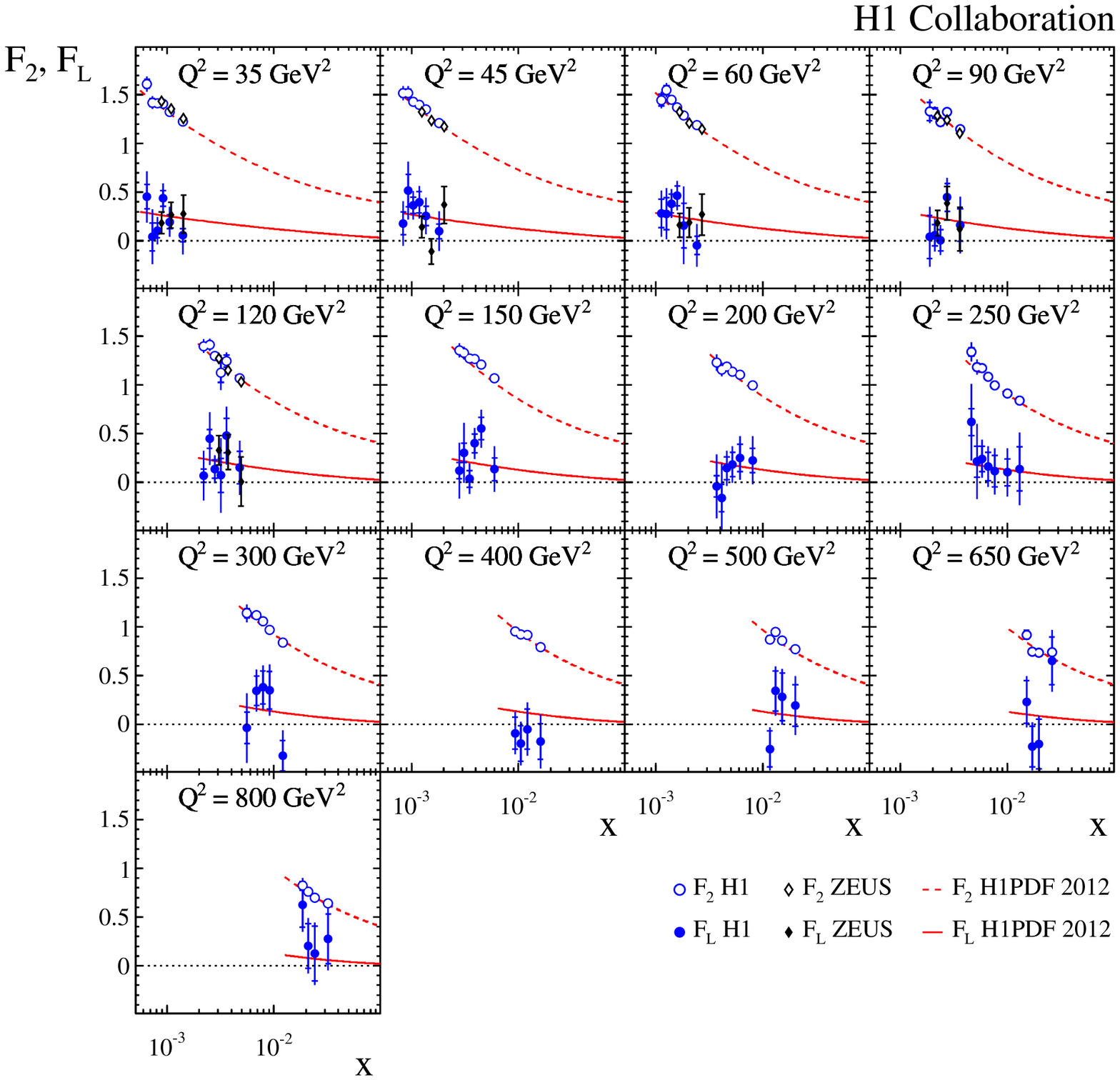}
\end{center}
\caption{\sl The proton structure functions $F_L(x,Q^2)$ (solid
  symbols) and $F_2(x,Q^2)$ (open symbols) measured by H1 (circles)
  and ZEUS (diamonds) in the region $35\leq Q^2 \leq
  800$~GeV$^2$. \kd{ Only the $F_2(x,Q^2)$ measurements obtained in the
  determinations of $F_L$ by H1 and ZEUS are shown.}  The inner error
  bars represent the statistical uncertainties, the full error bars
  include the statistical and systematic uncertainties added in
  quadrature, including all correlated and uncorrelated
  uncertainties. The curves represent the prediction from the
  H1PDF2012 NLO QCD fit. }
\label{fig:FLxq2-b}
\end{figure}

\begin{figure}[htbp]
\begin{center}
\includegraphics[width=\columnwidth]{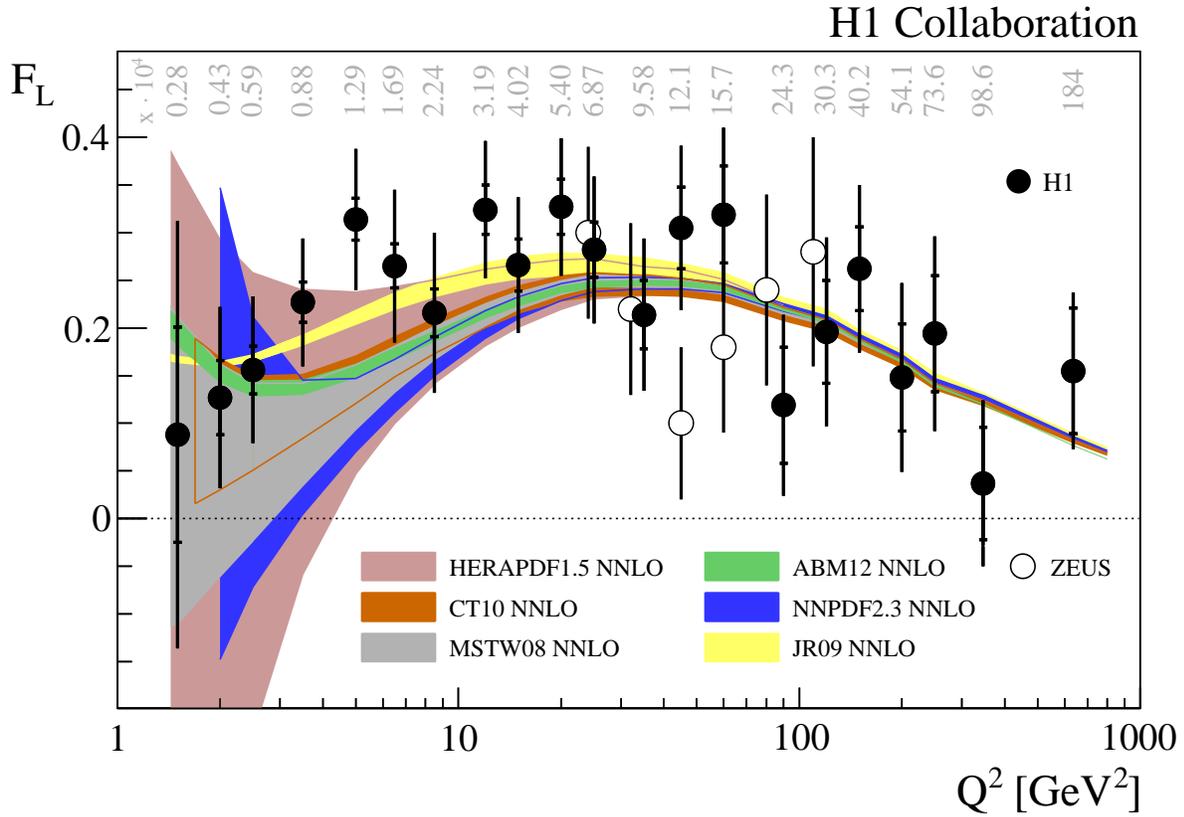}
\end{center}
\caption{\sl The proton structure function $F_L$ averaged over $x$ at
  different $Q^2$ (solid points). The average value of $x$ for each $Q^2$
  is given above each data point.  The inner error bars represent
  the statistical uncertainties, the full error bars include the
  statistical and systematic uncertainties added in quadrature,
  including all correlated and uncorrelated uncertainties. The $F_L$
  measurements by ZEUS are also shown (open points).  The data are
  compared to NNLO predictions from a selection of PDF sets as
  indicated.}
\label{fig:FLq2}
\end{figure}


\begin{figure}[htbp]
\begin{center}
\includegraphics[width=\columnwidth]{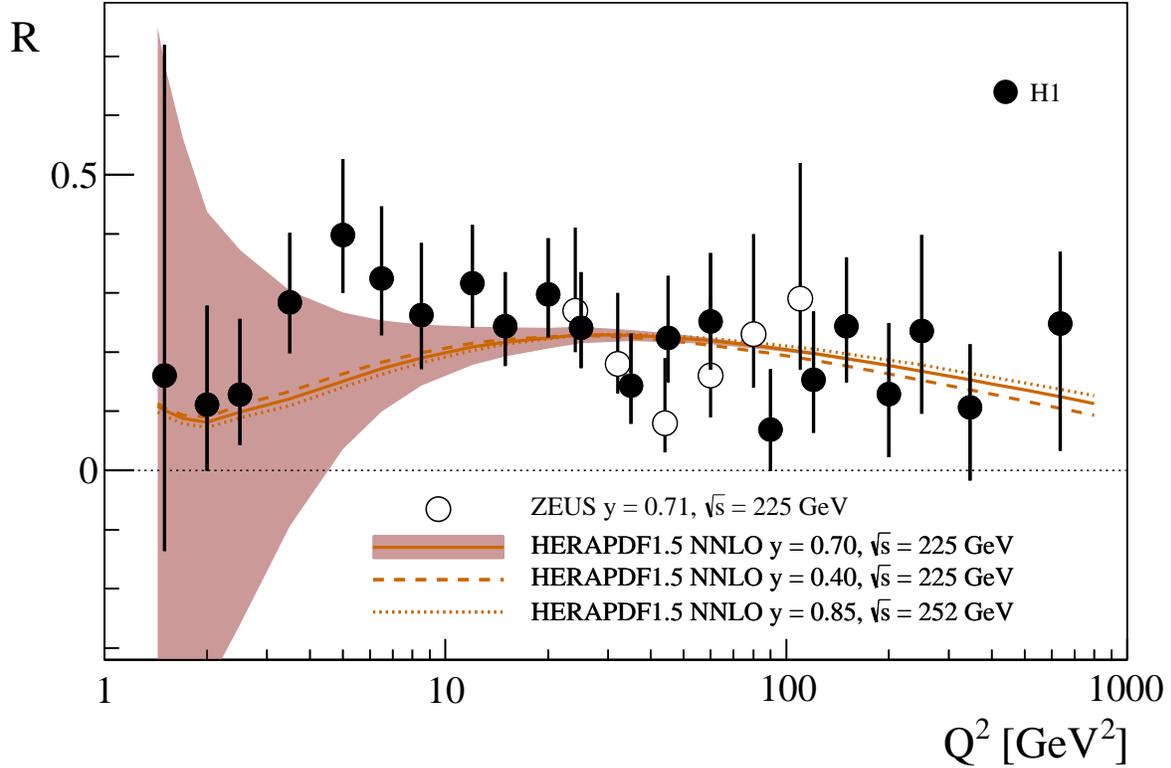}
\end{center}
\caption{\sl The ratio $R(Q^2)$ averaged over $x$ in the region
  $1.5\leq Q^2 \leq 800$~GeV$^2$ (solid points).  The error bars represent the full errors as obtained by the Monte Carlo procedure described in the text. 
  The ZEUS
  data are also shown (open symbols). The ZEUS data point at $Q^2=45$~GeV$^2$ is slightly shifted for better visibility of the erros. The solid curve represents the
  prediction from the HERAPDF1.5 NNLO QCD fit and its uncertainty for $\sqrt{s}=225$~Gev$^2$ and $y=0.7$. The
  additional dashed and dotted curves show the variations of $R$ in
  the region of $x$ where the data are sensitive to this quantity.
}
\label{fig:Rq2}
\end{figure}

\begin{figure}[htbp]
\begin{center}
\includegraphics[width=\columnwidth]{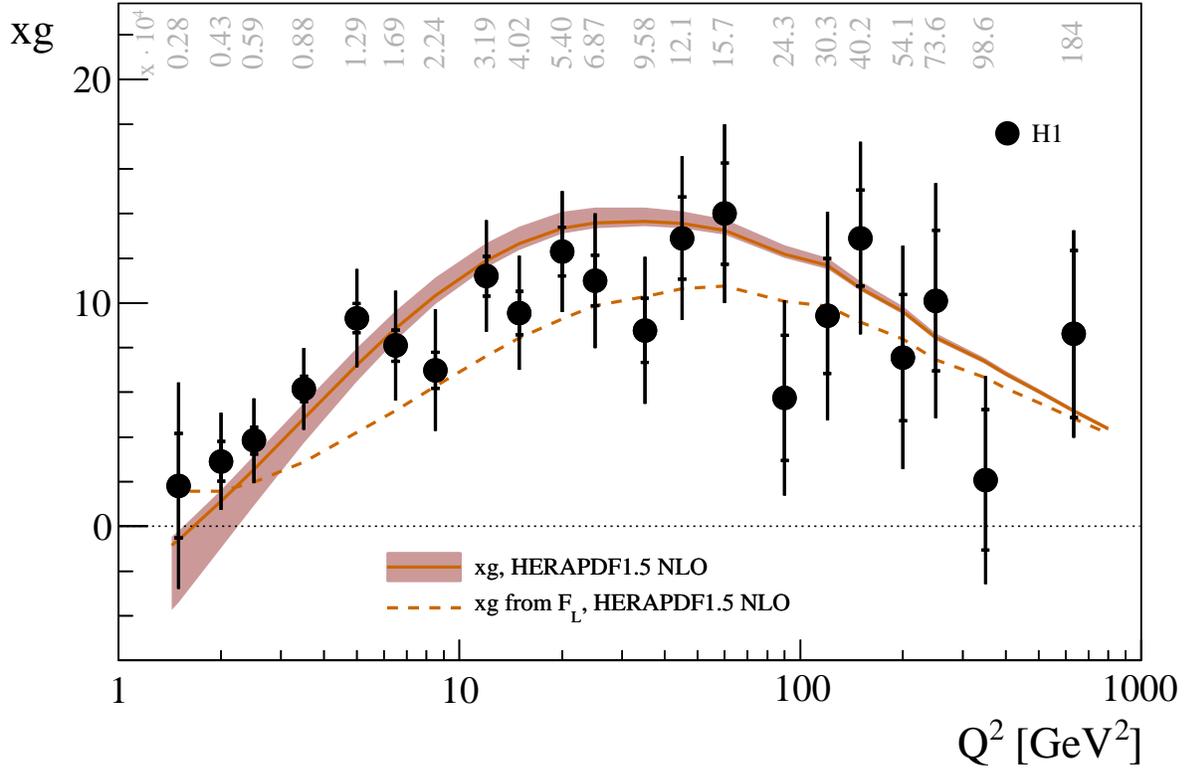}
\end{center}
\caption{\sl The gluon density $xg(x,Q^2)$ averaged over $x$ in the
  region $1.5\leq Q^2 \leq 800$~GeV$^2$ (solid points). The average
  value of $x$ for each $Q^2$ is given above each data point.  The
  inner error bars represent the statistical uncertainties, the full
  error bars include the statistical and systematic uncertainties
  added in quadrature, including all correlated and uncorrelated
  uncertainties. The shaded regions represent the prediction from the
  HERAPDF1.5 NLO QCD fit. 
  The dashed line corresponds to $xg$ as obtained by applying equation \ref{eq:xgapprox} to the $F_L$ prediction based on the HERAPDF1.5 NLO QCD fit.  }
\label{fig:gluon}
\end{figure}

\end{document}